\documentclass{article}
\usepackage{ijcai24}

\usepackage{times}
\usepackage{soul}
\usepackage{url}
\usepackage[hidelinks]{hyperref}
\usepackage[utf8]{inputenc}
\usepackage[small]{caption}
\usepackage{graphicx}
\usepackage{amsmath}
\usepackage{amsthm}
\usepackage{booktabs}
\usepackage{algorithm}
\usepackage{algorithmic}
\usepackage[switch]{lineno}

\usepackage{import}
\usepackage{ijcai24_preamble}

\title{Imperfect-Recall Games: Equilibrium Concepts and Their Complexity\footnote{Published in the Proceedings of the Thirty-Third International Joint Conference on Artificial Intelligence (IJCAI 2024).}}

\iftrue
\author{
Emanuel Tewolde$^{1,4}$
\and
Brian Hu Zhang$^1$
\and
Caspar Oesterheld$^{1,4}$
\and
\\
Manolis Zampetakis$^2$
\and
Tuomas Sandholm$^{1,5,6,7}$
\and
Paul W. Goldberg$^{3}$
\And
Vincent Conitzer$^{1,3,4}$
\affiliations
$^1$Carnegie Mellon University\\ 
$^2$Yale University\\ 
$^3$University of Oxford\\  
$^4$Foundations of Cooperative AI Lab (FOCAL)\\
$^5$Strategic Machine, Inc., 
$^6$Strategy Robot, Inc., 
$^7$Optimized Markets, Inc.
\emails
emanueltewolde@cmu.edu,
bhzhang@cs.cmu.edu,
oesterheld@cmu.edu,
manolis.zampetakis@yale.edu,
sandholm@cs.cmu.edu,
paul.goldberg@cs.ox.ac.uk,
conitzer@cs.cmu.edu
}
\fi

\begin{document}

\maketitle

\begin{abstract}
    We investigate optimal decision making under imperfect recall, that is, when an agent forgets information it once held before. An example is the absentminded driver game, as well as team games in which the members have limited communication capabilities. In the framework of extensive-form games with imperfect recall, we analyze the computational complexities of finding equilibria in multiplayer settings across three different solution concepts: Nash, \emph{multiselves} based on evidential decision theory (EDT), and multiselves based on causal decision theory (CDT). We are interested in both exact and approximate solution computation. As special cases, we consider (1) single-player games, (2) two-player zero-sum games and relationships to maximin values, and (3) games without exogenous stochasticity (chance nodes).
    We relate these problems to the complexity classes \P, \PPAD, \PLS, $\Sigma_2^\P$, $\exists \R$, and $\exists \forall \R$.  
\end{abstract}

\section{Introduction}
\label{sec:intro}

In game theory, it is common to restrict attention to games of {\em perfect recall}, that is, games in which no player ever forgets anything.  At first, it seems that this assumption is even better motivated for AI agents than for human agents: humans %tend to
forget things, but AI does not have to.  However, we argue this view is mistaken: there are often reasons to design AI agents to forget, or to structure them so that they can be modeled as forgetful.  Moreover, such forgetting-by-design follows predictable rules and is thereby easier to model formally than idiosyncratic human forgetting.  Thus, games of imperfect recall are receiving renewed attention from AI researchers.  

Imperfect recall is already being used for state-of-the-art \emph{abstraction} algorithms for larger games of perfect recall~\cite{Waugh09:Practical,GanzfriedS14,BrownGS15}. 
The idea is that by forgetting unimportant aspects of the past, the AI can afford to conduct equilibrium-approximation computations with a game model that has a more refined abstraction of the present. Indeed, imperfect-recall abstractions were a key component in the first superhuman AIs in no-limit Texas hold'em poker~\cite{BrownS18,BrownS19}.

\begin{figure} 
    \tikzset{
        every path/.style={-},
        every node/.style={draw},
    }
    \forestset{
    subgame/.style={regular polygon,
    regular polygon sides=3,anchor=north, inner sep=1pt},
    }

    \begin{minipage}{.6\linewidth}
     \centering
     \begin{subfigure}{\textwidth}
        \begin{forest}
            [,p1,name=p0
            [,p1,name=p1a
                [,p2,name=p2a
                    [\util1{-1},terminal,name=t1] [\util1{3},terminal]]
                    [,p2 [\util1{-1},terminal] [\util1{-1},terminal]
                ]
            ]
            [,p1,name=p1b
                [,p2
                    [\util1{-1},terminal] [\util1{-1},terminal]]
                    [,p2,name=p2b [\util1{3},terminal] [\util1{-1},terminal,name=t8]
                ]
            ]
            ]
            \node[above=0pt of p0,draw=none,p1color]{$I_1$};
            \draw[infoset1] (p1a) to node[below,draw=none,p1color,]{$I_{2}$} (p1b);
            \draw[infoset2] (p2a) -- (p2b);
        \end{forest}
     \caption{Forgetful penalty shoot-out. This game has no Nash equilibrium.} 
     \label{fig:forgetting kicker game}
    \end{subfigure}
    \end{minipage}
    \begin{minipage}{.05\linewidth}
    \,
    \end{minipage}
    \begin{minipage}{.3\linewidth}
      \centering
      \begin{subfigure}{\textwidth}
        \begin{forest}
            [,p1,name=p0,s sep=25pt,l sep=21pt
                [\util1{0},terminal,el={1}{e}{},yshift=-3.3pt]
                [,p1,name=p1b,el={1}{c}{},s sep=25pt,l sep=21pt
                    [\util1{0},terminal,el={1}{e}{},yshift=-3.3pt]
                    [,p1,name=p1c,el={1}{c}{},s sep=25pt,l sep=21pt
                        [\util1{6},terminal, el={1}{e}{},yshift=-3.3pt]
                        [\util1{0},terminal, el={1}{c}{},yshift=-3.3pt]
                    ]
                ]
            ]
            \draw[infoset1] (p0) to [bend left=90] (p1b) to[bend left=90] (p1c);
        \end{forest}
      \caption{Extended absentminded driver.}
      \label{fig:ext absentminded driver}
    \end{subfigure}
    \end{minipage}
    \caption{Games with imperfect recall. P1's (\pone) utility payoffs are labeled on each terminal node. If P2 (\ptwo) is present, the game is zero sum. Infosets are joined by dotted lines.}
    \label{fig:intro games}
\end{figure}
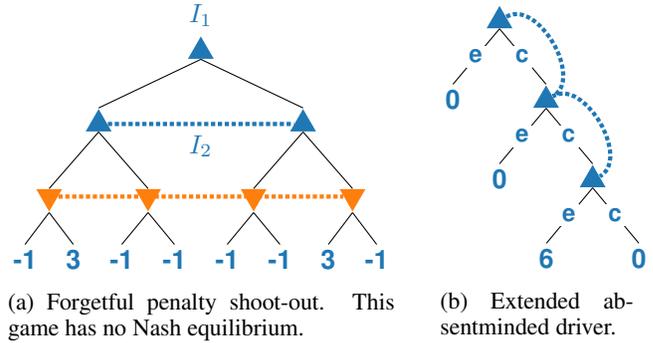

Imperfect recall also naturally models settings in which forgetting is deliberate for other reasons, such as privacy of sensitive data \cite{Conitzer19:Designing,Zhang22:Polynomial}. \citeauthor{Conitzer19:Designing} provides the example of an AI driving assistant designed to intervene whenever the human car driver makes a significant error. In such instances, the AI must assess the overall skill level of the human driver, despite not being allowed to store information about the individual.

\begin{table*}[ht]
\definecolor{newcolor}{RGB}{192,255,192}
\centering
\renewcommand\arraystretch{1.1}
\scalebox{0.90}{
    \begin{tabular}{|c|c|c|c|}
    \multicolumn{4}{c}{\bf Multi-player} \\\hline
& \bf Nash (D) & \bf EDT (D) & \bf CDT (S) \\\hline
\multirow{2}{*}{\bf exact} & \multicolumn{2}{c|}{\cellcolor{newcolor} $\exists \mathbb{R}$-hard and in $\exists \forall \mathbb{R}$} & \multirow{2}{*}{---}\\
&  \multicolumn{2}{c|}{\cellcolor{newcolor} (Thms.~\ref{thm:Nash-D etr hard} \& \ref{thm:EDT exact ETR hard})} & \\\hline
\multirow{2}{*}{\bf 1/exp} & \multicolumn{2}{c|}{\cellcolor{newcolor}} & \cellcolor{newcolor} \\
& \multicolumn{2}{c|}{\cellcolor{newcolor}} & \cellcolor{newcolor}  \\\cline{1-1}
\bf 1/poly & \multicolumn{2}{c|}{\cellcolor{newcolor}\multirow{-3}{*}{\makecell{$\Sigma_2^\P{}$-complete \\ (Thms.~\ref{th:apx-nash-hard} and~\ref{thm:apx EDT sigma2p compl})}}} &  \cellcolor{newcolor} \multirow{-3}{*}{\makecell{\PPAD{}-complete \\ (Thm.~\ref{thm:CDT is PPAD})}} \\\hline
\end{tabular}
\begin{tabular}{|c|c|c|c|}
   \multicolumn{4}{c}{\bf Single-player} \\\hline
& \bf Optimal (D) & \bf EDT (S) & \bf CDT (S) \\\hline
\multirow{2}{*}{\bf exact} & $\exists \mathbb{R}$-complete&  \multirow{2}{*}{---} & \multirow{2}{*}{---} \\
& \cite{Gimbert20} & & \\\hline
\multirow{2}{*}{\bf 1/exp} & \NP{}-complete  & \cellcolor{newcolor}\PLS{}-complete & \CLS{}-complete\\
& {[}\citealp{KollerM92}; & \cellcolor{newcolor}(Thm.~\ref{thm:EDT PLS-C}$^*$) & \cite{TewoldeOCG23} 
 \\\hhline{-|~|-|-}
\bf 1/poly & \citealp{TewoldeOCG23}{]} & \multicolumn{1}{c|}{\cellcolor{newcolor}\P{} (Cor.~\ref{cor:EDT SPIR inv poly in P}$^*$)} & \cellcolor{newcolor}\P{} (Cor.~\ref{cor:1PL CDT has FPTAS}) \\\hline
\end{tabular}
}
\caption{Summary of complexity results. New results from this paper are shown with a light green background. (S) stands for search problem, which is when we ask for a solution strategy profile. In multi-player, (D) stands for deciding whether such an equilibrium even exists. In single-player, Optimal (D) decides whether some target utility can be achieved. Citations are given for results found in the literature. All of the hardness results even hold for highly restricted game instances, such as, \eg, for games with no chance nodes or two-player zero-sum games where one player has perfect recall. $^*$: The number of actions per infoset is required to be constant for the membership result. `---': No results exist for these settings to our knowledge. Also note the technical complication that arises here from the fact that there exist single-player games in which every exact EDT or CDT equilibrium involves irrational values~\protect\cite{TewoldeOCG23}. 
}
\label{table:result summary}
\end{table*}

It can also model {\em teams} of agents with common goals and limited ability to communicate. Each team, represented by one agent with imperfect recall, is then striving for some notion of optimality among team members~\cite{vonStengel97:Team,Celli18:Computational,EmmonsOCC022,Zhang23:Team_DAG}. 
Highly distributed agents are similarly well-described by imperfect recall: such an agent may take an action at one node based on information at that node, and then need to take another action at a second node without yet having learned yet what happened at the first node. Thus, effectively, the distributed agent has forgotten what it knew before. Finally, a single agent can be instantiated multiple times in the same environment, where one copy does not know what another copy just knew \cite{conitzer23:focal}. For example, we might want to test goal-oriented AI agents in simulation to ensure that they will later act in a trustworthy fashion in the real world \cite{KovarikOC23,KovarikOC24}. Then, the AI agent will have to act in the real world without knowing how it acted in simulation.

Perfect recall is a common technical assumption in game theory because it implies many simplifying properties, such as polynomial-time solvability of single-player and two-player zero-sum settings \cite{KollerM92}. In multi-player settings with \textit{imperfect} recall, Nash equilibria may not exist anymore \cite{Wichardt08}; in fact, we %will
show that deciding existence is computationally hard. To give an illustrative running example, consider a variation of \citeauthor{Wichardt08}'s game in \Cref{fig:forgetting kicker game}, which we call the forgetful (soccer) penalty shoot-out. The shooter (P1) decides whether to shoot left or right, once before the whistle, and once again right before kicking the ball. At the second decision point, P1 has forgotten which direction they chose previously.
P1 only succeeds in shooting in any direction if she chooses that direction at both decision points. Upon succeeding, it becomes a matching pennies game with the goalkeeper (P2) who chooses to jump left or right to block the ball. A similar analysis to the one of matching pennies implies that in a potential Nash equilibrium, none of the two players can play one side more often than the other. However, both players randomizing $50/50$ at each infoset is not a Nash equilibrium either: P1 is not best responding to P2 because she could instead deterministically shoot towards one side to avoid miscoordination with herself altogether which would achieve a payoff of $1$ instead of $0$.

Indeed, many of our intuitions fail for imperfect-recall games -- to the point that a significant body of work in philosophy and game theory addresses conceptual questions about probabilistic reasoning and decision making in imperfect-recall games, such as in the Sleeping Beauty problem~\cite{Elga00:Self} or the absentminded driver game of \Cref{fig:ext absentminded driver} \cite{PiccioneR73}. From this literature, several distinct and coherent ways to approach games of imperfect recall have emerged. We will discuss these in detail in \Cref{sec: intro MS eqs}.

In this paper, we study the computational complexity of solving imperfect-recall extensive-form games. We focus on three solution concepts: (1) Nash equilibria where players play mutual best response strategies (or simply optimal strategies in single-player domains), (2) multiselves equilibria based on evidential decision theory, in which each infoset plays a best-response action to all other infosets and players, and (3) multiselves equilibria based on causal decision theory, in which each infoset plays a \textit{Karush-Kuhn-Tucker (KKT)} point action for the current strategy profile. The latter two are relaxations of the first. \Cref{sec:ir games,sec: intro MS eqs} cover preliminaries on imperfect-recall games and on multiselves equilibria, respectively. \Cref{sec: NEs,sec:mse main} analyze the computation of Nash equilibria and of multiselves equilibria, respectively, in various setting. Our complexity results for these are summarized in \Cref{table:result summary}. Last but not least, \Cref{sec:perf info IR} shows that games with imperfect recall stay computationally equally hard even in the absence of exogenous stochasticity (\ie, chance nodes). 

\section{Imperfect-Recall Games}
\label{sec:ir games}

We first define extensive-form games, allowing for imperfect recall. The concepts we use in doing so are standard; for more detail and background, see, \eg, \citet{Fudenberg91:Game_theory} and \citet{PiccioneR73}. In this section, we follow the exposition of \citet{TewoldeOCG23}, with the addition of introducing multi-player notation.

\begin{defn}
An extensive-form \emph{game with imperfect recall}, denoted by $\Gamma$, consists of:
\begin{enumerate}[nolistsep,leftmargin=*]
    \item A rooted tree, with nodes $\nds$ and where the edges are labeled with \emph{actions}. The game starts at the root node $h_0$ and finishes at a leaf node, also called \emph{terminal node}. We denote the terminal nodes in $\nds$ as $\term$ and the set of actions available at a nonterminal node $h \in \nds \setminus \term$ as $A_h$.
    \item A set of $N+1$ players $\pls \cup \{c\}$, for $N \in \N$, and an assignment of nonterminal nodes to a player that shall choose an action at that node. Player $c$ stands for \emph{chance} and represents exogenous stochasticity that chooses an action.  
    With $\nds^{(i)}$ we denote all nodes associated to player $i \in \pls$.
    \item A fixed distribution $\Prob^{(c)}(\cdot \mid h)$ over $A_h$ for each chance node $h \in \nds^{(c)}$, with which an action is determined at $h$.
    \item For each $i \in \pls$, a \emph{utility function} $u^{(i)} : \term \to \R$ that specifies the payoff that player $i$ receives from finishing the game at a terminal node.
    \item For each $i \in \pls$, a partition $\nds^{(i)} = \sqcup_{I \in \infs^{(i)}} I$ of player $i$'s decision nodes into information sets (\emph{infosets}). We require $A_h = A_{h'}$ for all nodes $h, h'$ of the same infoset. Therefore, infoset $I$ has a well-defined action set $A_I$.
\end{enumerate}
\end{defn}

\paragraph{Imperfect Recall.} 

Nodes of the same infoset are assumed to be indistinguishable to the player during the game even though the player is always aware of the full game structure. This may happen even in perfect-recall games due to \emph{imperfect information}, that is, when it is unobservable to the player what another player (or chance) has played. This effect is present in \Cref{fig:forgetting kicker game} for P2. In contrast, infoset $I_2$ of P1 exhibits \emph{imperfect recall} because once arriving there, the player has forgotten information about the history of play that she once held when leaving $I_1$, namely whether she chose left or right back then. In \Cref{fig:ext absentminded driver}, the player is unable to recall whether she has been in the same situation before or not. This phenomenon is a special kind of imperfect recall called \emph{absentmindedness}. 
The \emph{degree of absentmindedness} of an infoset shall be defined as the maximum number of nodes of the same game trajectory that belong to that infoset. In \Cref{fig:ext absentminded driver}, it is $3$. The {\em branching factor} of a game is the maximum number of actions at any infoset.

In contrast to that, games with \emph{perfect} recall have every infoset reflect that the player remembers the sequence of infosets she visited and the actions she took. We note that any node $h \in \nds$ uniquely corresponds to a history path $\hist(h)$ in the game tree, consisting of alternating nodes and actions from root $h_0$ to $h$. Let $\textnormal{exp}^{(i)}(h)$ be the experienced sequence of infosets visited and actions taken by player $i$ on the path $\hist(h)$. Then, formally, a game has perfect recall if for all players $i \in \pls$, all infosets $I \in \infs^{(i)}$, and all nodes $h, h' \in I$, we have $\textnormal{exp}^{(i)}(h) = \textnormal{exp}^{(i)}(h')$.

\paragraph{Strategies.} Let $\Delta(A_I)$ denote the set of probability distributions over the actions in $A_I$. These will also be referred to as \emph{randomized actions}. A (behavioral) \emph{strategy} $\mu^{(i)} : \infs^{(i)} \to \sqcup_{I \in \infs^{(i)}} \Delta(A_I)$ of a strategic player $i$ assigns to each of her infosets $I$ a probability distribution $\mu^{(i)} ( \cdot \mid I) \in \Delta(A_I)$. Upon reaching $I$, the player draws an action randomly from $\mu^{(i)} ( \cdot \mid I)$. A \emph{pure} strategy maps deterministically\footnote{Other work has also considered \emph{mixed} strategies, 
that is, probability distributions over all pure strategies. In the presence of imperfect recall, mixed strategies are not realization-equivalent to behavioral strategies \cite{Kuhn53}. Mixed strategies require the agent to coordinate her actions across infosets (\eg, access to a correlation device): For example, in contrast to our introductory discussion on the forgetful penalty shoot-out (\Cref{fig:forgetting kicker game}), this  game does admit a Nash equilibrium \emph{in mixed strategies} since P1 can now choose to kick left twice in a row $50\%$ of the time and to kick right twice in a row the other $50\%$ of the time. As this would imply a form of memory, it does not fit the motivation of this paper.} to $\sqcup_{I \in \infs^{(i)}} A_I$. A strategy profile, or \emph{profile}, $\mu = ( \mu^{(i)} )_{i \in \pls}$ specifies a behavioral strategy for each player. We may write $\big( \mu^{(i)}, \mu^{(-i)} \big)$ to emphasize the influence of $i \in \pls$ on $\mu$. Denote the strategy set of player $i \in \pls$ with $\strats^{(i)}$, and the set of profiles with $\strats$. 

For a computational analysis, we identify a randomized action set $\Delta(A_I)$ with the simplex $\Delta^{|A_I|-1}$, where $\Delta^{n-1} := \{ x \in \R^{n} \, : \, x_k \geq 0 \, \forall k\, , \sum_{k = 1}^{n} x_k = 1\}$. 
Therefore, the strategy sets are Cartesian products of simplices: 
\begin{center}
    $\strats \equiv \bigtimes_{i \in \pls} \bigtimes_{I \in \infs^{(i)}} \Delta^{|A_I| - 1} \, \textnormal{ and } \, \strats^{(i)} \equiv \bigtimes_{I \in \infs^{(i)}} \Delta^{|A_I| - 1}$.% \, .
\end{center}

\paragraph{Reach Probabilities and Utilities.} Let $\Prob(\bar{h} \mid \mu, h)$ be the probability of reaching node $\bar{h} \in \nds$ given that the current game state is at $h \in \nds$ and that the players are playing profile $\mu$. It evaluates as $0$ if $h \notin \hist(\bar{h})$, and as the product of probabilities of the actions on the path from $h$ to $\bar{h}$ otherwise. The expected utility payoff of player $i \in \pls$ at node $h \in \nds \setminus \term$ if profile $\mu$ is being followed henceforth is $\U^{(i)}(\mu \mid h) := \sum_{z \in \term} \Prob(z \mid \mu, h) \cdot u^{(i)}(z)$. We overload notation by defining $\Prob(h \mid \mu) := \Prob(h \mid \mu, h_0)$ for root $h_0$ of $\Gamma$, and by defining the function $\U^{(i)}$ as $\U^{(i)}(\mu) := \U^{(i)}(\mu \mid h_0)$, mapping a profile~$\mu$ to its expected utility from game start. In \Cref{fig:ext absentminded driver}, this is $\U^{(1)}(\mu) = 6c^2e$ -- or, to follow our notation more precisely, $\U^{(1)}(\mu) = 6 \mu^{(1)}(c \mid I)^2 \mu^{(1)}(e \mid I)$.

\paragraph{Polynomials.}

Each summand $\Prob(z \mid \mu, h) \cdot u^{(i)}(z)$ in \, $\U^{(i)}(\mu \mid h)$ is a monomial in $\mu$ times a scalar, and the expected utility function $\U^{(i)}$ is a polynomial function in the profile $\mu$. All these polynomials $\U^{(i)}$ can be constructed in polynomial time (polytime) in the encoding size of $\Gamma$.

One might also ask how general those polynomial utility functions may be. Indeed, imperfect-recall games can be very expressive. We give a polytime construction in \Cref{app:poly fcts to IR game} that, given a collection of $N$ multivariate polynomials $p^{(i)} : \bigtimes_{i = 1}^{ N } \bigtimes_{j = 1}^{ \ninfs^{(i)} } \R^{m_j^{(i)}} \to \R$, yields an associated $N$-player game~$\Gamma$ with imperfect recall such that its expected utility functions satisfy $\U^{(i)}(\mu) = p^{(i)}(\mu)$ on $\bigtimes_{i = 1}^{ N } \bigtimes_{j = 1}^{ \ninfs^{(i)} } \R^{m_j^{(i)}}$.

\paragraph{Approximate Solutions.}
The solution concepts we investigate will have a definition of the abstract form ``Strategy $\mu$ is a \emph{solution} if for all $y \in Y$ we have $f(\mu) \geq f_{\mu}(y)$'' for some set $Y$ of alternatives and some utility/objective functions $f$ and $f_{\mu}$. Then, we call a strategy $\mu$ an $\epsilon$-solution if $\forall \, y \in Y : f(\mu) \geq f_{\mu}(y) - \epsilon$.

\paragraph{Computational Considerations.}
In this paper, we 
discuss \emph{decision problems} and \emph{search problems}. The former ask for a yes/no answer; the latter ask for a solution point. The input to these computational problems may be a game $\Gamma$, a precision parameter $\epsilon > 0$, and/or a target value $t$. Values in $\Gamma$, as well as $\epsilon$ and $t$ are assumed to be rational. We assume that a game $\Gamma$ is represented by its game tree structure, which has size $\Theta(|\nds|)$, and by a binary encoding of its chance node probabilities and its utility payoffs. If there is a target $t$, then it shall be given in binary as well. 

If there is no precision parameter $\epsilon$, then we are dealing with problems involving \emph{exact} solutions. In our settings, such problems are usually beyond \NP{} because equilibria may require irrational probabilities and may therefore not be representable in finite bit length. In fact, \citet{TewoldeOCG23}[Figure~6] give a simple single-player example in which the unique equilibrium takes on irrational values. That is, in part, why we will also be interested in approximations up to a small precision error $\epsilon > 0$. Here, we mean `small' relative to the range of utility payoffs, which -- by shifting and rescaling utilies -- we can w.l.o.g.\ assume to be $[0,1]$.

\begin{rem*}
    By default, $\epsilon > 0$ will be given in binary, in which case we require \emph{inverse-exponential} (\invexpo{}) precision.
\end{rem*} 

Here, the term `inverse-exponential' indicates that $1/\epsilon$ can be exponentially larger than the tree size $|\nds|$. Occasionally, we may instead require \emph{inverse-polynomial} (\invpoly{}) precision, which is when $\epsilon$ is given in unary, or require constant precision, which is when $\epsilon$ is fixed to a constant $> 0$. Naturally, \invexpo{} precision is hardest to achieve.

\paragraph{Complexity Classes.} 
We give a brief overview of the complexity classes appearing in this paper, and refer to \Cref{app:complexity classes} for references and more details. The subset relationships of the complexities classes we present here are believed to be strict. \P{} describes the decision problems that can be solved in polytime. \NP{} describes the decision problems that can be solved in non-deterministic polytime. $\Sigma_2^{\P}$ describes the decision problems that can be solved in non-deterministic polytime if given oracle access to an NP solver, such as a {\sc SAT} oracle. We have \P{} $\subseteq$ \NP{} $\subseteq \Sigma_2^{\P} \subseteq$ \PSPACE{}. \NP{} and $\Sigma_2^{\P}$ are classes for decision problems that can be formulated as one over discrete variables (w.l.o.g.\ Boolean variables). Their counterparts for real-valued decision problems are the \emph{first-order-of-the-reals} classes $\exists \mathbb{R}$ and $\exists \forall \mathbb{R}$: A $\exists \mathbb{R}$ problem asks whether a sentence of the form $\exists x_1 \ldots \exists x_n F(x_1, \ldots, x_n)$ is true, where the $x_i$ represent real-valued variables and $F$ represents a quantifier-free formula of (in-)equalities of real polynomials in rational coefficients. $\exists \forall \mathbb{R}$ is defined analogously, except for sentences of the form $\exists x \in \R^{n_1} \forall y \in \R^{n_2} F(x, y)$. We have \NP{} $\subseteq \exists \mathbb{R} \subseteq$ \PSPACE{} $\cap \exists \forall \mathbb{R}$.

The complexity classes \FP{} and \FNP{} are the search problem analogues of \P{} and \NP{}, and as such, essentially have the same complexity. The landscape between \FP{} and \FNP{}, however, is rich. Total NP search problems are those problems in \FNP{} for which one knows that each problem instance admits a solution. The complexity classes in it can be characterized by the natural, but exponential-time method with which one can show that each problem instance admits a solution. For the class \PPAD{} the method is that of a fixed point argument, as is the case, \eg , for the existence of a Nash equilibrium. For the class \PLS{} the method is that of a local optimization argument on a directed acyclic graph. For the class \PLS{} the method is that of a \CLS{} a local optimization argument on a bounded polyhedral (continuous) domain. We have \FP{} $\subseteq$ \CLS{} $=$ \PPAD{} $\cap$ \PLS{} and \PPAD{}, \PLS{} $\subseteq$ \FNP{}.

\section{Nash Equilibria and Optimal Play}
\label{sec: NEs}

In this section, we present our computational results for the classic and most important solution concept in game theory -- the Nash equilibrium \cite{Nash48}.

\begin{defn}
\label{def: NE}
    A profile $\mu$ is said to be a \emph{Nash equilibrium} (in behavioral strategies) for game $\Gamma$ if for all player $i \in \pls$, and all alternative strategies $\pi^{(i)} \in \strats^{(i)}$, we have
    \begin{center}
        $\U^{(i)}(\mu^{(i)}, \mu^{(-i)}) \geq \U^{(i)}(\pi^{(i)}, \mu^{(-i)})$. % \, .
    \end{center}
\end{defn}
In a Nash equilibrium, no player has any utility incentives to deviate unilaterally to another strategy.  \citeauthor{Nash48} showed that any finite perfect-recall game admits at least one Nash equilibrium.  In contrast, some finite imperfect-recall games have no Nash equilibrium, as discussed in the introduction. If there is only a single player, however, finding a Nash equilibrium -- \ie, finding an \emph{optimal} strategy -- reduces to maximizing a polynomial utility function over a compact strategy space. Such a solution is guaranteed to exist, and its value is unique. Therefore, one may ask instead whether some target value $t$ can be achieved in a given game. In \Cref{fig:ext absentminded driver}, this would result in the $\exists \mathbb{R}$-sentence
    $\exists e,c: \, 6c^2e \geq t \, \land \, c \geq 0 \land e \geq 0 \land c + e = 1$. % \, .
This is an easier task than \emph{finding} an optimal strategy. Nonetheless, we have:

\begin{prop}[\citealp{Gimbert20}]
\label{lem:SPIR is etr complete}
    Deciding whether a single-player game with imperfect recall admits a strategy with value $\ge t$ is $\exists \mathbb{R}$-complete.
\end{prop}

For approximation, consider problem {\sc Opt-D} that asks to distinguish between whether $\exists \mu \in S : \, \U^{(1)}(\mu) \geq t$ and whether $\forall \mu \in S : \, \U^{(1)}(\mu) \leq t - \epsilon$.

\begin{prop}[\citealp{KollerM92}; \citealp{TewoldeOCG23}]
\label{lem:OPT-D NPC}
    {\sc Opt-D} is \NP{}-complete.
\end{prop}

Technically, \citeauthor{KollerM92} establish hardness for the {\em exact} decision problem. We shall merely add the observation that their proof also implies \NP{}-hardness of the approximate problem; and via the PCP theorem~\cite{Hastad01}, even for a constant precision $\eps < 1 / 8$.

\subsection{Two-Player Zero-Sum Games}
\label{sec:2p0s background}
A \textit{two-player zero-sum (2p0s)} game is a two-player game where $\U^{(2)} = - \U^{(1)}$. In that case utilities can be given in terms of P1, and P2 simply minimizes that utility. 

\citet{KollerM92} prove $\Sigma_2^{\P{}}$-completeness of deciding in 2p0s games with imperfect recall whether the max-min value in pure-strategy play exceeds some utility target $\geq t$. We will consider behavioral strategies instead.
\begin{defn}
    In a 2p0s game $\Gamma$, the (behavioral) {\em max-min value} and {\em min-max value} are defined as
        \begin{center}
        $\ubar{U} := \max_{\mu^{(1)} \in S^{(1)}} \min_{\mu^{(2)} \in S^{(2)}} \U^{(1)}(\mu^{(1)}, \mu^{(2)})$, 
        \\
        $\bar U := \min_{\mu^{(2)} \in S^{(2)}} \max_{\mu^{(1)} \in S^{(1)}}  \U^{(1)}(\mu^{(1)}, \mu^{(2)})$. 
        \end{center}
\end{defn}
    
\citet{Gimbert20} prove that deciding $\ubar{U} \geq t$ is in $\exists \forall \mathbb{R}$ and is $\exists \mathbb{R}$-hard. For approximation, we know the following.

\begin{lemma}[\citealp{Zhang23:Team_DAG}]
\label{lem:maxmin-hard}
    It is $\Sigma_2^\P$-complete to distinguish $\ubar U \ge 0$ from $\ubar U \le -\eps$ in 2p0s games with imperfect recall. Hardness holds even with no absentmindedness and \invpoly{} precision.
\end{lemma}

To leverage this result in the subsequent sections, we will first show a tight connection between the existence of Nash equilibria in a 2p0s game $\Gamma$, and $\Gamma$'s min-max and max-min values. Define the {\em duality gap} of $\Gamma$ as the difference
\begin{center}
    $\Delta := \bar U - \ubar U \geq 0$. % \, .
\end{center}
In \Cref{fig:forgetting kicker game} the duality gap is $1-0 = 1$.

\begin{prop}
\label{prop:gap-equilibrium}
    Let $\Gamma$ be a 2p0s game with imperfect recall. If $\Delta \le \eps$ then $\Gamma$ admits an $\eps$-Nash equilibrium. Conversely, if $\Gamma$ admits an $\eps$-Nash equilibrium, then $\Delta \le 2\eps$.
\end{prop}
In particular, there is an equivalence between Nash equilibrium existence and vanishing duality gap.  This result is not specific to behavioral strategies in imperfect-recall games; it holds for any family of strategies in any 2p0s game.

\subsection{Deciding Nash Equilibrium Existence}
\label{sec:decide NE exists}
\begin{figure} 
    \tikzset{
        every path/.style={-},
        every node/.style={draw},
    }
    \forestset{
  subgame/.style={regular polygon,
  regular polygon sides=3,anchor=north, inner sep=5pt},
    }
  \begin{center}
  \begin{forest}
    [,p2,s sep=50pt,l sep=20pt
        [\util1{0},terminal,el={2}{exit}{},yshift=-6.3pt]
        [,nat,s sep=60pt,el={2}{cont}{}
        [$\Gamma$,subgame]
        [G,subgame, name=g]
        ]
    ]
  \end{forest}
  \end{center}
   \caption{Game construction used to prove hardness of deciding equilibrium existence. We use boxes for chance nodes, at which chance plays uniformly at random. $\Gamma$ is a placeholder game. G is a game with no equilibrium; \Cref{sec:decide NE exists} for example uses \Cref{fig:forgetting kicker game}.
   }
   \label{fig:NE iff max min value nonneg}
\end{figure}
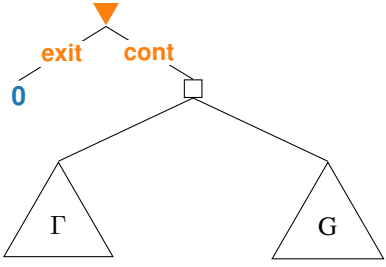

We observe that the existence of a Nash equilibrium can be formulated as ``there exists a profile $\mu$ such that for all other profiles $\pi$ the condition of \Cref{def: NE} are satisfied for all $i \in \pls$''. This puts the exact and approximate decision problems in $\exists \forall \mathbb{R}$ and $\Sigma_2^\P{}$ respectively. For an intuitive idea of our upcoming hardness results, consider the game in \Cref{fig:NE iff max min value nonneg} where subgame $G$ shall be that of \Cref{fig:forgetting kicker game} and where subgame $\Gamma$ is a game in which it is hard to decide what utility P1 can guarantee himself.
Then a profile cannot be a Nash equilibrium if P2 is supposed to continue at the root node, because in that case $G$ is reached with positive probability and the players cannot be in equilibrium in that subgame as we have discussed in the introduction. Note that exiting at the root node yields P2 a utility of $0$, and best-responding to P1 in subgame $G$ also yields P2 a utility of $\leq 0$ (recall that P2 is the minimizer). Thus, for a profile to be a Nash equilibrium in the overall game, P2 must exit at the root node as a best response, which is the case exactly if P1 cannot achieve a utility of at least $0$ in the subgame $\Gamma$. Using the problem instances of \Cref{lem:SPIR is etr complete} for the subgame $\Gamma$, we obtain
\begin{thm}
\label{thm:Nash-D etr hard}
    Deciding if a game with imperfect recall admits a Nash equilibrium is $\exists \mathbb{R}$-hard and in $\exists \forall \mathbb{R}$. Hardness holds even for 2p0s games where one player has a degree of absentmindedness of $4$ and the other player has perfect recall.
\end{thm}

Next, for the approximate case, we use the problem instances of \Cref{lem:maxmin-hard} for the subgame $\Gamma$. Define {\sc Nash-D} to ask to distinguish between whether an exact Nash equilibrium exists or whether no $\epsilon$-Nash equilibrium exists.

\begin{thm}
\label{th:apx-nash-hard}
    {\sc Nash-D} is $\Sigma_2^\P{}$-complete. Hardness holds for 2p0s games with no absentmindedness and \invpoly{} precision.
\end{thm}

With \Cref{prop:gap-equilibrium}, this immediately implies 

\begin{cor}
\label{cor:deciding positive duality gap is hard}
    It is $\Sigma_2^\P{}$-complete to distinguish $\Delta = 0$ from $\Delta \ge \eps$ in 2p0s games. Hardness holds for 2p0s games with no absentmindedness and \invpoly{} precision.
\end{cor}
Later in this paper, \Cref{thm:apx EDT sigma2p compl} will imply another $\Sigma_2^\P{}$-hardness for {\sc Nash-D} but with different restrictions.

\subsection{A Na\"ive Algorithm for Nash Equilibria}
\label{sec:NE naive algo}

For game $\Gamma$, let $|\Gamma|$ denote its representation size and $m := \sum_{i \in \pls} \sum_{ I \in \infs^{(i)} } |A_I|$ its the total number of pure actions.

\begin{prop}
\label{prop:Comp NE expo time}
    {\sc Nash-D} is solvable in time \\
    $\poly \Big( |\Gamma|, \log\frac{1}{\epsilon}, ( m \cdot |\nds| )^{m^2} \Big)$.
\end{prop}
In fact, our algorithm {\em finds} an $\eps$-Nash equilibrium whenever an exact Nash equilibrium exists. The idea is similar to that one of \citet{LiptonM04}[Theorem 2] for multi-player normal-form games: Namely, we iteratively subdivide the strategy space, and repeatedly decide with first-order-of-the-reals solvers whether a Nash equilibrium exists in this smaller region. Those solvers also give rise to the exponential time dependence on $m$. In particular, the algorithm becomes polytime if $m$ is bounded by a constant. This observation will aid us towards a \PLS{}-membership proof in \Cref{thm:EDT PLS-C}. Also note that such a bound on $m$ will not restrict the size of the game tree since the degree of absentmindedness can still grow arbitrarily (cf. \Cref{fig:ext absentminded driver}).
\section{Introducing Multiselves Equilibria}
\label{sec: intro MS eqs}

\Cref{sec: NEs} shows strong obstacles to finding Nash equilibria in games with imperfect recall. In light of these limitations, we relax the space of solutions and turn to the \emph{multiselves} approach (cf.\ the agent-form \cite{Kuhn53}), which we review in this section. This approach argues that, whenever a player finds herself in an infoset, she has no influence over which actions she chooses at other infosets. Therefore, at a multiselves equilibrium $\mu$, each player will play the best randomized action at each of their infosets, assuming that they themselves play according to $\mu$ at other infosets and assuming all other players also play according to $\mu$. 

\begin{figure}[t]
    \centering
    \begin{forest}
    [,p1,name=p0,s sep=20pt,l sep=20pt
    [,p1,name=p1a,el={1}{$l_1$}{},s sep=20pt,l sep=20pt
        [\util1{1},terminal,el={1}{$l_2$}{}]
        [\util1{0},terminal,el={1}{$r_2$}{}]
    ]
    [,p1,name=p1b,el={1}{$r_1$}{},s sep=20pt,l sep=20pt
        [\util1{0},terminal,el={1}{$l_2$}{}]
        [\util1{2},terminal,el={1}{$r_2$}{}]
    ]
    ]
    \node[above=0pt of p0,draw=none,p1color]{$I_1$};
    \draw[infoset1] (p1a) to node[below,draw=none,p1color,]{$I_{2}$} (p1b);
    \end{forest}

    \caption{A single-player game with imperfect recall where miscoordinating actions with yourself is punished most.}
    \label{fig:coord problem}
\end{figure}
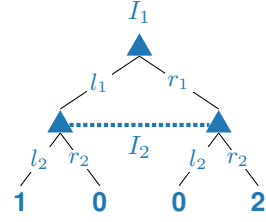

Consider \Cref{fig:coord problem}. 
The optimal strategy is to play $(r_1, r_2)$. This is also a multiselves equilibrium. However, $(l_1, l_2)$ is also a multiselves equilibrium, because if the player is at the top-level infoset $I_1$ and assumes that she will follow left at the bottom-level infoset $I_2$, then it is best for her to go left now. On the other hand, if the player is at $I_2$ and assumes that she played left at $I_1$, then it is again best for her to play left now. 

Multiselves equilibria can be arbitrarily bad in payoff in comparison to optimal strategies and Nash equilibria, as can be seen by replacing the payoff of $2$ in \Cref{fig:coord problem} with some $\lambda \to \infty$. This phenomenon is due to miscoordination across infosets, and it arises in the same manner across teams in team games: The corresponding normal-form game $\begin{pmatrix} \lambda,\lambda  & 0,0\\ 0,0 & 1,1 \end{pmatrix}$ shows that Nash equilibria can be arbitrarily worse relative to Pareto-optimal profiles. 

In games with absentmindedness it becomes controversial how to apply the multiselves idea. Specifically, how should a player reason about implications of a choice at the current decision point for her action choices at past and future decision points {\em within the same infoset}, and -- as a consequence -- compute incentives to deviate?  That is, in considering deviating, will the player assume they would perform the same deviation at other nodes in the same infoset, or that the deviation is a one-time-only event? We will handle this question using two well-motivated\footnote{The debate around decision theories is related to the approach for belief formation (cf. the Sleeping Beauty problem \cite{Elga00:Self}). Among other aspects, the literature has studied which combination of decision theories and belief formation avoid being Dutch-booked (money-pumped) \cite{PiccioneR73,Briggs10:Putting,Oesterheld22:Can}.} 
decision theories that correspond to these two cases: evidential decision theory and causal decision theory. We will see that Nash equilibria are multiselves equilibria under both decision theories.

This section is accompanied with an extensive \Cref{app:app3} that -- beyond proving the statements made in this section -- also introduces some additional observations and lemmas needed for the development of our main results.

\subsection{Evidential Decision Theory (EDT)} 
\label{sec: EDT}

Suppose a game $\Gamma$ is played with profile $\mu$, and a player $i$ arrives in one of her infosets $I \in \infs^{(i)}$. EDT postulates that if that player deviates to a randomized action $\alpha \in \Delta(A_I)$ at the current node, then she will have also deviated to $\alpha$ whenever she arrived in $I$ in the past, and that she will also deviate to $\alpha$ whenever she arrives in $I$ again in the future. This is because
EDT argues that
the choice to play $\alpha$ now is evidence for the player playing the same $\alpha$ in the past and future.

We denote the behavioral strategy that results from an EDT deviation as $\mu_{I\mapsto \alpha}^{(i)}$. It plays according to $\mu^{(i)}$ at every infoset except for at $I \in \infs^{(i)}$ where it plays according to $\alpha \in \Delta(A_I)$.

\begin{defn}
\label{defn:EDT eq}
    We call $\mu$ an \emph{EDT equilibrium} for game $\Gamma$ if for all players $i \in \pls$, all her infosets $I \in \infs^{(i)}$, and all %alternative
    randomized actions $\alpha \in \Delta(A_I)$, we have
    \begin{center}
     $\U^{(i)}(\mu) \geq \U^{(i)}( \mu_{I \mapsto \alpha}^{(i)}, \mu^{(-i)} )$. % \, .
    \end{center}
\end{defn}

In an EDT equilibrium, no player has an incentive to deviate at an infoset in an EDT fashion to another randomized action. This is because the right hand side of the inequality represents the expected \emph{ex-ante} utility of such an EDT deviation. \Cref{app:Perspectives} gives an extensive discussion on the ex-ante perspective %for
for multiselves equilibria. Regarding equilibrium computation, the following result is known:

\begin{prop}[\citealp{TewoldeOCG23}]
\label{lem:1PL EDT no FPTAS}
    Unless \NP{} = \ZPP, finding an $\epsilon$-EDT equilibrium in a single-player game for \invpoly{} precision is not in \P{}.
\end{prop}

\subsection{Causal Decision Theory (CDT)} 
\label{sec: CDT}

Say, again, game $\Gamma$ is played with profile $\mu$, and a player $i$ arrives in one of her infosets $I \in \infs^{(i)}$. Then CDT postulates that the player can deviate to an action $\alpha \in \Delta(A_I)$ at the current node without violating that she has been playing according to $\mu^{(i)}$ at past arrivals in $I$, or that she will be playing according to $\mu^{(i)}$ at future arrivals in $I$. This is in addition to assuming that all other players follow $\mu^{(-i)}$ as usual. The intuition behind CDT is that the player's choice to deviate from $\mu^{(i)}$ at the current node does not \textit{cause} any change in her behavior at any other node of the same infoset $I$. 

\begin{ex}
\label{ex:edt vs cdt in absentminded driver}
    Recall \Cref{fig:ext absentminded driver} in which -- as the story goes -- the absentminded driver has to exit a highway at the second highway exit to find home. Say the player enters the game with $\mu =$ `e' (exit), and upon arriving in the infoset, considers deviating to `c' (continue) at this point of time. EDT then argues that the player will always continue on the highway and arrive at the third ``0'' payoff of the game. CDT, on the other hand, argues that the player will continue on the highway once -- or more precisely, continue at the root node since that is the only decision node she could possibly be at given her strategy $\mu$ -- and then exit the highway at its second exit.
\end{ex}
For node $h \in \nds^{(i)}$ and pure action $a \in A_h$, let $ha$ denote the child node reached if player $i$ plays $a$ at $h$. Consequently, $\U^{(i)}(\mu \mid ha)$ is the expected utility of player $i$ from being at $h$, playing $a$, and everyone following profile $\mu$ afterwards. When at an infoset $I \in \infs^{(i)}$, the player does not know at which node of $I$ she currently is. Therefore, when computing her utility incentives for a CDT-style deviation to $a$, she scales each node by the probability of reaching that node under profile $\mu$. This yields utility incentives
\begin{center}
    $\sum_{h \in I} \Prob(h \mid \mu) \cdot \U^{(i)}(\mu \mid h a)$.
\end{center}
to CDT-deviate to pure action $a$ at infoset $I$. This value is known to be equal to the partial derivative $\nabla_{I,a} \, \U^{(i)}(\mu)$ of utility function $\U^{(i)}$ w.r.t.\ to action $a$ of $I \in \infs^{(i)}$ at point $\mu$ \cite{PiccioneR73,Oesterheld22:Can}. Hence, we can formulate the following definition.
\begin{defn}
\label{defn:CDT utility}
    Given a profile $\mu$ in game $\Gamma$, a player $i \in \pls$ determines her (ex-ante) utility from CDT-deviating at infoset $I \in \infs^{(i)}$ to randomized action $\alpha \in \Delta(A_I)$ as
    \\
    $\U_{\CDT}^{(i)}(\alpha \mid \mu, I) :=$ 
    \begin{center}
    $\U^{(i)}(\mu) + \sum_{a \in A_I} (\alpha(a) - \mu(a \mid I)) \cdot \nabla_{I,a} \, \U^{(i)}(\mu)$. 
    \end{center}
\end{defn}

In other words, this is the first-order Taylor approximation of $\U^{(i)}$ at $\mu$ in the subspace $\Delta(A_I)$. In \Cref{fig:dont go straight} of \Cref{app:app3}, we illustrate on a simple game that the ex-ante CDT-utility  -- as a first-order approximation -- may yield unreasonable utility payoffs for values $\alpha$ far away from $\mu(\cdot \mid I)$. Moreover, if $\alpha \neq \mu(\cdot \mid I)$, we observe that the resulting behavior of the deviating player cannot be captured by a \emph{behavioral strategy} anymore that the player could have chosen from the beginning. That is because the player is now acting differently at different nodes of the same infoset. 

\begin{defn}
\label{defn:CDT eq}
    A profile $\mu$ is said to be a \emph{CDT equilibrium} for game $\Gamma$ if for all player $i \in \pls$, all her infosets $I \in \infs^{(i)}$, and all alternative randomized actions $\alpha \in \Delta(A_I)$, we have
    \begin{center}
        $\U^{(i)}(\mu) = \U_{\CDT}^{(i)} \big( \mu^{(i)}( \cdot \mid I) \bigm\vert \mu, I \big) \geq \U_{\CDT}^{(i)} (\alpha \mid \mu, I)$. % \, .
    \end{center}
\end{defn}

Therefore, no player has any utility incentives to deviate at an infoset in a CDT fashion to another randomized action. CDT equilibria have received a more thorough treatment in the literature than EDT equilibria have. 

\begin{lemma}[\citealp{LambertMS19}]
\label{lem:CDT eq exists}
    Any game $\Gamma$ with imperfect recall admits a CDT equilibrium.
\end{lemma}

Thus, we shall define {\sc CDT-S} as the search problem that asks for an $\epsilon$-CDT equilibrium in the game (which always exists). Let {\sc 1P-CDT-S} be its restriction to single-player games.

\begin{prop}[\citealp{TewoldeOCG23}]
\label{lem:CDT KKT and CLS}   
\,
    \begin{enumerate}[nolistsep, leftmargin=*]
        \item A profile $\mu$ is a CDT equilibrium of $\Gamma$ if and only if for all player $i \in \pls$, strategy $\mu^{(i)}$ is a KKT-point of 
        \begin{center}
            $\max_{\pi^{(i)} \in S^{(i)}} \U^{(i)}(\pi^{(i)}, \mu^{(-i)})$. %\, .
        \end{center}

        \item The problem {\sc 1P-CDT-S} is \CLS{}-complete.
    \end{enumerate}
\end{prop}

The original formulation of \citeauthor{TewoldeOCG23} was not given for the multi-player setting and the ex-ante perspective. The advantages of the latter are discussed in \Cref{app:Perspectives}. Furthermore, we may also highlight a positive algorithmic implication which has not been stated before. It can be obtained analogously to \cite[Lemma C.4]{FGHS23}.

\begin{cor}
\label{cor:1PL CDT has FPTAS}
    {\sc 1P-CDT-S} for \invpoly{} precision is in \P{}.
\end{cor}

\subsection{Comparing the Solution Concepts}
\label{sec:connections}

The three solution concepts form an inclusion hierarchy. This result is known for single-player settings and extends straight-forwardly to multi-player settings.
\begin{prop}[\citealp{Oesterheld22:Can}]
\label{lem:EQ hierarchy}
    A Nash equilibrium is an EDT equilibrium. An EDT equilibrium is a CDT equilibrium.
\end{prop}

This also implies that any single-player game admits both EDT and CDT equilibria since it admits an optimal strategy (= Nash equilibrium). In general, neither statement in \Cref{lem:EQ hierarchy} holds in reverse. Indeed, we have seen in \Cref{fig:coord problem} that multiselves equilibria may not be the optimal strategy. Moreover, the strategy $\mu$ described in \Cref{ex:edt vs cdt in absentminded driver} forms a CDT equilibrium but not an EDT equilibrium (an EDT deviation to a uniformly randomized action achieves positive utility).

We will find in this paper that CDT equilibria are easier to compute than EDT equilibria. Indeed, \Cref{lem:1PL EDT no FPTAS} and \Cref{cor:1PL CDT has FPTAS} already serve as the first evidence towards such a separation. We can also find a hint towards such an insight by considering the easier problem of \emph{verifying} whether a given profile could be an equilibrium. For CDT, this can be done in polytime: since $\U_{\CDT}^{(i)}$ is linear in $\alpha$, we do not actually need to check \Cref{defn:CDT eq} for all $\alpha \in \Delta(A_I)$, but it suffices to only check it for \emph{pure} actions $a \in A_I$. For EDT equilibria, on the other hand, there is no simple-to-check characterization: $\U^{(i)}( \mu_{I \mapsto \cdot}^{(i)}, \mu^{(-i)} )$ is a polynomial function over $\Delta(A_I)$, for which no easy verification method is known. At least, this is true in general. As for special cases, we have:

\begin{rem}
\label{rem:w/o absentmindedness edt equals cdt}
    Without absentmindedness, deviation incentives of EDT and of CDT coincide, and so do the equilibrium concepts. Hence, complexity results such as \Cref{lem:CDT KKT and CLS} and \Cref{thm:CDT is PPAD} will apply to EDT equilibria as well.
\end{rem}

\begin{rem}
\label{rem: 1 infoset yields EDT eq NE}
    If each player has only one infoset in total, then the EDT equilibria coincide with the Nash equilibria.
\end{rem}

\subsection{On Utility Perspectives}
\label{app:Perspectives}

\begin{figure}[t]
    \centering
    \begin{forest}
    [,p1,name=p0,s sep=20pt,l sep=20pt
    [,p1,name=p1a,el={1}{$l_1$}{},s sep=20pt,l sep=20pt
        [\util1{1},terminal,el={1}{$l_2$}{}]
        [\util1{0},terminal,el={1}{$r_2$}{}]
    ]
    [,p1,name=p1b,el={1}{$r_1$}{},s sep=20pt,l sep=20pt
        [\util1{0},terminal,el={1}{$l_3$}{}]
        [\util1{2},terminal,el={1}{$r_3$}{}]
    ]
    ]
    \node[above=0pt of p0,draw=none,p1color]{$h_1$};
    \node[above left=0pt of p1a,draw=none,p1color]{$h_2$};
    \node[above right=0pt of p1b,draw=none,p1color]{$h_3$};
    \end{forest}

    \caption{Differences of the ex-ante and de-se utility perspective explained on a perfect-recall variant of \Cref{fig:coord problem}. Again, the only optimal strategy takes the path $r_1$ -- $r_3$. But what action can you choose at $h_2$?}
    \label{fig:utility perspectives}
\end{figure}
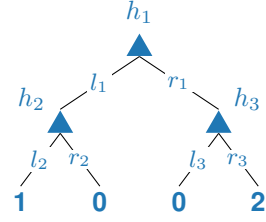

Let us discuss the \emph{ex-ante} and \emph{de-se} perspectives on utilities, and why we chose the former. Consider the game in \Cref{fig:utility perspectives}, which is the \emph{perfect-recall} version of \Cref{fig:coord problem},  
and consider the strategy $\mu = (\epsilon l_1 + (1-\epsilon) r_1, r_2, r_3)$ for some small $\epsilon > 0$. In this case, how much does it matter what randomized action the player chooses at node $h_2$? 
In the \emph{de-se} perspective, the player calculates her expected deviation gains for her current situation onwards. In our example, she would calculate an incentive of $1$ to deviate to $l_2$ assuming she is already at $h_2$. In the \emph{ex-ante} perspective, the player calculates her expected deviation gains on the ex-ante utility (from before the game started). In our example, she would calculate an incentive of $\eps$ to deviate to $l_2$ at $h_2$ since that node is rarely visited anyways.

Previous work in the literature has considered agents that maximize their de-se utilities, as in \citet{Strotz55} with the strategy of consistent planning or in \citet{PiccioneR73}. 
This might fit well for human agents who are interested in the impacts of their actions on their \emph{current} self. In this paper, however, we argue that for AI and team agents, the ex-ante perspective is more suitable. Indeed, such an agent should ground its optimization in the impact its actions has on the overall ex-ante utility; despite imperfect recall limiting the agent's decision or commitment powers to the current infoset (EDT) or decision node (CDT). 

There are also technical advantages supporting the ex-ante perspective. At infosets that are never reached, the action choices do not affect the ex-ante utility. Under de-se reasoning, however, the agent would have to generate beliefs on the impossible event of being at that infoset. In order to make such beliefs well-defined, one has to pick one of many possible options for equilibrium refinement. In optimization terms, the de-se utility functions are fractions of polynomials with possible singularities on the boundary of the strategy set due to vanishing denominators. \citeauthor{TewoldeOCG23}[\citeyear{TewoldeOCG23}, Theorem 2] circumvents this issue in their formulation of \Cref{lem:CDT KKT and CLS} 
by only considering games that come with universal lower bounds on the positive reach probabilities of all infosets. Unfortunately, many (simple) games such as \Cref{fig:utility perspectives} do not satisfy this property.

\section{Complexities of Multiselves Equilibria}
\label{sec:mse main}

In this section, we present our computational results for multiselves equilibria.

\subsection{EDT Equilibria}
\label{sec:EDT results}

\begin{figure} 
    \tikzset{
        every path/.style={-},
        every node/.style={draw},
    }
    \forestset{
    subgame/.style={regular polygon,
    regular polygon sides=3,anchor=north, inner sep=1pt},
    }
    \begin{center}
        \begin{forest}
            [,p1,name=p0
            [,p1,name=p1a
                [,p2,name=p2a
                    [\util1{$\lambda$},terminal,name=t1] [\util1{3},terminal]]
                    [,p2 [\util1{-1},terminal] [\util1{-1},terminal]
                ]
            ]
            [,p1,name=p1b
                [,p2
                    [\util1{-1},terminal] [\util1{-1},terminal]]
                    [,p2,name=p2b [\util1{3},terminal] [\util1{-1},terminal,name=t8]
                ]
            ]
            ]           
            \draw[infoset1,bend left=90] (p1a) to (p0) to (p1b);
            \draw[infoset2] (p2a) -- (p2b);
        \end{forest}
    \end{center}
    \caption{A variant of \Cref{fig:forgetting kicker game} where P1 has one single infoset with absentmindedness. It is parametrized by the payoff $\lambda \in \R$ from P1 shooting left and P2 blocking left.}
   \label{fig:absentminded kicker game}
\end{figure}
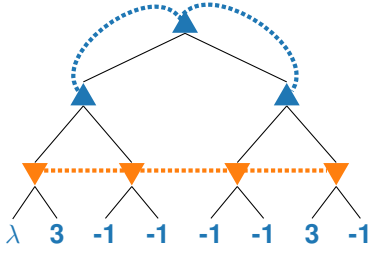

Consider the (parametrized) \emph{absentminded} penalty shoot-out in \Cref{fig:absentminded kicker game}. It shows that in multi-player settings, EDT equilibria may not exist. Absentmindedness is crucial for such an example due to \Cref{rem:w/o absentmindedness edt equals cdt} and \Cref{lem:CDT eq exists}.
\begin{lemma}
\label{lem:NE existence iff lambda geq 1}
    \Cref{fig:absentminded kicker game} has an EDT equilibrium if and only if $\lambda \geq 3$.
\end{lemma}

The next result establishes $\exists \mathbb{R}$-hardness again by similar arguments to \Cref{thm:Nash-D etr hard}. Except in this construction, we attach the single-player game $\Gamma$ from \Cref{lem:SPIR is etr complete} to the bottom left of \Cref{fig:absentminded kicker game}. Note here that by an appropriate payoff shift in $\Gamma$, we can w.l.o.g.\ assume the target $t$ for $\Gamma$ to be $3$.

\begin{thm}
\label{thm:EDT exact ETR hard}
    Deciding whether a game with imperfect recall admits an EDT equilibrium is $\exists \mathbb{R}$-hard and in $\exists \forall \mathbb{R}$. Hardness holds even for 2p0s games where one player has a degree of absentmindedness of $4$ and the other player has perfect recall.
\end{thm}

Now consider problem {\sc EDT-D} that asks to distinguish between whether an exact EDT equilibrium exists or whether no $\epsilon$-EDT equilibrium exists.
\begin{thm}
\label{thm:apx EDT sigma2p compl}
    {\sc EDT-D} is $\Sigma_2^\P{}$-complete. Hardness holds for \invpoly{} precision and 2p0s games with one infoset per player and a degree of absentmindedness of $4$.
\end{thm}

The technically involved proof casts the game construction for \Cref{thm:Nash-D etr hard} to a game where each player only has one infoset, in order to use \Cref{rem: 1 infoset yields EDT eq NE}. For that, we cannot reduce from \Cref{lem:maxmin-hard} this time, but we reduce directly from the $\Sigma_2^\P{}$-complete problem {\sc $\exists \forall$3-DNF-SAT} \cite{Stockmeyer76}. Moreover, we make use of the flexibility that EDT-utilities can represent arbitrary polynomial functions as long as they are only over a single simplex.

Next, we turn to the search problem. The algorithm of \Cref{prop:Comp NE expo time} can also find $\epsilon$-EDT equilibria if we adjust for its equilibrium conditions. In single-player settings, however, we can do better since EDT equilibria are guaranteed to exist. Let {\sc 1P-EDT-S} be the search problem that asks for an $\epsilon$-EDT equilibrium. This problem was left open by \citet{TewoldeOCG23}.

\begin{thm}
\label{thm:EDT PLS-C}
    {\sc 1P-EDT-S} is \PLS{}-complete when the branching factor is constant. Hardness holds even when the branching factor and the degree of absentmindedness are $2$.
\end{thm}

Before we touch on the proof idea, we shall highlight its contrast to \Cref{lem:CDT KKT and CLS} on the \CLS{}-membership of {\sc 1P-CDT-S}, since \CLS{} is believed to be a proper subset of \PLS{} (evidenced by conditional separations as discussed in \Cref{app:complexity classes}). Furthermore, we also get:

\begin{cor}
\label{cor:EDT SPIR inv poly in P}
    {\sc 1P-EDT-S} for \invpoly{} precision is in \P{} when the branching factor is constant.
\end{cor}

The proofs first establish that {\sc 1P-EDT-S} is computationally equivalent to the search problem that takes a polynomial function~$p$ over a product of simplices, and asks for an approximate ``Nash equilibrium point'' of it. In the special case where the branching factor is $2$, the domain becomes the hypercube $[0,1]^{\ninfs}$, and an $\epsilon$-Nash equilibrium $x$ is characterized by the property 
\begin{center}
    $\forall j \in [\ninfs] \, \forall y \in [0,1]: \quad p(x) \geq p(y, x_{-j}) - \epsilon$. 
\end{center}
We show that this problem is \PLS{}-complete. This result may be of independent interest for the optimization literature.

The \PLS{}-hardness follows from a reduction from the \PLS{}-complete problem {\sc MaxCut/Flip}~\cite{SchafferY91,Yannakakis2003}. For the positive algorithmic results of \PLS{} and \P{} membership respectively, we show that $\epsilon$-best-response dynamics converges to an $\epsilon$-EDT equilibrium. We run a similar method to \Cref{prop:Comp NE expo time} in order to compute an $\epsilon$-best response randomized action of an infoset to the other infosets. This takes polytime if the number of actions per infoset (= branching factor) is bounded. Without this restriction, we run into the impossibility result of \Cref{lem:1PL EDT no FPTAS}.

\subsection{CDT Equilibria}
\label{sec:CDT results}

How hard is {\sc CDT-S}, now that we allow for many players? We can get \PPAD{}-hardness straightforwardly because any normal-form game can be cast to extensive form, and because finding a Nash equilibrium in a normal-form game is \PPAD{}-complete \cite{DGP09,ChenDT09}.  Interestingly enough, we can also show \PPAD{}-membership. 
\begin{thm}
\label{thm:CDT is PPAD}
    {\sc CDT-S} is \PPAD{}-complete. Hardness holds even for two-player perfect-recall games with one infoset per player and for \invpoly{} precision.
\end{thm} 

For membership we investigate the existence proof of \Cref{lem:CDT eq exists} by \citeauthor{LambertMS19}. They first shows a connection to perfect-recall games with particular symmetries, and then give a Brouwer fixed point argument which resembles that of \citeauthor{Nash51}'s for symmetric games. However, the connection relies on a construction whose size blows up in the order of factorials, \ie, super-polynomially. Therefore, we modify the fixed point argument to one that works directly on CDT utilities: In a game of imperfect recall, given a profile $\mu$, define the advantage of a pure action $a$ at infoset $I$ of player $i$ as
\begin{center}
    $g_{I,a}^{(i)}(\mu) := \U_{\CDT}^{(i)} (a \mid \mu, I) - \U^{(i)}(\mu)$ .
\end{center}

Intuitively, if the advantage of an action $a$ over the current randomized action $\mu^{(i)}( \cdot \mid I)$ is large, then the player should increase its probability of play. Therefore, we may define the Brouwer function to map any profile $\mu$ to profile $\pi$ defined as
\[
    \pi^{(i)}( a \mid I) := \frac{ \mu^{(i)}( a \mid I) + \max\{0,g_{I,a}^{(i)}(\mu)\} }{ 1 + \sum_{a' \in I} \max\{0,g_{I,a'}^{(i)}(\mu) \} }\, .
\]

Then we show that this forms a valid a Brouwer function whose fixed points are indeed CDT equilibria of the underlying game, and that the Brouwer function and precision errors satisfy the computational requirements developed by \citet{EtessamiY10} to imply \PPAD{}-membership.

The \PPAD{}-membership result is a positive algorithmic result: it shows that we can find CDT equilibria with fixed point solvers and path-following methods, just as it is the case with Nash equilibria in normal-form games. In particular, we shall highlight the stark contrast to \Cref{thm:apx EDT sigma2p compl}. Finding a CDT equilibrium sits well within in the landscape of total \NP{} search problems, whereas even deciding whether an EDT equilibrium exists is already on higher levels of the polynomial hierarchy, let alone finding one.

\section{The Insignificance of Exogenous Stochasticity}
\label{sec:perf info IR}

As of now, the hardness results for single-player settings rely on the presence of chance nodes; see \Cref{lem:SPIR is etr complete,lem:OPT-D NPC} and \Cref{thm:EDT PLS-C}. In this section, we investigate the complexity of games without chance nodes. Of course, one might choose to add players to the game to simulate nature, even in games of perfect recall. However, adding players may add significantly to the computational complexity of the game, cf.\ \P{} vs \PPAD{} for Nash equilibria in single-player vs two-player settings under perfect recall, or \Cref{lem:CDT KKT and CLS} vs \Cref{thm:CDT is PPAD} for CDT equilibria under imperfect recall. Interestingly enough, we can show that in the presence of imperfect recall, chance nodes do not affect the complexity.

\begin{thm}
\label{thm:restr to no chance possible}
    All computational hardness results in this paper for the three equilibrium concepts \{Nash, EDT, CDT\} still hold even when the game has no chance nodes. They hold together with previously possible restrictions (\eg, on the branching factor), except that the restrictions on the number of infosets and the degree of absentmindedness increase by one and to $\mathcal{O}(\log|\nds|)$ respectively.
\end{thm}

\begin{figure} 
    \tikzset{
        every path/.style={-},
        every node/.style={draw},
    }
    \forestset{
    subgame/.style={regular polygon,
    regular polygon sides=3,anchor=north, inner sep=1pt},
    }
    \begin{center}
        \begin{minipage}{.35\linewidth}
        \centering
        \begin{forest}
        [,nat,s sep=20pt,l sep=20pt
            [{\textover[c]{$
            G$}{$G'$}},subgame,el={0}{$\nicefrac{1}{2}$}{}]
            [$G'$,subgame,el={0}{$\nicefrac{1}{2}$}{}]
        ]
        \end{forest}
        \end{minipage}
        \begin{minipage}{0.04\linewidth}
        \begin{center}
        {\Large $\rightarrow$}
        \end{center}
        \end{minipage}
        \begin{minipage}{.5\linewidth}
        \centering
        \begin{forest}
            [,p1,name=p0,s sep=15pt,l sep=1pt
                [,p1,name=p1a,s sep=20pt,l sep=15pt
                    [\util1{-1},terminal,name=t1] 
                    [{\textover[c]{$G$}{$G'$}},subgame, yshift=10pt]
                ]
                [,p1,name=p1b,s sep=20pt,l sep=15pt
                    [$G'$,subgame, yshift=10pt] 
                    [\util1{-1},terminal]
                ]
            ]          
            \node[below=10pt of p0,draw=none,p1color]{$I_c$};
            \draw[infoset1,bend left=90] (p1a) to (p0) to (p1b);
        \end{forest}
        \end{minipage}
    \end{center}
    \caption{How to remove a chance node if it is located at the root. Starting with the game on the left, replace it with infoset $I_c$. Assuming w.l.o.g.\ that the subgames $G$ and $G'$ always yield positive payoffs, the player of $I_c$ will want to randomize uniformly at $I_c$ -- independent of the play in $G$ and $G'$. For another example with more chance node children, see \Cref{fig:chance removal multi actions} in \Cref{app:perf info IR}.
    }
   \label{fig:chance removal}
\end{figure}
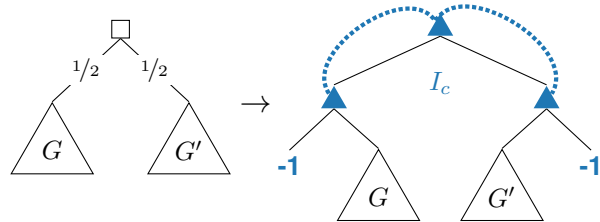

In other words, all exogenous stochasticity can be replaced by one infoset (of an arbitrary player, say P1) with absentmindedness, \ie, replaced by uncertainty that arises from forgetting one's past actions in an identical situation. The proof first transforms the game $\Gamma$ to an equivalent game $\tilde{\Gamma}$ that only has a single chance node $h_c$ that is located at the root. Next, we replace $h_c$ with an infoset $I_c$ with absentmindedness. We illustrate in \Cref{fig:chance removal} how to do it with a chance node that uniformly randomizes over two actions. The resulting game $\Gamma'$ has the same number of players and strategy sets as $\Gamma$, except for the additional infoset $I_c$ for P1. In equilibrium, the induced conditional probability distribution over the children of $h_c$ in $\Gamma$ and the nonterminal ``children'' of $I_c$ in $\Gamma'$ will be the same. Finally, there will be a polynomial relationship between the equilibrium precision errors in $\Gamma$ and $\Gamma'$.

Next, recall {\sc Opt-D} from \Cref{lem:OPT-D NPC} which asks whether an approximate target value can be achieved in a single-player game with imperfect recall. We improve on \Cref{thm:restr to no chance possible} in the specific problem {\sc Opt-D} via an independent proof.

\begin{prop}
\label{prop:SPIR no chance still NP-hard}
    {\sc Opt-D} is \NP{}-hard, even for games with no chance nodes, one infoset, a degree of absentmindedness of~$2$, and \invpoly{} precision.
\end{prop}

Due to \Cref{rem: 1 infoset yields EDT eq NE}, this hardness result also holds for deciding whether \emph{all} EDT equilibria achieve an approximate target value. The proof reduces from the {\sc 2-MinSAT} problem \cite{KohliKM94}.

\section{Conclusion}
\label{sec:conclusion}

Historically, games of imperfect recall have received only limited attention, as it is not clear that they cleanly model any strategic interactions between humans. However, as we argued in the introduction, they are more practically significant in the context of AI agents.  However, they also pose new challenges.
Optimal decision making under imperfect recall is hard due to its close connections to polynomial optimization. This and previous work has shown this for the single-player setting. Moreover, it holds even more so in multi-player settings, where we established that even deciding whether a Nash equilibrium (\ie, mutual best responses) exists is very hard. Therefore, we turned towards suitable relaxations that arose from the game theory and philosophy literature. We studied them, and their relationship to each other and to the Nash equilibrium concept, with a computational lens. 

We find that CDT equilibria stay relatively easy to find, joining the complexity class of finding a Nash equilibrium in \emph{perfect-recall} or \emph{normal-form} games. This is because CDT defines the most local form of deviation, affecting only one decision node at a time. EDT equilibria show a more convoluted picture. In single-player settings, we relate it to polynomial local search via best-response dynamics. Furthermore, without absentmindedness, EDT and CDT equilibria coincide and hence become equally easy (\Cref{rem:w/o absentmindedness edt equals cdt}). \emph{With} absentmindedness, on the other hand, the relevant decision problems for EDT equilibria (in single- or multi-player settings) tend to coincide in complexity with the analogous problems for Nash equilibria under \emph{imperfect} recall.

One conclusion, however, has presented itself in all settings considered throughout this paper: (assuming well-accepted complexity assumptions), CDT equilibria are in general strictly easier to find and decide than EDT and Nash equilibria (\Cref{lem:CDT KKT and CLS} vs \Cref{thm:EDT PLS-C}, \Cref{cor:1PL CDT has FPTAS} vs \Cref{lem:1PL EDT no FPTAS}, and \Cref{thm:CDT is PPAD} vs \Cref{thm:apx EDT sigma2p compl}). Does this imply that CDT-based reasoning is more suitable for computationally-bounded agents?

Finally, the computational differences between EDT equilibria and Nash equilibria have yet to be properly understood, that is, the differences between global optimization of polynomials over a single simplex versus a product of simplices. We leave this open for future work, with a particular interest in the search complexities of these two equilibrium concepts.

\appendix

\section*{Acknowledgments}

We are grateful to Ioannis Anagnostides for the fruitful discussions during the development of this project, and to the anonymous reviewers for their valuable improvement suggestions for this paper. Emanuel Tewolde, Caspar Oesterheld, and Vincent Conitzer thank the Cooperative AI Foundation, Polaris Ventures (formerly the Center for Emerging Risk Research) and Jaan Tallinn's donor-advised fund at Founders Pledge for financial support. Caspar Oesterheld's work is also supported by an FLI PhD Fellowship. Paul Goldberg is supported by EPSRC Grant EP/X040461/1. The work of Prof. Sandholm’s group is supported by the Vannevar Bush Faculty Fellowship ONR N00014-23-1-2876, National Science Foundation grants RI-2312342 and RI-1901403, ARO award W911NF2210266, and NIH award A240108S001. Brian Hu Zhang's work is supported in part by the CMU Computer Science Department Hans Berliner PhD Student Fellowship.

\bibliographystyle{named}
\bibliography{ijcai24}

\begin{thebibliography}{}

\bibitem[\protect\citeauthoryear{Arora and Barak}{2009}]{AroraB09}
Sanjeev Arora and Boaz Barak.
\newblock {\em Computational Complexity - {A} Modern Approach}.
\newblock Cambridge University Press, 2009.

\bibitem[\protect\citeauthoryear{Briggs}{2010}]{Briggs10:Putting}
Rachael Briggs.
\newblock {Putting a Value on Beauty}.
\newblock In {\em {Oxford Studies in Epistemology: Volume 3}}, pages 3--34. {Oxford University Press}, 2010.

\bibitem[\protect\citeauthoryear{Brown and Sandholm}{2018}]{BrownS18}
Noam Brown and Tuomas Sandholm.
\newblock Superhuman {AI} for heads-up no-limit poker: Libratus beats top professionals.
\newblock {\em Science}, 359(6374):418--424, 2018.

\bibitem[\protect\citeauthoryear{Brown and Sandholm}{2019}]{BrownS19}
Noam Brown and Tuomas Sandholm.
\newblock Superhuman {AI} for multiplayer poker.
\newblock {\em Science}, 365(6456):885--890, 2019.

\bibitem[\protect\citeauthoryear{Brown \bgroup \em et al.\egroup }{2015}]{BrownGS15}
Noam Brown, Sam Ganzfried, and Tuomas Sandholm.
\newblock Hierarchical Abstraction, Distributed Equilibrium Computation, and Post-Processing, with Application to a Champion No-Limit {T}exas Hold'em Agent.
\newblock In {\em Proceedings of the 2015 International Conference on Autonomous Agents and Multiagent Systems, {AAMAS} '15}, pages 7--15.

\bibitem[\protect\citeauthoryear{Buresh-Oppenheim and Morioka}{2004}]{BureshM04-NP-search-problems}
Joshua Buresh-Oppenheim and Tsuyoshi Morioka.
\newblock Relativized {NP} search problems and propositional proof systems.
\newblock In {\em Proceedings of the 19th IEEE Conference on Computational Complexity (CCC)}, pages 54--67.

\bibitem[\protect\citeauthoryear{Buss and Johnson}{2012}]{buss2012propositional}
Samuel~R. Buss and Alan~S. Johnson.
\newblock Propositional proofs and reductions between {NP} search problems.
\newblock {\em Annals of Pure and Applied Logic}, 163(9):1163--1182, 2012.

\bibitem[\protect\citeauthoryear{Canny}{1988}]{Canny88}
John Canny.
\newblock Some Algebraic and Geometric Computations in {PSPACE}.
\newblock In {\em Proceedings of the Twentieth Annual ACM Symposium on Theory of Computing}, pages 460--467.

\bibitem[\protect\citeauthoryear{Celli and Gatti}{2018}]{Celli18:Computational}
Andrea Celli and Nicola Gatti.
\newblock Computational Results for Extensive-Form Adversarial Team Games.
\newblock In {\em Proceedings of the Thirty-Second {AAAI} Conference on Artificial Intelligence, (AAAI-18), the 30th innovative Applications of Artificial Intelligence (IAAI-18), and the 8th {AAAI} Symposium on Educational Advances in Artificial Intelligence (EAAI-18)}, pages 965--972.

\bibitem[\protect\citeauthoryear{Chen \bgroup \em et al.\egroup }{2009}]{ChenDT09}
Xi~Chen, Xiaotie Deng, and Shang{-}Hua Teng.
\newblock Settling the Complexity of Computing Two-Player Nash Equilibria.
\newblock {\em J. {ACM}}, 56(3):14:1--14:57, 2009.

\bibitem[\protect\citeauthoryear{Conitzer and Oesterheld}{2023}]{conitzer23:focal}
Vincent Conitzer and Caspar Oesterheld.
\newblock Foundations of Cooperative AI.
\newblock In {\em Thirty-Seventh {AAAI} Conference on Artificial Intelligence}, pages 15359--15367.

\bibitem[\protect\citeauthoryear{Conitzer}{2019}]{Conitzer19:Designing}
Vincent Conitzer.
\newblock Designing Preferences, Beliefs, and Identities for Artificial Intelligence.
\newblock In {\em Proceedings of the Thirty-Third AAAI Conference on Artificial Intelligence}, pages 9755--9759.

\bibitem[\protect\citeauthoryear{Daskalakis and Papadimitriou}{2011}]{DaskalakisP11}
Constantinos Daskalakis and Christos Papadimitriou.
\newblock Continuous Local Search.
\newblock In {\em Proceedings of the Twenty-Second Annual ACM-SIAM Symposium on Discrete Algorithms}, pages 790--804.

\bibitem[\protect\citeauthoryear{Daskalakis \bgroup \em et al.\egroup }{2009}]{DGP09}
Constantinos Daskalakis, Paul~W. Goldberg, and Christos~H. Papadimitriou.
\newblock The Complexity of Computing a {N}ash Equilibrium.
\newblock {\em {SIAM} J. Comput.}, 39(1):195--259, 2009.

\bibitem[\protect\citeauthoryear{Elga}{2000}]{Elga00:Self}
Adam Elga.
\newblock {Self-locating belief and the Sleeping Beauty problem}.
\newblock {\em Analysis}, 60(2):143--147, 2000.

\bibitem[\protect\citeauthoryear{Emmons \bgroup \em et al.\egroup }{2022}]{EmmonsOCC022}
Scott Emmons, Caspar Oesterheld, Andrew Critch, Vincent Conitzer, and Stuart Russell.
\newblock For Learning in Symmetric Teams, Local Optima are Global {N}ash Equilibria.
\newblock In {\em International Conference on Machine Learning, {ICML} 2022}, pages 5924--5943.

\bibitem[\protect\citeauthoryear{Etessami and Yannakakis}{2010}]{EtessamiY10}
Kousha Etessami and Mihalis Yannakakis.
\newblock On the Complexity of {N}ash Equilibria and Other Fixed Points.
\newblock {\em {SIAM} J. Comput.}, 39(6):2531--2597, 2010.

\bibitem[\protect\citeauthoryear{Fearnley \bgroup \em et al.\egroup }{2023}]{FGHS23}
John Fearnley, Paul Goldberg, Alexandros Hollender, and Rahul Savani.
\newblock The Complexity of Gradient Descent: {CLS} = {PPAD} {\(\cap\)} {PLS}.
\newblock {\em J. {ACM}}, 70(1):7:1--7:74, 2023.

\bibitem[\protect\citeauthoryear{Fudenberg and Tirole}{1991}]{Fudenberg91:Game_theory}
Drew Fudenberg and Jean Tirole.
\newblock {\em Game Theory}.
\newblock MIT Press, October 1991.

\bibitem[\protect\citeauthoryear{Ganzfried and Sandholm}{2014}]{GanzfriedS14}
Sam Ganzfried and Tuomas Sandholm.
\newblock Potential-Aware Imperfect-Recall Abstraction with Earth Mover's Distance in Imperfect-Information Games.
\newblock In {\em Proceedings of the Twenty-Eighth {AAAI} Conference on Artificial Intelligence}, pages 682--690.

\bibitem[\protect\citeauthoryear{Gimbert \bgroup \em et al.\egroup }{2020}]{Gimbert20}
Hugo Gimbert, Soumyajit Paul, and B.~Srivathsan.
\newblock A Bridge between Polynomial Optimization and Games with Imperfect Recall.
\newblock In {\em Proceedings of the 19th International Conference on Autonomous Agents and Multiagent Systems, {AAMAS} '20}, pages 456--464.

\bibitem[\protect\citeauthoryear{H{\aa}stad}{2001}]{Hastad01}
Johan H{\aa}stad.
\newblock Some Optimal Inapproximability Results.
\newblock {\em J. {ACM}}, 48(4):798--859, 2001.

\bibitem[\protect\citeauthoryear{Johnson \bgroup \em et al.\egroup }{1988}]{JPY88}
David~S. Johnson, Christos~H. Papadimitriou, and Mihalis Yannakakis.
\newblock How Easy is Local Search?
\newblock {\em Journal of Computer and System Sciences}, 37(1):79--100, 1988.

\bibitem[\protect\citeauthoryear{Kohli \bgroup \em et al.\egroup }{1994}]{KohliKM94}
Rajeev Kohli, Ramesh Krishnamurti, and Prakash Mirchandani.
\newblock The Minimum Satisfiability Problem.
\newblock {\em {SIAM} J. Discret. Math.}, 7(2):275--283, 1994.

\bibitem[\protect\citeauthoryear{Koller and Megiddo}{1992}]{KollerM92}
Daphne Koller and Nimrod Megiddo.
\newblock The Complexity of Two-Person Zero-Sum Games in Extensive Form.
\newblock {\em Games and Economic Behavior}, 4(4):528--552, 1992.

\bibitem[\protect\citeauthoryear{Kovar{\'{\i}}k \bgroup \em et al.\egroup }{2023}]{KovarikOC23}
Vojtech Kovar{\'{\i}}k, Caspar Oesterheld, and Vincent Conitzer.
\newblock Game Theory with Simulation of Other Players.
\newblock In {\em Proceedings of the Thirty-Second International Joint Conference on Artificial Intelligence, {IJCAI} 2023}, pages 2800--2807.

\bibitem[\protect\citeauthoryear{Kovar{\'{\i}}k \bgroup \em et al.\egroup }{2024}]{KovarikOC24}
Vojtech Kovar{\'{\i}}k, Caspar Oesterheld, and Vincent Conitzer.
\newblock Recursive Joint Simulation in Games.
\newblock {\em CoRR}, abs/2402.08128, 2024.

\bibitem[\protect\citeauthoryear{Kuhn}{1953}]{Kuhn53}
Harold~W. Kuhn.
\newblock Extensive Games and the Problem of Information.
\newblock In {\em Contributions to the Theory of Games (AM-28), Volume II}, chapter~11, pages 193--216. Princeton University Press, 1953.

\bibitem[\protect\citeauthoryear{Lambert \bgroup \em et al.\egroup }{2019}]{LambertMS19}
Nicolas~S. Lambert, Adrian Marple, and Yoav Shoham.
\newblock On equilibria in games with imperfect recall.
\newblock {\em Games Econ. Behav.}, 113:164--185, 2019.

\bibitem[\protect\citeauthoryear{Lipton and Markakis}{2004}]{LiptonM04}
Richard~J. Lipton and Evangelos Markakis.
\newblock Nash Equilibria via Polynomial Equations.
\newblock In {\em {LATIN} 2004: Theoretical Informatics, 6th Latin American Symposium}, pages 413--422.

\bibitem[\protect\citeauthoryear{Morioka}{2001}]{Morioka01-Mthesis-PLS}
Tsuyoshi Morioka.
\newblock Classification of Search Problems and Their Definability in Bounded Arithmetic.
\newblock Master's thesis, University of Toronto, 2001.

\bibitem[\protect\citeauthoryear{Nash}{1950}]{Nash48}
John~F. Nash.
\newblock Equilibrium Points in N-Person Games.
\newblock {\em Proceedings of the National Academy of Sciences}, 36(1):48--49, 1950.

\bibitem[\protect\citeauthoryear{Nash}{1951}]{Nash51}
John Nash.
\newblock Non-Cooperative Games.
\newblock {\em Annals of Mathematics}, 54(2):286--295, 1951.

\bibitem[\protect\citeauthoryear{Oesterheld and Conitzer}{2022}]{Oesterheld22:Can}
Caspar Oesterheld and Vincent Conitzer.
\newblock {Can {\em de se} choice be {\em ex ante} reasonable in games of imperfect recall?}
\newblock \url{https://www.andrew.cmu.edu/user/coesterh/DeSeVsExAnte.pdf}, 2022.
\newblock Working paper. Accessed: 2022-12-14.

\bibitem[\protect\citeauthoryear{Papadimitriou}{1994}]{Pap94}
Christos~H. Papadimitriou.
\newblock On the Complexity of the Parity Argument and Other Inefficient Proofs of Existence.
\newblock {\em Journal of Computer and System Sciences}, 48(3):498--532, 1994.

\bibitem[\protect\citeauthoryear{Piccione and Rubinstein}{1997}]{PiccioneR73}
Michele Piccione and Ariel Rubinstein.
\newblock On the Interpretation of Decision Problems with Imperfect Recall.
\newblock {\em Games and Economic Behavior}, 20(1):3--24, 1997.

\bibitem[\protect\citeauthoryear{Renegar}{1992}]{RENEGAR1992255}
James Renegar.
\newblock On the Computational Complexity and Geometry of the First-Order Theory of the Reals. Part I: Introduction. Preliminaries. The Geometry of Semi-Algebraic Sets. The Decision Problem for the Existential Theory of the Reals.
\newblock {\em Journal of Symbolic Computation}, 13(3):255--299, 1992.

\bibitem[\protect\citeauthoryear{Schaefer and Stefankovic}{2017}]{SchaeferS17}
Marcus Schaefer and Daniel Stefankovic.
\newblock Fixed Points, {N}ash Equilibria, and the Existential Theory of the Reals.
\newblock {\em Theory Comput. Syst.}, 60(2):172--193, 2017.

\bibitem[\protect\citeauthoryear{Sch{\"{a}}ffer and Yannakakis}{1991}]{SchafferY91}
Alejandro~A. Sch{\"{a}}ffer and Mihalis Yannakakis.
\newblock Simple Local Search Problems That are Hard to Solve.
\newblock {\em {SIAM} J. Comput.}, 20(1):56--87, 1991.

\bibitem[\protect\citeauthoryear{Shor}{1990}]{Shor90}
Peter~W. Shor.
\newblock Stretchability of Pseudolines is {NP}-Hard.
\newblock In {\em Applied Geometry And Discrete Mathematics, Proceedings of a {DIMACS} Workshop}, pages 531--554.

\bibitem[\protect\citeauthoryear{Stockmeyer}{1976}]{Stockmeyer76}
Larry~J. Stockmeyer.
\newblock The Polynomial-Time Hierarchy.
\newblock {\em Theor. Comput. Sci.}, 3(1):1--22, 1976.

\bibitem[\protect\citeauthoryear{Strotz}{1955}]{Strotz55}
R.~H. Strotz.
\newblock Myopia and Inconsistency in Dynamic Utility Maximization.
\newblock {\em The Review of Economic Studies}, 23(3):165--180, 1955.

\bibitem[\protect\citeauthoryear{Tewolde and Conitzer}{2024}]{Tewolde21}
Emanuel Tewolde and Vincent Conitzer.
\newblock Game Transformations That Preserve Nash Equilibria or Best-Response Sets.
\newblock In {\em Proceedings of the Thirty-Third International Joint Conference on Artificial Intelligence, {IJCAI} 2024}.

\bibitem[\protect\citeauthoryear{Tewolde \bgroup \em et al.\egroup }{2023}]{TewoldeOCG23}
Emanuel Tewolde, Caspar Oesterheld, Vincent Conitzer, and Paul~W. Goldberg.
\newblock The Computational Complexity of Single-Player Imperfect-Recall Games.
\newblock In {\em Proceedings of the Thirty-Second International Joint Conference on Artificial Intelligence, {IJCAI} 2023}, pages 2878--2887.

\bibitem[\protect\citeauthoryear{von Stengel and Koller}{1997}]{vonStengel97:Team}
Bernhard von Stengel and Daphne Koller.
\newblock Team-Maxmin Equilibria.
\newblock {\em Games and Economic Behavior}, 21(1-2):309--321, 1997.

\bibitem[\protect\citeauthoryear{Waugh \bgroup \em et al.\egroup }{2009}]{Waugh09:Practical}
Kevin Waugh, Martin Zinkevich, Michael Johanson, Morgan Kan, David Schnizlein, and Michael~H. Bowling.
\newblock A Practical Use of Imperfect Recall.
\newblock In {\em Eighth Symposium on Abstraction, Reformulation, and Approximation, {SARA} 2009}.

\bibitem[\protect\citeauthoryear{Wichardt}{2008}]{Wichardt08}
Philipp~C. Wichardt.
\newblock Existence of Nash Equilibria in Finite Extensive Form Games with Imperfect Recall: A Counterexample.
\newblock {\em Games Econ. Behav.}, 63(1):366--369, 2008.

\bibitem[\protect\citeauthoryear{Yannakakis}{2003}]{Yannakakis2003}
Mihalis Yannakakis.
\newblock {\em Computational Complexity}, pages 19--56.
\newblock Princeton University Press, 2003.

\bibitem[\protect\citeauthoryear{Zhang and Sandholm}{2022}]{Zhang22:Polynomial}
Brian~Hu Zhang and Tuomas Sandholm.
\newblock Polynomial-Time Optimal Equilibria with a Mediator in Extensive-Form Games.
\newblock In {\em Annual Conference on Neural Information Processing Systems, NeurIPS 2022}.

\bibitem[\protect\citeauthoryear{Zhang \bgroup \em et al.\egroup }{2023}]{Zhang23:Team_DAG}
Brian~Hu Zhang, Gabriele Farina, and Tuomas Sandholm.
\newblock Team Belief {DAG:} Generalizing the Sequence Form to Team Games for Fast Computation of Correlated Team Max-Min Equilibria via Regret Minimization.
\newblock In {\em International Conference on Machine Learning, {ICML} 2023}, pages 40996--41018.

\end{thebibliography}

\clearpage

\clearpage
\section{On \Cref{sec:ir games}}

In this section, we expand on the technical background needed on games with imperfect recall to develop our results. Again, this section closely follows the notation of \citet{TewoldeOCG23} and simply extends their exposition to multi-player settings.

\subsection{Notation} 

Players in $\pls$ will be referred to as \emph{strategic} players. Recall that we can identify their strategy sets with Cartesian products of simplices
\[
    \strats \equiv \bigtimes_{i \in \pls} \bigtimes_{I \in \infs^{(i)}} \Delta^{|A_I| - 1} \quad \textnormal{and} \quad \strats^{(i)} \equiv \bigtimes_{I \in \infs^{(i)}} \Delta^{|A_I| - 1} \, .
\]

For that purpose, we have to fix an ordering of the infosets and actions in a considered game $\Gamma$. Denote the number of infosets of strategic player $i \in \pls$ as $\ninfs^{(i)} := |\infs^{(i)}|$, and fix an ordering $I_1^{(i)}, \ldots, I_{\ninfs^{(i)}}^{(i)}$ of infosets in $\infs^{(i)}$. Similarly, denote the number of actions at infoset $I_j^{(i)} \in \infs^{(i)}$ as $m_j^{(i)} := |A_{I_j^{(i)}}|$, and fix an ordering $a_{j 1}^{(i)}, \ldots, a_{j m_j^{(i)}}^{(i)}$ of actions in $A_{I_j^{(i)}}$.

Then, a strategy $\mu \in \strats$ is uniquely identified with the vector 
\[
    \mu = \big( \mu_{jk}^{(i)} \big)_{i,j,k} \in \bigtimes_{i \in [N]} \bigtimes_{j \in [\ninfs^{(i)}] } \Delta^{m_j^{(i)} - 1} \subset \bigtimes_{i \in [N]} \bigtimes_{j \in [\ninfs^{(i)}] } \R^{m_j^{(i)}}
\]
where $\mu_{jk}^{(i)} \in [0,1]$ is the probability $\mu^{(i)}\big( a_{j k}^{(i)} \mid I_j^{(i)} \big)$ that strategic player $i$ assigns to action $a_{j k}^{(i)}$ at infoset $I_j^{(i)}$.

\subsection{Representation of Polynomials} 

Since we draw some of our complexity results from known results in polynomial optimization, we first note that polynomials shall be represented in the Turing (bit) model, which will be described below. Say, we have a general polynomial function $p : \bigtimes_{i \in [N]} \bigtimes_{j \in [\ninfs^{(i)}] } \R^{m_j^{(i)}} \to \R$ in variables $x = (x_{jk}^{(i)})_{i,j,k}$ and of total degree $d \in \N$. Let $\bm{m} := (m_j^{(i)})_{i,j=1}^{N,\ninfs^{(i)}}$ denote the associated vector of dimensions. Then, the relevant standard monomial basis to $p$ is $\big\{ \prod_{i,j,k = 1}^{N, \ninfs^{(i)}, m_j^{(i)}} (x_{jk}^{(i)})^{D_{ijk}} \big\}_{D \in \MB( d, \bm{m} )}$. Here, each vector $D$ indicates one unique way to distribute the total degree $d$ to the variables, and $\MB( d, \bm{m} )$ denotes the collection of them. That is
\begin{align*}
    \MB( d, \bm{m} ) := &\{ D = (D_{ijk})_{ijk} \in \bigtimes_{i \in [N]} \bigtimes_{j \in [\ninfs^{(i)}] } ( \N \cup \{0\} )^{m_j^{(i)}} : 
    \\
    &\, \sum_{i,j,k = 1}^{N,\ninfs^{(i)},m_j^{(i)}} D_{ijk} \leq d \} \, .
\end{align*}
By abuse of notation, let $x^D := \prod_{i,j,k} (x_{jk}^{(i)})^{D_{ijk}}$. Then, polynomial $p$ can be uniquely represented as $p(x) = \sum_{D \in \MB( d, \bm{m} )} \lambda_D \cdot x^D$ where each $\lambda_D \in \Q$ for computational considerations. Finally, any general polynomials $p$ in this paper are assumed to be represented as a binary encoding of these values $(\ninfs^{(i)})_{i=1}^N, \bm{m}$ and coefficients $(\lambda_D)_{D \in \MB( d,\bm{m} )}$. 

\subsection{Lipschitz Constants} 
\label{app:lps constant}
As polynomial functions, utility functions $\U^{(i)}$ are Lipschitz continuous over strategy space $S$. We will sometimes need access to these Lipschitz constants. \citet{TewoldeOCG23}[Lemma 22] describe how, given a polynomial function $p:\R^n \to \R$ in the Turing (bit) model, one can obtain a Lipschitz constant $L_{\infty}$ of $p$ over the hypercube $[0,1]^n$ w.r.t. the infinity norm within polytime.

One possible Lipschitz constant is the maximum gradient norm over the hypercube
\begin{align*}
    &\max_{x \in [0,1]^n} \{ || \nabla p (x) ||_{\infty} \} = \max_{x \in [0,1]^n} \max_{j \in [n]} | \nabla_j \, p (x) | 
    \\
    &= \max_{j \in [n]} \max_{x \in [0,1]^n} | \nabla_j \, p (x) | \, .
\end{align*}
Consider a dimension $j \in [n]$ and suppose polynomial $\nabla_j \, p (x)$ has monomial coefficients $(\lambda_D)_{D}$. Since all variables $x_{j'}$ are bounded by $1$, we get 
\[
    \max_{x \in [0,1]^n} | \nabla_j \, p (x) | \leq \max_{x \in [0,1]^n} | \sum_D \lambda_D \cdot x^D | \leq \sum_{\lambda_D} |\lambda_D| =: L_j \, ,
\]
which is polytime computable. Hence, we can set
\begin{align*}
    L_{\infty} := \max \{ 1, \max_{j \in [n]} L_j \}  \, .
\end{align*}

Now say, we start with a game $\Gamma$ with imperfect recall. Observe that the strategy space $S$ is a subset of one high-dimensional hypercube $\bigtimes_{i \in [N]} \bigtimes_{j \in [\ninfs^{(i)}] } [0,1]^{m_j^{(i)}}$. By the method described above, we can get a Lipschitz constant $L_{\infty}^{(i)}$ of each player's utility function $\U^{(i)}$, and a Lipschitz constant $L_{\infty}^{(i,j,k)}$ of each partial derivative $\nabla_{jk} \, \U^{(i)}$. Summarize them to one Lipschitz constant 
\begin{align*}
    L_{\infty} := \max \{ 1, \max_{i \in [N]} L_{\infty}^{(i)}, \max_{i \in [N], j \in [\ninfs^{(i)}], k \in [m_j^{(i)}]} L_{\infty}^{(i,j,k)} \}
\end{align*}
for the game $\Gamma$.

Furthermore, if the utility payoffs in $\Gamma$ are bounded, say in $[-2,2]$, then the coefficients $\lambda_D$ will be (at worst) a monomial coefficient from $\U^{(i)}$ times two monomial degrees from $\U^{(i)}$ (due to second derivative). These values are bounded for the by the maximum absolute utility value $2$ and by the squared number of nodes $|\nds|^2$ respectively. Moreover, there will be at most as many monomials as there are terminal nodes in $\Gamma$, which itself is bounded by $|\nds|$ again. Hence, each $L_j \leq 2 |\nds|^3$, and thus $L_{\infty} = \poly(|\nds|)$.

\subsection{From Polynomials to Imperfect-Recall Games} 
\label{app:poly fcts to IR game}

Given a set of polynomials $\bm{p} = (p^{(1)}, \ldots, p^{(N)}) : \bigtimes_{i \in [N]} \bigtimes_{j \in [\ninfs^{(i)}] } \R^{m_j^{(i)}} \to \R^N$, we can construct an associated $N$-player game with imperfect recall $\Gamma$ such that $\Gamma$'s its expected utility functions satisfy $\U^{(i)}(\mu) = p^{(i)}(\mu)$ on $\bigtimes_{i = 1}^{ N } \bigtimes_{j = 1}^{ \ninfs^{(i)} } \R^{m_j^{(i)}}$.

Denote $\supp(\bm{p}) := \{ D \in \MB( d,\bm{m} ) \, : \, \lambda_D^{(i)} \neq 0 \textnormal{ for some } i \in [N] \})$. The constructed game $\Gamma$ shall have $N$ players and an infoset $I_j^{(i)}$ for each $i \in [N]$ and $j \in [\ninfs^{(i)}]$. At $I_j^{(i)}$, there shall $m_j^{(i)}$ action choices. The game tree will have a depth of up to $d+1$. The root $h_0$ will be a chance node that has one outgoing edge to a subtree $T_D$ for each monomial index $D \in \supp(\bm{p})$. An outgoing edge is drawn uniformly at random. Let us build $T_D$ associated to $D$, which, in turn, is associated to monomial $\prod_{i,j,k} (x_{jk}^{(i)})^{D_{ijk}}$. Let $\supp^{\textnormal{ms}}(D)$ be a lexicographically ordered version of the multiset which contains $D_{ij}$ many copies of element $(i,j,k)$ if $D_{ijk} > 0$. Then, going through the list $\supp^{\textnormal{ms}}(D)$ means that we will encounter a variable $x_{jk}^{(i)}$ that degree $D_{ijk}$ exactly $D_{ijk}$-many times back to back. Therefore, starting with the edge that goes into $T_D$, do the following loop:
\begin{enumerate}
    \item Take the next element $(i,j,k)$ of $\supp^{\textnormal{ms}}(D)$.
    \item Add a nonterminal node $h$ to the current edge and assign $h$ to info set $I_j^{(i)}$ of player $i$.
    \item Create $m_j^{(i)}$ outgoing edges from $h$, one for each action at $I_j^{(i)}$. 
    \item At the end of edges $\neq k$, add a terminal node with utility payoff $0$ for all $i \in [N]$.
    \item Go to the $k-th$ edge.
\end{enumerate}
Lastly, once we are through with $\supp^{\textnormal{ms}}(D)$, add a final terminal node $z_D$ to the current edge, with utility payoff $\lambda_D^{(1)} \cdot |\supp(\bm{p})|, \ldots, \lambda_D^{(N)} \cdot |\supp(\bm{p})|$. With this procedure, subtree $T_D$ will have depth $|\supp^{\textnormal{ms}}(D)| = ||D||_1$.

In this reduction, we have that any point $x \in \bigtimes_{i \in [N]} \bigtimes_{j \in [\ninfs^{(i)}] } \R^{m_j^{(i)}}$ that is also in $\bigtimes_{i \in [N]} \bigtimes_{j \in [\ninfs^{(i)}]} \Delta^{m_j^{(i)}-1}$ naturally comprises a strategy in $\Gamma$ that selects the $k$-th action at $I_j^{(i)}$ with probability $x_{jk}^{(i)}$. Moreover, each expected utility function $\U^{(i)}$ of player $i$ in $\Gamma$ satisfies for such a point $x$:
\begin{align*}
    &\U^{(i)}(x) = \sum_{z \in \term} \Prob(z \mid x) \cdot u^{(i)}(z)
    \\
    &= \sum_{D \in \supp(\bm{p})} \Prob(z_D \mid x) \cdot \lambda_D^{(i)} \cdot |\supp(\bm{p})|
    \\
    &= \sum_{D \in \supp(\bm{p})} \bigg[ \Big( \frac{1}{|\supp(\bm{p})|} \cdot \prod_{D_{ijk} > 0} (x_{jk}^{(i)})^{D_{ijk}} \Big) 
    \\
    &\, \quad \quad \cdot \lambda_D^{(i)} \cdot |\supp(\bm{p})| \bigg] 
    \\
    &= \sum_{D \in \MB( d,\bm{m} ) } \lambda_D \cdot \prod_{i,j} (x_{jk}^{(i)})^{D_{ijk}}
    \\
    &= p(x) \, .
\end{align*}
This identity extends to the whole space $\bigtimes_{i \in [N]} \bigtimes_{j \in [\ninfs^{(i)}] } \R^{m_j^{(i)}}$. Finally, the construction of $\Gamma$ takes polytime in the encoding size of $p^{(1)}, \ldots, p^{(N)}$.

\subsection{Further Comments on the Complexity Classes}
\label{app:complexity classes}

\paragraph{Decision Problems} 
\P{}, \NP{}, and $\Sigma_2^{\P}$ are parts of the lower levels of the polynomial hierarchy. For a more detailed treatment we refer to \cite{AroraB09}[Section 5]. The first order of the reals, its subclasses $\exists \forall \mathbb{R}$ and the existential theory of the reals $\exists \mathbb{R}$, as well as algorithms to solve those decision problems are discussed in \cite{RENEGAR1992255,SchaeferS17}. The chain \NP{} $\subseteq \exists \mathbb{R} \subseteq$ \PSPACE $\cap \exists \forall \mathbb{R}$ is due to \cite{Shor90,Canny88}, and it ties into the previous discussion that the solutions of an $\exists \mathbb{R}$ sentence may take on irrational solution values.

Let us discuss the running time of deciding a $\exists \forall \mathbb{R}$-sentence $\exists x \in \R^{n_1} \forall y \in \R^{n_2} F(x, y)$, using the standard variable notation that may overlap with our variable notation in this paper. In its standard form, quantifier-free formula $F$ is assumed to be a Boolean formula $P$ where the atomic predicates can be of the form $g_i \Delta_i 0$. Here, $g_i$ is a polynomial function in integer coefficients and $\Delta_i \in \{>, \geq, =, \neq, \leq, <\}$. Let $n_1$ and $n_2$ be the number of variables of the $\exists$ and $\forall$ quantifiers, $a$ be the length of $P$, $m$ be the number of atomic predicates $g_i \Delta_i 0$, $d$ be an upper bound on the degree of polynomials $g_i$, and $L$ be the bit length to represent the coefficients in the $g_i$. Then, \citet{RENEGAR1992255} gives an algorithm $\mathcal{A}$ that takes time $poly(L,a) \cdot (m d)^{\mathcal{O}(n_1 \cdot n_2)}$ to decide whether such a sentence is true or not. In order for this to be polynomial time, we may require the number of variables $n_1$ and $n_2$ to be constant. For deciding an $\exists \mathbb{R}$-sentence, just set $n_2 = 1$ in the above running time bound. 
    
\paragraph{Total NP search problems} 
Decision problems ask for a yes/no answer. Search problems can ask for more sophisticated answers, usually they ask directly for solutions (if existent) with which we can verify a ``yes'' instances of the associated decision problem. The complexity classes \FP{} and \FNP{} are the search problem analogues of \P{} and \NP{}. We have \P{} $=$ \NP{} if and only if \FP{} $=$ \FNP{}. However, the landscape between \FP{} and \FNP{} has been characterized better than the landscape between \P{} and \NP{}. More specifically, there is a special interest in search problems for which one knows that each problem instance admits a solution (the landscape of total NP search problems). Within that, we will be interested in the three complexity classes \CLS{}, \PLS{}, \PPAD{}, which can be characterized by the method with which one can show that each problem instance admits a solution. For the class \PPAD{} (``Polynomial Parity Arguments on Directed graphs'', \citealp{Pap94}), that is if one can show that the existence of a solution can be proven by a fixed point argument. This is the case for example for the existence of an approximate Nash equilibrium \cite{Nash51,DGP09}. For the class \PLS{} (``Polynomial Local Search'', \citealp{JPY88}), that is if one can show that the existence of a solution can be proven by a local optimization argument on a directed acyclic graph. Since we will prove \PLS{}-membership directly, we will give a precise definition further below. For the class \CLS{} (``Continuous Local Search'', \citealp{DaskalakisP11}), that is if one can show that the existence of a solution can be proven by a local optimization argument on a bounded polyhedral (continuous) domain. Alternatively, \CLS{} can be characterized as the intersection of \PPAD{} and \PLS{} \cite{FGHS23}.

In \Cref{sec:EDT results}, we discuss the differences in complexity of finding an approximate EDT vs CDT equilibrium in single-player settings, and relate it to \PLS{} versus \CLS{}. As of yet, \CLS{} is believed to be a proper subclass of \PLS{}, a belief supported by separation oracles (a kind of conditional separation) for \PPAD{} and \PLS{} \cite{BureshM04-NP-search-problems,buss2012propositional,Morioka01-Mthesis-PLS}. An unconditional separation of these classes would imply \P{} $\subsetneq$ \NP{}.

\paragraph{Definition of \PLS{}}

A \emph{local search problem} is given by a set of instances $\mathcal{J}$. For every instance $J \in \mathcal{J}$ we are given a finite set of feasible solutions $\mathcal{S}(J)$, an objective function $c : \mathcal{S}(J) \to \Q$, and for every feasible solution $s \in \mathcal{S}(J)$ a \emph{neighborhood} $N(s, J) \subseteq \mathcal{S}(J)$. Given an instance $J$, the goal is to compute a \emph{locally optimal solution} $s^*$, that is, a solution that does not have a strictly better neighbor in terms of the objective value.
\begin{defn}[\citet{JPY88}]
The complexity class polynomial local search (\PLS) consists of all local search problems that admit a polytime algorithm for each of the following tasks:
\begin{enumerate}
    \item Given an instance $J$, compute an initial feasible solution;
    \item Given an instance $J$ and a feasible solution $s \in \mathcal{S}(J)$, compute the objective value $C(s)$;
    \item Given an instance $J$ and a feasible solution $s \in \mathcal{S}(J)$, determine if $s$ is a local optimum, or otherwise determine a feasible solution $s \in N(s, J)$ with a strictly higher objective value.
\end{enumerate}
\end{defn}

\section{On \Cref{sec: NEs}}
\label{app: NEs}

In this section, we prove the results in \Cref{sec: NEs}. To that end, we restate results taken from the main body, and give new numbers to results presented first in this appendix.

\begin{prop*}[Restatement of \Cref{prop:gap-equilibrium}]
    Let $\Gamma$ be a 2p0s game with imperfect recall. If $\Delta \le \eps$ then $\Gamma$ admits an $\eps$-Nash equilibrium. Conversely, if $\Gamma$ admits an $\eps$-Nash equilibrium, then $\Delta \le 2\eps$.
\end{prop*}
\begin{proof} \,
    \paragraph{1.}
    Suppose $\Delta \le \eps$. Let 
    \[
        \pi^{(1)} = \argmax_{\mu^{(1)} \in S^{(1)}} \min_{\mu^{(2)} \in S^{(2)}} \U^{(1)}(\mu^{(1)}, \mu^{(2)}) \, ,
    \]
    and     
    \[
        \pi^{(2)} = \argmin_{\mu^{(2)} \in S^{(2)}} \max_{\mu^{(1)} \in S^{(1)}}  \U^{(1)}(\mu^{(1)}, \mu^{(2)}) \, .
    \]
    Then we can show that $(\pi^{(1)}, \pi^{(2)})$ is an $\eps$-Nash equilibrium. Indeed, we have 
    \begin{align*}
        \ubar U &= \min_{\mu^{(2)} \in S^{(2)}} \U^{(1)}(\pi^{(1)}, \mu^{(2)}) \le \U^{(1)}(\pi^{(1)}, \pi^{(2)}) 
        \\
        &\le \max_{\mu^{(1)} \in S^{(1)}}  \U^{(1)}(\mu^{(1)}, \pi^{(2)}) = \bar U \, .
    \end{align*}
    Thus, using $\bar U - \ubar U = \Delta \le \eps$, we obtain the $\eps$-Nash equilibrium conditions
    \[
        \U^{(1)}(\pi^{(1)}, \pi^{(2)}) \geq \max_{\mu^{(1)} \in S^{(1)}}  \U^{(1)}(\mu^{(1)}, \pi^{(2)}) - \epsilon \, ,
    \]
    and 
    \[
        \U^{(1)}(\pi^{(1)}, \pi^{(2)}) \leq \min_{\mu^{(2)} \in S^{(2)}} \U^{(1)}(\pi^{(1)}, \mu^{(2)}) + \epsilon \, .
    \]

    \paragraph{2.}
    Suppose $\mu^*$ is an $\eps$-Nash equilibrium. Then
    \begin{align*}
    \Delta &= \min_{\mu^{(2)} \in S^{(2)}} \max_{\mu^{(1)} \in S^{(1)}}  \U^{(1)}(\mu^{(1)}, \mu^{(2)}) 
    \\
    &\, \quad - \max_{\mu^{(1)} \in S^{(1)}} \min_{\mu^{(2)} \in S^{(2)}} \U^{(1)}(\mu^{(1)}, \mu^{(2)})
    \\
    &\le \max_{\mu^{(1)} \in S^{(1)}}  \U^{(1)}(\mu^{(1)}, (\mu^*)^{(2)}) 
    \\
    &\, \quad - \min_{\mu^{(2)} \in S^{(2)}} \U^{(1)}((\mu^*)^{(1)}, \mu^{(2)})
    \\
    &= \max_{\mu^{(1)} \in S^{(1)}}  \U^{(1)}(\mu^{(1)}, (\mu^*)^{(2)}) - \U^{(1)}(\mu^*)
    \\
    &\, \quad + \U^{(1)}(\mu^*) - \min_{\mu^{(2)} \in S^{(2)}} \U^{(1)}((\mu^*)^{(1)}, \mu^{(2)})
    \\
    &\leq \epsilon + \epsilon = 2 \epsilon \, . \qedhere
\end{align*}
\end{proof}

\begin{figure} 
    \tikzset{
        every path/.style={-},
        every node/.style={draw},
    }
    \forestset{
  subgame/.style={regular polygon,
  regular polygon sides=3,anchor=north, inner sep=5pt},
    }
  \begin{center}
  \begin{forest}
    [,p2,s sep=50pt,l sep=20pt
        [\util1{0},terminal,el={2}{exit}{},yshift=-6.3pt]
        [,nat,s sep=60pt,el={2}{cont}{}
        [$\Gamma$,subgame]
        [,p1,name=p0
        [,p1,name=p1a%,tier=x,el={2}{L}{}
            [,p2,name=p2a [\util1{-1},terminal,name=t1] [\util1{3},terminal]]
            [,p2 [\util1{-1},terminal] [\util1{-1},terminal]]
        ]
        [,p1,name=p1b%,el={2}{R}{}
            [,p2 [\util1{-1},terminal] [\util1{-1},terminal]]
            [,p2,name=p2b [\util1{3},terminal] [\util1{-1},terminal,name=t8]]
        ]
        ]]
    ]
    \draw[infoset1] (p1a) -- (p1b);
    \draw[infoset2] (p2a) -- (p2b);
    \node[fit=(t1)(t8)(p0), draw=p3color, densely dashed, rounded corners,name=box] {};
    \node[above right=0cm of box.north, draw=none, p3color]{$\Gamma'$ := \Cref{fig:forgetting kicker game}};
  \end{forest}
  \end{center}
   \caption{Game used in \Cref{lem:NE iff max min value nonneg} and the proofs of \Cref{thm:Nash-D etr hard,th:apx-nash-hard}. $\Gamma$ is a placeholder game. We use boxes for chance nodes, at which chance plays uniformly at random.}
   \label{app fig:NE iff max min value nonneg}
\end{figure}
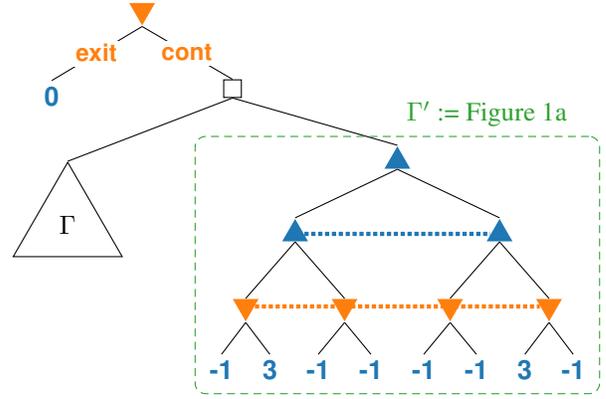

Next, consider the game of \Cref{app fig:NE iff max min value nonneg} where $\Gamma$ is a 2p0s game. Let $\ubar U$ be the max-min value in the subgame $\Gamma$. Moreover, let $v := \min_{z \in \term} u^{(1)}(z) \leq -1$ be the minimum terminal payoff in \Cref{app fig:NE iff max min value nonneg}.

\begin{lemma}
\label{lem:NE iff max min value nonneg}
    If max-min value $\ubar U \geq 0$ in $\Gamma$, then the game in \Cref{app fig:NE iff max min value nonneg} has an exact Nash equilibrium. If $\ubar U \leq - \eps$ in $\Gamma$ for $\eps > 0$ sufficiently small ($\eps \leq -\frac{1}{8v}$), then \Cref{app fig:NE iff max min value nonneg} has no $\eps/2$-Nash equilibrium.
\end{lemma}
\begin{proof} \,
    First we note that the max-min value of the forgetful penalty shoot-out subgame $\Gamma'$ (cf. \Cref{fig:forgetting kicker game}) is $0$, realized for example by P1 randomizing $50/50$ at her infosets, and P2 going left deterministically. The min-max value in $\Gamma'$, on the other hand, is $1$, realized for example by P2 randomizing $50/50$, and P1 going left deterministically. Therefore, $\Delta = 1$, and by \Cref{prop:gap-equilibrium}, there is no $1/4$-Nash equilibrium in $\Gamma'$.

    The proof idea is that given a strategy of P1, P2 would be interested in continuing at the root node if and only if he can achieve negative utility after that. This comes down to if he can achieve negative utility in $\Gamma$ because in $\Gamma'$, there is always a best response with which he can achieve non-positive utility. Suppose that is the case and, thus, P2 is continuing at the root node with sufficiently high probability. Then the profile is already guaranteed to not be an (approximate) Nash equilibrium because the players will never be in a $1/4$-Nash equilibrium in the subgame $\Gamma'$.
     
    \paragraph{1.}
    Suppose $\ubar U \geq 0$ in $\Gamma$. Consider the profile where P2 exits $100\%$ of the time, and where P1 plays her max-min strategies in $\Gamma$ and $\Gamma'$, and P2 plays anything in $\Gamma$ and $\Gamma'$. This makes an exact Nash equilibrium: The only play that has an effect on the ex-ante utility is P2 exiting at his first node. And this is optimal for P2, because upon continuing he would receive a loss of $\geq 0$ instead (recall the 2p0s assumption).
    
    \paragraph{2.}
    Suppose $\ubar U \leq - \eps$ for some $0 < \eps < -\frac{1}{8v}$. For the sake of a contradiction, assume \Cref{app fig:NE iff max min value nonneg} also has an $\eps^2/4$-Nash equilibrium $\pi$. Let $\rho$ be the probability with which P2 continues at the root in profile $\pi$. 
    Then observe that P2 can deviate to the strategy $\mu^{(2)}$ that deterministically continues at the root and that plays a best response to $\pi^{(1)}$ in $\Gamma$ and $\Gamma'$. Therefore, since $\pi$ is an $\eps^2/4$-Nash equilibrium, we get
    \begin{align*}
        \rho v &\leq \U^{(1)}(\pi) \leq \U^{(1)}(\pi^{(1)}, \mu^{(2)}) + \frac{\eps^2}{4}
        \\
        &\leq \frac{1}{2} \ubar U + \frac{1}{2}  0 + \frac{1}{4}\eps  \leq \frac{1}{2} \ubar U + \frac{1}{4} \cdot (-\ubar U) = \frac{1}{4} \ubar U < 0 \, .
    \end{align*}

    Since $v < 0$, we get $\rho > 0$ and hence $\rho \geq \frac{1}{4v} \ubar U$.

    Again, $\pi$ is an $\eps^2/4$-Nash equilibrium. Thus, in particular, no player has an incentive of more than $\eps^2/4$ to deviate to another strategy in the game $\Gamma'$. Hence $\pi|_{\Gamma'}$ would make an approximate Nash equilibrium in $\Gamma'$ (considered as its own game) with rescaled approximation error
    \begin{align*}
        &\frac{1}{\rho \cdot \frac{1}{2}} \cdot \frac{\eps^2}{4} \leq \frac{\eps^2}{\frac{1}{2v} \ubar U} = \frac{2 \cdot (-v) \cdot \eps^2}{- \ubar U} \leq \frac{2 \cdot (-v) \cdot \eps^2}{\eps} 
        \\
        &= - 2 v \eps \leq - 2 v \frac{-1}{8v} = \frac{1}{4} \, .
    \end{align*} 
    This contradicts that $\Gamma'$ has no $1/4$-Nash equilibrium as discussed in the beginning of the proof.

\end{proof}

\begin{thm*}[Restatement of \Cref{thm:Nash-D etr hard}]
    Deciding if a game with imperfect recall admits a Nash equilibrium is $\exists \mathbb{R}$-hard and in $\exists \forall \mathbb{R}$. Hardness holds even for 2p0s games where one player has a degree of absentmindedness of $4$ and the other player has perfect recall.
\end{thm*}
\begin{proof}
    $\exists \forall \mathbb{R}$-membership follows because we can formulate the question of Nash equilibrium existence as the sentence
    \begin{align*}
        \exists \mu \forall \pi : &\, \mu \in S \wedge \bigg[ \pi \notin S \, \lor 
        \\
        &\wedge_{i \in [N]} \Big( \U^{(i)}(\mu) \geq \U^{(i)}(\pi^{(i)}, \mu^{(-i)}) \Big) \bigg]\, .
    \end{align*}
    $\exists \mathbb{R}$-hardness follows from a reduction from \Cref{lem:SPIR is etr complete}. Let $(\Gamma, t)$ be an instance of that decision problem. W.l.o.g.\ we can assume $t = 0$ (otherwise first shift the payoffs in $\Gamma$ by $1 - t$). Insert $\Gamma$ into \Cref{app fig:NE iff max min value nonneg} by letting P1 play in that subgame. Payoffs of P2 in that subgame shall simply be the negative of the payoffs of P1. Call this new game construction $G$. This is a polytime construction. Moreover, asking whether a utility of $0$ can be achieved in original $\Gamma$ is equivalent to asking whether $\ubar U \geq 0$ in subgame $\Gamma$ of $G$.
    
    The equivalence of those decision problems follows by \Cref{lem:NE iff max min value nonneg}: If $\ubar U \geq 0$ then $G$ has a Nash equilibrium. If $G$ has a Nash equilibrium, then it also has an $\epsilon/2$-Nash equilibrium for arbitrary small $\epsilon > 0$. Hence, $\ubar U \geq 0$.

    About the hardness restrictions: \citet{Gimbert20} show hardness of \Cref{lem:SPIR is etr complete} even for degree $4$ polynomials, that is, games with degree of absentmindedness $4$. Moreover, $G$ is a 2p0s game in which P2 has perfect recall.
\end{proof}

\begin{thm*}[Restatement of \Cref{th:apx-nash-hard}]
    {\sc Nash-D} is $\Sigma_2^\P{}$-complete. Hardness holds for 2p0s games with no absentmindedness and \invpoly{} precision.
\end{thm*}
\begin{proof}
    $\Sigma_2^\P{}$-membership: Given an instance $(\Gamma, \eps)$, we can guess a profile $\mu$ and verify in non-deterministic polytime whether it is an $\eps$-Nash equilibrium. Namely, guess $\mu$ to have action probabilities with values with denominators $\leq \frac{2 L_{\infty}}{\eps}$, where $L_{\infty}$ is obtained as in \Cref{app:lps constant}. Note that this is a polysized guess in $(\Gamma,\eps)$. Then, to verify, guess another profile $\pi$ in the same way. Finally, check for each player $i \in \pls$ whether $\U^{(i)}(\mu^{(i)}, \mu^{(-i)}) \geq \U^{(i)}(\pi^{(i)}, \mu^{(-i)}) - \eps$. If so, then this serves as a verification that $\Gamma$ has an $\eps$-Nash equilibrium, and therefore, an exact Nash equilibrium.
    
    This works, because if $\Gamma$ has an exact Nash equilibrium $\mu^*$, then this method is able to find an $\eps$-Nash equilibrium. Namely, $\mu^*$ will be at most $\frac{\epsilon}{2 L_{\infty}}$ away (in $|| \cdot ||_{\infty}$) from a profile $\mu$ that could have been guessed by the method above. And this profile will satisfy for all player $i \in \pls$, and all alternative strategies $\pi^{(i)} \in \strats^{(i)}$
    
    \begin{align*}
        \U^{(i)}(\mu) &= \U^{(i)}(\mu) - \U^{(i)}(\mu^*) + \U^{(i)}(\mu^*) 
        \\
        &\geq \U^{(i)}(\mu^*) - L_{\infty} ||\mu - \mu^*||_{\infty}
        \\
        &\geq \U^{(i)}(\pi^{(i)}, (\mu^*)^{(-i)}) - L_{\infty} \frac{\epsilon}{2 L_{\infty}} 
        \\
        &\, \quad - \U^{(i)}(\pi^{(i)}, \mu^{(-i)}) + \U^{(i)}(\pi^{(i)}, \mu^{(-i)})
        \\
        &\geq \U^{(i)}(\pi^{(i)}, \mu^{(-i)}) - \frac{\epsilon}{2} 
        \\
        &\, \quad - L_{\infty} \Big|\Big|\big(\pi^{(i)}, (\mu^*)^{(-i)}\big) - \big(\pi^{(i)}, \mu^{(-i)}\big)\Big|\Big|_{\infty} 
        \\
        &\geq \U^{(i)}(\pi^{(i)}, \mu^{(-i)}) - \frac{\epsilon}{2} - L_{\infty} \frac{\epsilon}{2 L_{\infty}}
        \\
        &\geq \U^{(i)}(\pi^{(i)}, \mu^{(-i)}) - \epsilon \, .
    \end{align*}    
    
    $\Sigma_2^\P{}$-hardness: We reduce from \Cref{lem:maxmin-hard}, and the idea is analogous to the proof of \Cref{thm:Nash-D etr hard}. Given an instance $(\Gamma, \eps)$ for it, insert  $\Gamma$ in \Cref{app fig:NE iff max min value nonneg} and call the construction $G$. Let $\ubar U$ be the min-max value in $\Gamma$, and $v$ be the minimum terminal payoff in $G$. Set $\eps' := \frac{1}{2} \min\{\epsilon, -\frac{1}{8v}\}$. Then, by \Cref{lem:NE iff max min value nonneg}, we have for corresponding {\sc Nash-D} instance $(\Gamma, \eps')$: If $\ubar U \geq 0$, then $G$ has an exact Nash equilibrium, which will be correctly identified as such by a {\sc Nash-D} solver. If $\ubar U \leq -\eps$, then also $\ubar U \leq -\min\{\epsilon, -\frac{1}{8v}\}$. Hence, $G$ has no $\eps'$-Nash equilibrium, which will be correctly identified as such by a {\sc Nash-D} solver.

    The restrictions on the hardness result follow directly from \Cref{lem:maxmin-hard} or \Cref{th:apx-nash-hard}.
\end{proof}

\begin{cor*}[Restatement of \Cref{cor:deciding positive duality gap is hard}]
    It is $\Sigma_2^\P{}$-complete to distinguish $\Delta = 0$ from $\Delta \ge \eps$ in 2p0s games. Hardness holds for 2p0s games with no absentmindedness and \invpoly{} precision.
\end{cor*}
\begin{proof}
    Reduce from \Cref{th:apx-nash-hard} using \Cref{prop:gap-equilibrium}.
\end{proof}

\begin{prop*}[Restatement of \Cref{prop:Comp NE expo time}]
    {\sc Nash-D} is solvable in time \\
    $\poly \Big( |\Gamma|, \log\frac{1}{\epsilon}, ( m \cdot |\nds| )^{m^2} \Big)$.
\end{prop*}

\begin{proof}

    Let $(\Gamma, \epsilon)$ be an instance of {\sc Nash-D}. Let $m := \sum_{i \in [N]} \sum_{j \in [\ninfs^{(i)}]} m_j^{(i)}$ be the total number of pure actions in the game. This will be the number of variables $n_1$ and $n_2$ in the resulting $\exists \forall \mathbb{R}$-sentences (recall \Cref{app:complexity classes}). By abuse of notation, let $S(\mu)$ be the system of linear (in-)equalities in a profile $\mu$ that describes whether $\mu$ lies in the profile set, that is, the conjunctions of 
    \begin{align*}
        &\mu_{jk}^{(i)} \geq 0 \quad \forall i \in [N], \forall j \in [\ninfs^{(i)}], \forall k \in [m_j^{(i)}]
        \\    
        &\textnormal{and } \sum_{k = 1}^{m_j^{(i)}} \mu_{jk}^{(i)} = 1 \quad \forall i \in [N], \forall j \in [\ninfs^{(i)}] \, .
    \end{align*}

    Notice that the profile set $S$ lies in the standard hypercube $[0,1]^m$ which can be described as the (conjuction) system $B(\mu)$ of linear (in-)equalities
    \begin{align*}
        &\mu_{jk}^{(i)} \geq 0 := y_{jk}^{(i)} \quad \forall i \in [N], \forall j \in [\ninfs^{(i)}], \forall k \in [m_j^{(i)}] \, ,
        \\
        \textnormal{and } \, &\mu_{jk}^{(i)}  \leq 1 := z_{jk}^{(i)} \quad \forall i \in [N], \forall j \in [\ninfs^{(i)}], \forall k \in [m_j^{(i)}] \, .
    \end{align*}
    
     We will make use and adjust the values $y_{jk}^{(i)}$ and $z_{jk}^{(i)}$ later on.
    
    First, we decide the sentence whether there exists ($\exists$) $\mu \in \R^m$ such that for all ($\forall$) $\pi \in \R^m$ we have
    \begin{align}
    \label{NE existence in box}
    \begin{aligned}
         &S(\mu) \wedge B(\mu) 
         \\
         &\wedge \bigg[ \neg S(\pi) \lor \wedge_{i \in [N]} \Big( \U^{(i)}(\mu) \geq \U^{(i)}(\pi^{(i)}, \mu^{(-i)}) \Big) \bigg] \, .
    \end{aligned}
    \end{align}

    If this is false, then we can return that no ($\eps$-)Nash equilibrium exists in $\Gamma$ (since {\sc Nash-D} is a promise problem). If the sentence is true, then we can work on finding an $\eps$-Nash equilibrium. We do it by shrinking the region of profile space $S$ that we may consider further and further, until any point of that region is an $\eps$-Nash equilibrium. Namely, compute a Lipschitz constant $L_{\infty}$ of $\Gamma$ as described in \Cref{app:lps constant}, and run the subdivision method in \Cref{alg:subdivision NEs}.
    
    \begin{algorithm}[H]
    \caption{Subdivison Search for a Nash Equilibrium}
    \label{alg:subdivision NEs}
    \begin{algorithmic}[1]
        \WHILE{$\textnormal{diam} \geq \frac{\epsilon}{2 L_{\infty}}$ }
            \FOR{$i \in [N], j \in [\ninfs^{(i)}], k \in [m_j^{(i)}]$}
                \IF{$\exists \mu \forall \pi : \, (\ref{NE existence in box}) \wedge \, \mu_{jk}^{(i)} \leq \frac{y_{jk}^{(i)} + z_{jk}^{(i)}}{2}$}
                    \STATE $z_{jk}^{(i)} \gets \frac{y_{jk}^{(i)} + z_{jk}^{(i)}}{2}$
                \ELSE
                    \STATE $y_{jk}^{(i)} \gets \frac{y_{jk}^{(i)} + z_{jk}^{(i)}}{2}$
                \ENDIF
                \STATE Update $B$ accordingly
            \ENDFOR
            \STATE $\textnormal{diam} \gets \textnormal{diam} / 2$
        \ENDWHILE
    \end{algorithmic}
    \end{algorithm}
    
    After each for loop, the box $B$ shrinks by $1/2$ along each dimension, while making sure that the the sentence to (\ref{NE existence in box}) remains true. Therefore, once the while loop terminates, there (still) is a profile $\hat{\mu} \in B$ that is an exact Nash equilibrium for $\Gamma$. Select any point $\mu$ that satisfies the linear (in-)equality system $S(\mu) \wedge B(\mu)$. Then, due to termination condition, we have $||\mu - \hat{\mu}||_{\infty} < \frac{\epsilon}{2 L_{\infty}}$. All in all, we can therefore derive by analogous reasoning to the $\Sigma_2^\P{}$-membership proof of \Cref{th:apx-nash-hard} that $\mu$ is an $\eps$-Nash equilibrium of $\Gamma$.

    Running time: Let us assume oracle access to an $\exists \forall \R$ solver for a second. The diameter of $B$ w.r.t. the infinity norm starts at $\textnormal{diam} = 1$, and it halves once after each finished for loop. Any for loop takes $\mathcal{O}(m)$ time. This makes the subdivison algorithm take $\mathcal{O}\Big(m \cdot \big( \log L_{\infty} + \log\frac{1}{\epsilon} \big) \Big)$ time, which is polytime in instance $(\Gamma, \eps)$. Therefore, the bounds $y_{jk}^{(i)}$ and $z_{jk}^{(i)}$ remain polysized. On the other hand, observe that the maximal degree of the polynomial functions in (\ref{NE existence in box}) is bounded by the maximal tree depth in $\Gamma$, which is in turn bounded by $|\nds|$. Hence, by the discussion in \Cref{app:complexity classes}, each $\exists \forall \R$-sentence can be decided in running time 
    \begin{align*}
        &\poly \Big( |\Gamma|, \log\frac{1}{\epsilon} \Big) \cdot \Big( \mathcal{O}(m) \cdot |\nds| \Big)^{\mathcal{O}(m^2)} 
        \\
        &= \poly \Big( |\Gamma|, \log\frac{1}{\epsilon}, ( m \cdot |\nds| )^{m^2} \Big) \, .    
    \end{align*}    
    Finally, solving a linear (in-)equality system to get the point $\mu$ takes $\poly \Big( |\Gamma|, \log\frac{1}{\epsilon} \Big)$-time. This gives the overall running time bound
    \begin{align*}
        &\mathcal{O}\Big(m \cdot \big(\log L_{\infty} + \log\frac{1}{\epsilon}  \big) \Big) \cdot \poly \Big( |\Gamma|, \log\frac{1}{\epsilon}, ( m \cdot |\nds| )^{m^2} \Big)
        \\
        &= \poly \Big( |\Gamma|, \log\frac{1}{\epsilon}, ( m \cdot |\nds| )^{m^2} \Big) \, .
    \end{align*}

\end{proof}
\section{On \Cref{sec: intro MS eqs}}
\label{app:app3}

In this section, we expand on the technical background needed on multiselves equilibria and prove the claims made in \Cref{sec: intro MS eqs}. To that end, we restate results taken from the main body, and give new numbers to results presented first in this appendix.

\subsection{On CDT Utilities and Equilibria}
\label{app:CDT utils and eqs}
\paragraph{CDT Utilities and Derivatives}
\begin{lemma}[\citealp{PiccioneR73,Oesterheld22:Can}]
\label{lem:CDT util equals derivative}
    For all player $i \in \pls$, infosets $I \in \infs^{(i)}$, \emph{pure} actions $a \in A_I$, and strategy $\mu \in S$, we have 
        \[
            \nabla_{I,a} \, \U^{(i)}(\mu) = \sum_{h \in I} \Prob(h \mid \mu) \cdot \U^{(i)}(\mu \mid h a) \, ,
        \]
        where the l.h.s. denotes the partial derivative of utility function $\U^{(i)}$ w.r.t.\ to action $a$ of $I \in \infs^{(i)}$ at point $\mu$.
\end{lemma}
\begin{proof}
    Take a player $i \in \pls$. Recall that $\U^{(i)}$ is a polynomial function over strategy set $S \subset \bigtimes_{i' \in [N]} \bigtimes_{j' \in [\ninfs^{(i')}] } \R^{m_{j'}^{(i')}}$. Fix an infoset $j \in [\ninfs^{(i)}]$ and pure action $k \in [m_j^{(i)}]$. Let $e_{jk}^{(i)}$ denote the unit vector in direction of dimension $(i,j,k)$ of $S$. The partial derivative of $\U^{(i)}$ in that dimension and at a strategy $\mu$ is then defined as
    \begin{align*}
        \nabla_{j k} \, \U^{(i)}|_{\mu} := \lim_{\epsilon \to 0} \frac{1}{\epsilon} \cdot \Big( \U^{(i)}( \mu + \epsilon \cdot e_{j k}^{(i)} ) - \U^{(i)}(\mu) \Big) \, .
    \end{align*}

    Note that $x := \mu + \epsilon \cdot e_{j k}^{(i)}$ is not a profile since its action values at infoset $I_j$ sum up to $1 + \epsilon$. Nonetheless, utility $\U^{(i)}$ and reach probability $\Prob$ as polynomials are still well defined there. Thus
    \begin{align*}
        &\U^{(i)}(x) = \sum_{z \in \term} \Prob(z \mid x ) \cdot u^{(i)}(z) = (\dagger) \, .
    \end{align*}
    Here, product $\Prob(z \mid x)$ is equal to the product $\Prob(z \mid \mu)$, except that factor $\mu_{jk}^{(i)}$ is replaced by factor $\mu_{jk}^{(i)} + \epsilon$. This factor occurs as often in that product as $a_k$ needs to be played in the history of $z$. Multiply out this product and sort the resulting sum by their order in $\epsilon$:
    \begin{align*}
        (\dagger) &= \sum_{z \in \term} \Prob(z \mid \mu) \cdot u^{(i)}(z)
        \\
        &\, \quad + \sum_{z \in \term} \Big( u^{(i)}(z) \cdot \sum_{h \in \hist(z) \cap I_j} \Prob( h \mid \mu) \cdot \epsilon \cdot \Prob (z \mid \mu, h a_k) \Big)   
        \\
        &\, \quad + \mathcal{O}(\epsilon^2)
        \\
        &= \U^{(i)}(\mu) + (\star) + \mathcal{O}(\epsilon^2)
    \end{align*}
    
    Therefore, 
    \begin{align*}
        \nabla_{j k} \, \U^{(i)}|_{\mu} &= \lim_{\epsilon \to 0} \frac{1}{\epsilon} \cdot \Big( \U^{(i)}(\mu) + (\star) + \mathcal{O}(\epsilon^2) - \U^{(i)}(\mu) \Big) 
        \\
        &= \lim_{\epsilon \to 0} \frac{(\star)}{\epsilon}
        \\
        &= \sum_{z \in \term} \sum_{h \in I_j} u^{(i)}(z) \cdot \Prob(h \mid \mu) \cdot \Prob (z \mid \mu, h a_k ) 
        \\
        &= \sum_{h \in I_j} \Prob(h \mid \mu) \cdot \sum_{z \in \term} \Prob (z \mid \mu, h a_k ) \cdot u^{(i)}(z)
        \\
        &= \sum_{h \in I_j} \Prob(h \mid \mu) \cdot \U^{(i)}(\mu \mid h a_k) \, . \qedhere
        \end{align*}
\end{proof}

\paragraph{Alternative Characterizations}

\begin{rem}
\label{rem:CDT util linearity}
    The CDT utility of a player $i \in \pls$ in $\Gamma$ at infoset $I \in \infs^{(i)}$ from randomized action $\alpha \in \Delta(A_I)$ under profile $\mu$  satisfies the linearity property
    \begin{align*}
        \U_{\CDT}^{(i)} (\alpha \mid \mu, I) = \sum_{a \in A_I} \alpha(a) \cdot \U_{\CDT}^{(i)} (a \mid \mu, I) \, .   
    \end{align*}
\end{rem}
\begin{proof}
    We have
    \begin{align*}
        &\sum_{a \in A_I} \alpha(a) \cdot \U_{\CDT}^{(i)} (a \mid \mu, I) 
        \\
        &= \sum_{a \in A_I} \alpha(a) \cdot \Big( \U^{(i)}(\mu) + \nabla_{I,a} \, \U^{(i)}(\mu) 
        \\
        &\, \quad \quad - \sum_{a' \in A_I} \mu(a' \mid I) \cdot \nabla_{I,a'} \, \U^{(i)}(\mu) \Big)
        \\
        &= \sum_{a \in A_I} \alpha(a) \cdot \Big( \U^{(i)}(\mu) - \sum_{a' \in A_I} \mu(a' \mid I) \cdot \nabla_{I,a'} \, \U^{(i)}(\mu) \Big)
        \\
        &\, \quad \quad + \sum_{a \in A_I} \alpha(a) \cdot \nabla_{I,a} \, \U^{(i)}(\mu)
        \\
        &= \U^{(i)}(\mu) - \sum_{a' \in A_I} \mu(a' \mid I) \cdot \nabla_{I,a'} \, \U^{(i)}(\mu)
        \\
        &\, \quad \quad + \sum_{a \in A_I} \alpha(a) \cdot \nabla_{I,a} \, \U^{(i)}(\mu)
        \\
        &= \U_{\CDT}^{(i)} (\alpha \mid \mu, I) \, . \qedhere
    \end{align*}
\end{proof}

\begin{lemma}
\label{lem:CDT charact pure actions}
    A profile $\mu$ is a CDT equilibrium for game $\Gamma$ if and only if for all player $i \in \pls$, all her infosets $I \in \infs^{(i)}$, and all alternative \emph{pure} actions $a \in A_I$, we have
    \begin{align*}
    %\label{exact well-supp CDT cond}
        \U_{\CDT}^{(i)} (a \mid \mu, I) \geq \max_{a' \in A_I} \U_{\CDT}^{(i)} (a' \mid \mu, I) \, .   
    \end{align*}
\end{lemma}
\begin{proof}
    Using \Cref{rem:CDT util linearity}, we have for all $i \in [N]$ and $j \in [\ninfs^{(i)}]$
    \begin{align*}
        \U_{\CDT}^{(i)} ( \mu_{j \cdot}^{(i)} \mid \mu, I_j ) &= \sum_{k \in [m_j^{(i)}] } \mu_{jk}^{(i)}  \cdot \U_{\CDT}^{(i)} (a_k \mid \mu, I_j)
        \\
        &= \sum_{k \in \supp(\mu_{j \cdot}^{(i)}) } \mu_{jk}^{(i)}  \cdot \U_{\CDT}^{(i)} (a_k \mid \mu, I_j) \, ,
    \end{align*}

    and analogously for randomized action $\alpha$ instead of $\mu_{j \cdot}^{(i)}$. This allows us to see that randomized action $\mu_{j \cdot}^{(i)}$ is not optimal in $\Delta(A_{I_j})$ if and only if it does not solely randomize over optimal pure actions.
\end{proof}

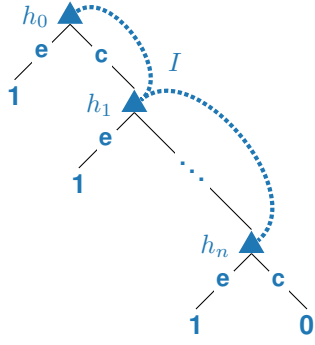
\begin{figure}[t]
    \centering
    \begin{forest}
        [,p1,name=p0,s sep=25pt,l sep=21pt, calign=fixed edge angles, calign primary angle=135, calign secondary angle=225
            [\util1{1},terminal,el={1}{e}{}]
            [,p1,name=p1b,el={1}{c}{},s sep=25pt,l sep=21pt, calign=fixed edge angles, calign primary angle=135, calign secondary angle=225
                [\util1{1},terminal,el={1}{e}{}]
                [,p1,name=p1d,el={1}{\raisebox{1ex}{$\ddots$}}{},s sep=25pt,l sep=21pt, yshift=-20pt, xshift=30pt, calign=fixed edge angles, calign primary angle=135, calign secondary angle=225
                    [\util1{1},terminal, el={1}{e}{}]
                    [\util1{0},terminal, el={1}{c}{}]
                ]
            ]
        ]
        \draw[infoset1] (p0) to [bend left=90] (p1b) to[bend left=90] (p1d);
        \node[above right=10pt of p1b,draw=none,p1color]{$I$};
        \node[left=1pt of p0,draw=none,p1color]{$h_0$};
        \node[left=1pt of p1b,draw=none,p1color]{$h_1$};
        \node[left=1pt of p1d,draw=none,p1color]{$h_n$};
    \end{forest}

    \caption{A single-player single-infoset game with $n$ nodes. Exiting at any of them gives the player a utility of $1$.}
    \label{fig:dont go straight}
\end{figure}

\paragraph{Unreasonable CDT Utilities In Distance}
\, \\
As a first-order Taylor approximation of $\U^{(i)}$, the ex-ante CDT-utility may yield unreasonable utility payoffs for values $\alpha$ far away from $\mu(\cdot \mid I)$. Consider \Cref{fig:dont go straight}. Say, the player enters the game with the strategy $\mu$ that always continues at $I$, the player arrives at $I$, and considers a deviation to action $\alpha$ that deterministically exits now. Then

\begin{align*}
    \U_{\CDT}^{(1)}(\alpha \mid \mu, I) &= 0 + \nabla_{I,\textnormal{exit}} \, \U^{(1)}(\mu) - \nabla_{I,\textnormal{continue}} \, \U^{(1)}(\mu)
    \\
    &= \sum_{h_i \in I} \Prob(h_i \mid \mu) \cdot \U^{(1)}(\mu \mid h_i \circ \textnormal{exit} ) - 
    \\
    &\, \quad \sum_{h_i \in I} \Prob(h_i \mid \mu) \cdot \U^{(1)}(\mu \mid h_i \circ \textnormal{continue} )
    \\
    &= n \cdot 1 - n \cdot 0 = n \, .
\end{align*}

This is unreasonable to expect because the game only has payoffs between $0$ and $1$. However, we can give the following bound on CDT utilities:
\begin{align}
\label{CDT util bound}
\begin{aligned}
    &|\U_{\CDT}^{(i)}(\alpha \mid \mu, I)| 
    \\
    &= |\U^{(i)}(\mu) + \sum_{a \in A_I} (\alpha(a) - \mu(a \mid I)) \cdot \nabla_{I,a} \, \U^{(i)}(\mu)|
    \\
    &\leq |\U^{(i)}(\mu)| + ||\alpha - \mu(\cdot \mid I)||_{\infty} ||\nabla_{I, \cdot} \, \U^{(i)}(\mu)||_{\infty}
    \\
    &\overset{(*)}{\leq} |\U^{(i)}(\mu)| + 1 \cdot |\nds| |\U^{(i)}(\mu)|
    \\
    &\leq 2 |\nds| \max_{z \in \term} |u^{(i)}(z)| \, ,
\end{aligned}
\end{align}

where in $(*)$ we used \Cref{lem:CDT util equals derivative}.

\paragraph{Computational Results}

Inserting \Cref{defn:CDT utility} into \Cref{defn:CDT eq} immediately yields
\begin{rem}
\label{rem:CDT charact as updated ex ante CDT utils}
    A profile $\mu$ is an $\eps$-CDT equilibrium ($\eps \geq 0$) for game $\Gamma$ if and only if for all player $i \in \pls$, all her infosets $I \in \infs^{(i)}$, and all alternative randomized actions $\alpha \in \Delta(A_I)$, we have
    \[
        \sum_{a \in A_I} \mu(a \mid I) \cdot \nabla_{I,a} \, \U^{(i)}(\mu) \geq \sum_{a \in A_I} \alpha(a) \cdot \nabla_{I,a} \, \U^{(i)}(\mu) - \eps \, .
    \]
\end{rem}

\begin{prop*}[Restatement of \Cref{lem:CDT KKT and CLS}; \citealp{TewoldeOCG23}]
\,
    \begin{enumerate}[nolistsep, leftmargin=*]
        \item A profile $\mu$ is a CDT equilibrium of $\Gamma$ if and only if for all player $i \in \pls$, strategy $\mu^{(i)}$ is a KKT-point of 
        \begin{center}
            $\max_{\pi^{(i)} \in S^{(i)}} \U^{(i)}(\pi^{(i)}, \mu^{(-i)})$. %\, .
        \end{center}

        \item Problem {\sc 1P-CDT-S} is \CLS{}-complete.
    \end{enumerate}
\end{prop*}

\begin{proof} \,
    We refer to \citet{TewoldeOCG23}[Theorem 1 and 2] for these results. We shall use characterization \Cref{rem:CDT charact as updated ex ante CDT utils} which is analogous to their Definition~$9$, except that ours is in the ex-ante perspective, so make use of \Cref{lem:CDT util equals derivative}. \citeauthor{TewoldeOCG23} prove the KKT correspondence for single-player settings. Our proof works analogously because the single-player result then implies the multi-player result: A profile $\mu$ is a CDT-equilibrium of $\Gamma$ if and only if for each player $i \in \pls$, their strategy $\mu^{(i)}$ is a CDT-equilibrium of the single-player version of $\Gamma$ where all players $i' \neq i$ play fixed the strategy $\mu^{(i')}$.

    We highlight that all KKT conditions together for a point $\mu \in \bigtimes_{i \in [N]} \bigtimes_{j \in [\ninfs^{(i)}] } \R^{m_j^{(i)}}$ become that there exist KKT multipliers $\{ \tau_{jk}^{(i)} \in \R \}_{i,j,k=1}^{N,\ninfs^{(i)},m_j^{(i)}}$ and $\{ \kappa_j^{(i)} \in \R \}_{i,j = 1}^{N,\ninfs^{(i)}}$ such that
    \begin{align}
    \label{CDT eq KKT conditions}
    \begin{aligned}
        &\mu_{jk}^{(i)} \geq 0 \quad \forall i \in [N], \forall j \in [\ninfs^{(i)}], \forall k \in [m_j^{(i)}] %\nonumber
        \\    
        &\sum_{k = 1}^{m_j^{(i)}} \mu_{jk}^{(i)} = 1 \quad \forall i \in [N], \forall j \in [\ninfs^{(i)}] %\nonumber
        \\
        &\tau_{jk}^{(i)} \geq 0 \quad \forall i \in [N], \forall j \in [\ninfs^{(i)}], \forall k \in [m_j^{(i)}] %\nonumber
        \\
        &\tau_{jk}^{(i)} = 0 \quad \textnormal{or} \quad \mu_{jk}^{(i)} = 0 \quad \forall i \in [N], \forall j \in [\ninfs^{(i)}], \forall k \in [m_j^{(i)}] %\nonumber
        \\
        &\nabla_{jk} \, \U^{(i)}(\mu) + \tau_{jk}^{(i)} - \kappa_j^{(i)} = 0 \, \forall i \in [N], \forall j \in [\ninfs^{(i)}], \forall k \in [m_j^{(i)}]% \, .
        \end{aligned}
    \end{align}
\end{proof}

Next, we remark that \Cref{lem:CDT charact pure actions} motivates another notion of approximate CDT equilibrium: A profile $\mu$ is said to be an $\epsilon$-well-supported CDT equilibrium for a game if it satisfies inequality \Cref{lem:CDT charact pure actions} %(\ref{exact well-supp CDT cond}) 
up to $\eps$ relaxation on the r.h.s..

This approximation concept is polynomially precision-related to $\eps$-CDT equilibria, and it has a close connection to approximate KKT points. 

\begin{lemma}[\citealp{TewoldeOCG23}]
\label{lem:approximation in CDT well supported and KKT}
    Let $\mu$ be a profile of a game $\Gamma$ with imperfect recall. Then:
    \begin{enumerate}
        \item If $\mu$ is an $\eps$-well-supported CDT equilibrium of $\Gamma$ then it is also an $\eps$-CDT equilibrium of $\Gamma$
        \item If $\mu$ is an $\eps$ CDT equilibrium of $\Gamma$ then we can compute a $(3 L_{\infty} |\nds| \sqrt{\eps})$-well-supported CDT equilibrium of $\Gamma$, where $L_{\infty}$ is the Lipschitz constant obtained as in \Cref{app:lps constant}.
        \item If $\mu$ is an $\eps$-well-supported CDT equilibrium of $\Gamma$ then it is an $\eps$-KKT point to \Cref{lem:CDT KKT and CLS}.1, that is, it satisfies KKT conditions (\ref{CDT eq KKT conditions}), except the last one that is replaced by
        \[
            | \nabla_{jk} \, \U^{(i)}(\mu) + \tau_{jk}^{(i)} - \kappa_j^{(i)} | \leq \eps \, .
        \]
    \end{enumerate}
\end{lemma}
\begin{proof}
    Analogous \citet{TewoldeOCG23}[Section H.3].
\end{proof}

\begin{cor*}[Restatement of \Cref{cor:1PL CDT has FPTAS}]
    {\sc 1P-CDT-S} for \invpoly{} precision is in \P{}.
\end{cor*}
\begin{proof}
    Let the reward range of the game be in $[0, 1]$. Then the utility function $\U^{(1)}$ and its gradient are both $L_{\infty}$-Lipschitz continuous for $L_{\infty} = \poly(|\Gamma|)$ (\Cref{app:lps constant}). Thus, it suffices to run projected gradient descent for $\poly(|\Gamma|, 1/\eps)$ steps to obtain an $\eps$-KKT point, which, in return, makes an $\eps$-CLS equilibrium of $\Gamma$. A formal analysis of can be found, for example, in \cite[Lemma C.4]{FGHS23}.)
\end{proof}

\subsection{On Comparing the Solution Concepts}
\label{app:comparing solution concepts}

\begin{prop*}[Restatement of \Cref{lem:EQ hierarchy}; \citealp{Oesterheld22:Can}]
    A Nash equilibrium is an EDT equilibrium. An EDT equilibrium is a CDT equilibrium.
\end{prop*}
\begin{proof}
    If $\mu$ is a Nash equilibrium, then every player $i \in \pls$ plays their optimal strategy in $S^{(i)}$ in response to the profile of the others. In particular, for each of her infosets $I_j$, she plays the optimal randomized action in $\Delta^{ m_j^{(i)} - 1 }$ in response to her own strategy at other infosets $j' \neq j$ and to the profile of the others. Therefore, $\mu$ is an EDT-equilibrium.

    If $\mu$ is an EDT-equilibrium, then for every player $i \in \pls$ and each of her infosets $I_j$, we have that $\mu_{j \cdot}^{(i)}$ is the global optimum of 
    \[
        \max_{\alpha \in \Delta^{ m_j^{(i)} - 1 }} \U^{(i)}( \mu_{I_j \, \mapsto \alpha}^{(i)}, \mu^{(-i)} ) \, .
    \]
    In particular, it will be a KKT point of this maximization problem. These KKT conditions, grouped together for all infosets $I_j$, coincide with with the KKT conditions of \Cref{lem:CDT KKT and CLS}. Therefore, $\mu$ is a CDT-equilibrium. 
\end{proof}

\begin{rem*}[Restatement of \Cref{rem:w/o absentmindedness edt equals cdt}]
    Without absentmindedness, deviation incentives of EDT and of CDT coincide, and so do the equilibrium concepts. Hence, complexity results such as \Cref{lem:CDT KKT and CLS} and \Cref{thm:CDT is PPAD} will apply to EDT equilibria as well.
\end{rem*}

\begin{proof}
    If there is no absentmindedness in a game $\Gamma$, then each player enters each of her infosets at most once within the same game play. Therefore, utility function $\U^{(i)}(\mu) = \U^{(i)}( \mu_{I \mapsto \mu(\cdot \mid I)}^{(i)}, \mu^{(-i)} )$ is linear in an action probability $\mu(a \in I)$. In particular, the first order approximation $\U_{\CDT}^{(i)}(\alpha \mid \mu, I)$ of $\U^{(i)}( \mu_{I \mapsto \alpha}^{(i)}, \mu^{(-i)} )$ becomes exact. Hence, EDT utilities and CDT utilities coincide, yielding coinciding equilibrium sets.

\end{proof}

\begin{rem*}[Restatement of \Cref{rem: 1 infoset yields EDT eq NE}]
    If each player has only one infoset in total, then the EDT equilibria coincide with the Nash equilibria.
\end{rem*}

\begin{proof}
    If a player $i \in \pls$ has only one infoset $I$, then $S^{(i)} = \Delta(A_I)$ as well as $\U^{(i)}( \mu_{I \mapsto \alpha}^{(i)}, \mu^{(-i)} ) = \U^{(i)}( \alpha, \mu^{(-i)} )$ for $\alpha \in \Delta(A_I)$. Hence, the \Cref{def: NE,defn:EDT eq} coincide.
\end{proof}

\section{On \Cref{sec:mse main}}
\label{app:app4}

In this section, we prove the results in \Cref{sec:mse main}. To that end, we restate results taken from the main body, and give new numbers to definitions and results presented first in this appendix.

\subsection{On the Existence of EDT Equilibria}
\label{app:EDT eq existence}

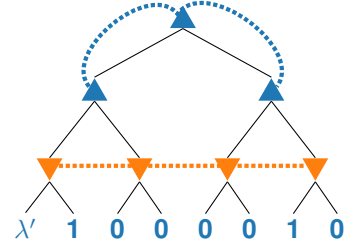
\begin{figure} 
    \tikzset{
        every path/.style={-},
        every node/.style={draw},
    }
    \forestset{
    subgame/.style={regular polygon,
    regular polygon sides=3,anchor=north, inner sep=1pt},
    }
    \begin{center}
        \begin{forest}
            [,p1,name=p0
            [,p1,name=p1a%,el={1}{L}{}
                [,p2,name=p2a%,el={1}{L}{}
                    [\util1{$\lambda'$},terminal,name=t1] [\util1{1},terminal]]
                    [,p2 [\util1{0},terminal] [\util1{0},terminal]
                ]
            ]
            [,p1,name=p1b%,el={1}{R}{}
                [,p2%,el={1}{L}{}
                    [\util1{0},terminal] [\util1{0},terminal]]
                    [,p2,name=p2b [\util1{1},terminal] [\util1{0},terminal,name=t8]
                ]
            ]
            ]           
            \draw[infoset1,bend left=90] (p1a) to (p0) to (p1b);
            \draw[infoset2] (p2a) -- (p2b);
        \end{forest}
    \end{center}
    \caption{As in \Cref{fig:absentminded kicker game} but with payoffs first shifted by $1$ and then scaled with $1/4$, such that $\lambda' \equiv (\lambda + 1)/4$.}
   \label{fig:transformed absentminded kicker game}
\end{figure}

Recall the absentminded penalty shoot-out in \Cref{fig:absentminded kicker game}, and also consider the variation of \Cref{fig:transformed absentminded kicker game} in which its payoffs are shifted and scaled. 

\begin{lemma}
\label{lem:transform eqvl absentminded kicker game}
    In the games of \Cref{fig:absentminded kicker game} for value $\lambda$ and \Cref{fig:transformed absentminded kicker game} for value $\lambda' = (\lambda + 1)/4$, we have coinciding EDT and Nash equilibrium sets.
\end{lemma}
\begin{proof}
    First note that EDT and Nash equilibria coincide within each game due to \Cref{rem: 1 infoset yields EDT eq NE}. Thus we may turn our attention to Nash equilibria. The Nash equilibrium sets of the two games coincide because they are positive affine transforms of each other \cite{Tewolde21}: P1 maximizes her utility function $\U^{(1)}( \cdot ,\mu^{(-1)})$. But its maximum stays the same even if one adds constant $1$ and scales it afterwards with positive $1/4$. Hence, P1's best response sets stay the same in both games. Analogously for P2 whose utility function was shifted by $-1$ and scaled by positive $1/4$. In particular, any Nash equilibria in one game (if existent at all) will remain Nash equilibria in the other game.
\end{proof}

\begin{lemma*}[Restatement of \Cref{lem:NE existence iff lambda geq 1}]
    \Cref{fig:absentminded kicker game} has an EDT equilibrium if and only if $\lambda \geq 3$.
\end{lemma*}
\begin{proof}
    Again, the first equivalence follows from \Cref{rem: 1 infoset yields EDT eq NE}. 
    
    For the second equivalence note that due to \Cref{lem:transform eqvl absentminded kicker game}, \Cref{fig:absentminded kicker game} has a Nash equilibrium for some $\lambda$ if and only if \Cref{fig:transformed absentminded kicker game} has a Nash equilibrium for $\lambda' = (\lambda + 1)/4$. Hence, we can show instead that \Cref{lem:transform eqvl absentminded kicker game} has a Nash equilibrium if and only if $\lambda' \geq 1$.
    
    Suppose $\lambda' \geq 1$ in \Cref{fig:transformed absentminded kicker game}. Then P2 has a (weakly) dominant strategy of playing right. P1 best responds to that with playing left. This profile forms a Nash equilibrium.

    Suppose $\lambda' < 1$ in \Cref{fig:transformed absentminded kicker game}. For the sake of contradiction, suppose furthermore that the game has a Nash equilibrium $\mu^*$. 
    
    First, note that this game has the following best response cycle: P1 plays left $\implies$ P2 plays left $\implies$ P1 plays right $\implies$ P2 plays right $\implies$ P1 plays left. Therefore, $\mu^*$ cannot contain a pure strategy because this would lead to a contradiction due to the best response cycle.

    Therefore, $\mu^*$ is fully randomized. Let $0 < s^*, t^* < 1$ be the probabilities with which P1 and P2 play left in $\mu^*$ respectively. Note that the strategy spaces of P1 and P2 are fully determined by the probability put on left, which is the interval $[0,1]$. As a Nash equilibrium of the game, $s^*$ and $t^*$ have to simultaneously maximize $\U^{(1)}(\cdot,t^*)$ and $\U^{(2)}(s^*, \cdot)$ over $[0,1]$ respectively. Since $s^*$ and $t^*$ both lie in the interior, they must therefore, in particular, be stationary points. We obtain
    \begin{align*}
        \U^{(1)}(s,t^*) &= s^2 \cdot \big( t^* \lambda' + (1-t^*) \big) + (1-s)^2 t^* 
        \\
        &= s^2 \cdot \big( t^* \lambda' + 1 \big) - 2st^* + t^* \, . 
    \end{align*}    
    Hence for P1's perspective
    \begin{align*}
        0 &= \frac{d}{ds} \U^{(1)}(\cdot,t^*) \Big|_{s = s^*} = \Big( 2s \cdot \big( t^* \lambda' + 1 \big) - 2 t^* \Big)\Big|_{s = s^*}
        \\
        &= 2s^* \cdot \big( t^* \lambda' + 1 \big) - 2 t^* \, .
    \end{align*} 
    In particular, $t^* \lambda' + 1 \neq 0$, 
    because otherwise would imply $t^* = 0$. Therefore,
    \begin{align}
    \label{s* value}
        s^* = \frac{t^*}{t^* \lambda' + 1} \, .
    \end{align}
    We can now deduce that actually $t^* \lambda' + 1 > 0$ is the case because otherwise
    \begin{align*}
        &t^* \lambda' + 1 < 0 \implies 0 > \frac{t^*}{t^* \lambda' + 1} = s^* \implies \textnormal{Contradiction.}
    \end{align*}
    Finally, we can deduce that $s^*$ could not have been a best response to $t^*$ because $s = 0$ performs better than $s^*$. We get $\U^{(1)}(0,t^*) = t^*$ and therefore
    \begin{align*}
        &\U^{(1)}(0,t^*) - \U^{(1)}(s^*,t^*) 
        \\
        &= t^* - (s^*)^2 \cdot \big( t^* \lambda' + 1 \big) + 2s^*t^* - t^*
        \\
        &\overset{(\ref{s* value})}{=} - \frac{(t^*)^2}{t^* \lambda' + 1} + \frac{2(t^*)^2}{t^* \lambda' + 1} = \frac{(t^*)^2}{t^* \lambda' + 1} > 0 \, .
    \end{align*}
    Hence, $(s^*, t^*)$ could not have been a Nash equilibrium.    
    
\end{proof}

\begin{thm*}[Restatement of \Cref{thm:EDT exact ETR hard}]
    Deciding whether a game with imperfect recall admits an EDT equilibrium is $\exists \mathbb{R}$-hard and in $\exists \forall \mathbb{R}$. Hardness holds even for 2p0s games where one player has a degree of absentmindedness of $4$ and the other player has perfect recall.
\end{thm*}
\begin{proof}
    $\exists \forall \mathbb{R}$-membership follows because we can formulate the question of EDT equilibrium existence as the sentence
    \begin{align*}
        \exists \mu \forall \pi : &\, \mu \in S \wedge \bigg[ \pi \notin S \, \lor 
        \\
        &\wedge_{i \in [N], j \in [\ninfs^{(i)}]} \U^{(i)}(\mu) \geq \U^{(i)}( \mu_{I_j^{(i)} \, \mapsto \pi_{j \cdot}^{(i)}}^{(i)}, \mu^{(-i)} ) \bigg]\, .
    \end{align*}

    $\exists \mathbb{R}$-hardness follows from a reduction from \Cref{lem:SPIR is etr complete}. Let $(\Gamma, t)$ be an instance of that decision problem. W.l.o.g.\ we can assume $t = 3$ (otherwise first shift the payoffs in $\Gamma$ by $1 - t$). Create a new game $G$ by attaching $\Gamma$ to the game $G'$ of \Cref{fig:absentminded kicker game}, by replacing the terminal node on the bottom left of $G'$ with action history (left, left, left) with the root node of $\Gamma$. Let P1 play the subgame $\Gamma$ in $G$ and receive its payoffs. Payoffs of P2 in that subgame shall simply be the negative of the payoffs of P1. This is a polytime construction. Let us show equivalence of the decision problems. 
    
    Suppose a utility $\geq 3$ can be achieved in $\Gamma$. Then, there is also an optimal strategy $\pi$ for $\Gamma$ with utility $\geq 3$. By \Cref{lem:EQ hierarchy}, this is also an EDT equilibrium of $\Gamma$. Then, the profile ((left, $\pi$), right) makes an EDT equilibrium in $G$: P1 cannot improve at her first (and only relevant) infoset because P2 is going right, and P2 cannot improve at his infoset because he would receive a utility (loss) $\geq 3$ upon going left. 

    Suppose $G$ has an EDT equilibrium $\mu$. Let $\lambda$ be the de-se utility that $\mu$ achieves in subgame $\Gamma$. Observe that the infosets of subgame $\Gamma$ in $G$ are separated from the other infosets in $G$. In particular, P1 knows when she is in subgame $\Gamma$ of $G$, and P2 does not get to act in that subgame. Therefore, $\mu$ restricted to the first infoset of each player must make an EDT equilibrium of \Cref{fig:absentminded kicker game} for that value $\lambda$. By \Cref{lem:NE existence iff lambda geq 1}, we obtain $\lambda > 3$. Hence, a utility of $\geq 3$ can be achieved in $\Gamma$.

    About the hardness restrictions: \citet{Gimbert20} show hardness of \Cref{lem:SPIR is etr complete} even for degree $4$ polynomials, that is, games with degree of absentmindedness $4$. Moreover, $G$ is a 2p0s game in which P2 has perfect recall.
\end{proof}

\begin{thm*}[Restatement of \Cref{thm:apx EDT sigma2p compl}]
    {\sc EDT-D} is $\Sigma_2^\P{}$-complete. Hardness holds for \invpoly{} precision and 2p0s games with one infoset per player and a degree of absentmindedness of $4$.
\end{thm*}

\begin{proof}
    $\Sigma_2^\P{}$-membership: {\sc EDT-D} is the special case of {\sc Nash-D} in which each infoset is played by a new additional player. Thus membership follows from \Cref{th:apx-nash-hard}.

    $\Sigma_2^\P{}$-hardness: We reduce from the $\Sigma_2^\P$-complete problem $\exists \forall$3-DNF-SAT \cite{Stockmeyer76}[Section 4], which is the following problem: given a $3$-DNF formula $\phi(x, y)$ with $k-2$ clauses where $x \in \{0, 1\}^{m-1}$ and $y \in \{0, 1\}^{n-1}$, decide whether $\exists x~\forall y~\phi(x, y)$. We consider the standard multilinear form of a DNF formula $\phi : \R^{m-1} \times \R^{n-1} \to \R$ given by replacing $\land$ with multiplication, $\lor$ with addition, and $\neg z$ with $1-z$, so that, for $x \in \{0, 1\}^{m-1}$ and $y \in \{0, 1\}^{n-1}$, $\phi(x, y)$ is the number of clauses satisfied by $(x, y)$. In particular, formula $\phi(x, y)$ is satisfied if and only if $\phi(x, y) \geq 1$. 
    
    \paragraph{Construction (Part 1)} First, we add variables $x_m$ and $y_n$ and the two clauses $x_m \land \neg y_n$ and $\neg x_m \land y_n$ to $\phi$, to get a formula 
    \begin{align*}
        \phi' &: \R^m \times \R^n \to \R  \\
        (x, y) &\mapsto \phi(x, y) + x_m (1 - y_n) + (1 - x_m) y_n.
    \end{align*}
    Note that $\phi'$ is also a 3-DNF formula with $m+n$ variables and $k$ clauses, and $\phi'$ (as a SAT formula) is $\exists\forall$-satisfiable if and only if $\phi$ is. That is because $y_n$ can always be set to $x_m$ in order to not satisfy the last two added clauses. The sole purpose of these two clauses will be, in our game construction later on, to ensure that if there is no equilibrium, there is also no $\epsilon$-equilibrium. Next, construct a 2p0s game as follows. P1 and P2 each have one infoset of $m+1$ and $n+2$ actions. For randomized action profile $(x,y) \in \Delta(m+1) \times \Delta(n+2) =: X \times Y$, we set P1's utility function as
    \begin{align*}
    \begin{aligned}
        U^{(1)}(x, y) := {}& (1 - y_{n+2}) \qty(\phi'(mx_{1:m}, ny_{1:n}) - \frac12)
        \\&- L \psi_{m}(x) + L\psi_{n}(y) \, .
    \end{aligned}
    \end{align*}
    Here, $x_{1:m}$ denotes indexing, that is, the subvector of $x$ comprising of the first $m$ elements, and analogous for $y_{1:n}$. For simplicity of notation, we will hereafter write $\tilde x = m x_{1:m}$ and $\tilde y = ny_{1:n}$. Moreover, $\psi_{m}$ is defined as
    \begin{align*}
        \psi_{m}(x) := \sum_{i=1}^{m} (mx_i)^2 (1-mx_i)^2,
    \end{align*}
    and $\psi_{n}$ is defined analogously. Lastly, $L$ is a large number to be picked later. Note that we have $\psi_{m} \ge 0$, and equality if and only if $\tilde x_i \in \{0, 1\}$ for all $i \in \{1, \dots, m\}$. Also note that $U^{(1)}$ has polynomial degree at most $4$.

    \paragraph{Intuition} With this utility function, the two players play the multilinear form of $\phi'$ against each other, determined by a rescaling of randomization $x_{1:m}$ and $ny_{1:n}$. In order to make those rescalings practicall boolean, we harshly punish with $L \cdot \psi$ any values in $(x_{1:m}, ny_{1:n})$ that do not rescale to close to boolean. Then, $\qty(\phi'(mx_{1:m}, ny_{1:n}) - \frac12)$ will be strictly positive / negative depending on if $\phi'$ is satisfied or not. P2 has an additional action alternative $e_{n+2}$ that also allows him to not play this game $\phi'$ in the first place if he -- as the minimizer -- does not see how to satisfy $\phi'$. Finally, actions $e_{m+1}$ of P1 and $e_{n+1}$ of P2 have the sole purpose to fill up the vector sum to $1$ in order to represent an element of the simplex, i.e., a probability vector.

    \paragraph{Construction (Part 2)}
    Use \Cref{app:poly fcts to IR game} to construct a 2p0s imperfect-recall game $\Gamma$ whose utility function is $U^{(1)}$ (that is, $U^{(2)}=-U^{(1)}$). In particular, $\Gamma$ has a degree of absentmindedness at most $4$ and one infoset for each player. Due to the latter, we can speak of an $\eps$-equilibrium which entails both, $\eps$-EDT-equilibria and $\eps$-nash equilibria. Next, set
    \begin{align*}
        % \label{values in s2p compl}
        R &:= k \cdot (\max \{ m, n \} )^3 \\
        \eps &:= \Big( \frac{1}{28k} \Big)^2 \\
        L &:= \frac{8R}{\eps} \, . 
    \end{align*}
    Then the construction of corresponding {\sc EDT-D} instance $(\Gamma,\eps)$ takes polytime and $\eps$ is of \invpoly{} precision. The remaining goal is to proof that 

        \begin{enumerate}[leftmargin=1.5cm]
        \item[Claim 1:] if $\phi'$ is $\exists\forall$-satisfiable, then $\Gamma$ admits an exact equilibrium.
        \item[Claim 2:] if $\Gamma$ admits an $\eps$-equilibrium, then from it, we can construct an assignment $\tilde x \in \{0,1\}^m$ that shows $\exists\forall$-satisfiability of $\phi'$.
    \end{enumerate}
    
    Those two claims together imply that $\Gamma$ either admits an exact equilibrium or no $\eps$-equilibrium. As a conclusion, $\phi$ will be a ``yes'' (and resp. ``no'') instance of $\exists \forall$3-DNF-SAT if and only if $\phi'$ is if and only if the corresponding {\sc EDT-D} instance $(\Gamma, \epsilon)$ has an exact equilibrium (and resp. has no $\eps$-equilibrium equilibrium), which is the decision question of {\sc EDT-D}. This completes the reduction.

    \paragraph{Claim 1:} Suppose $\phi'$ is $\exists\forall$-satisfiable, that is, there exists $\tilde x \in \{0, 1\}^m$ for which $\phi'(\tilde x, \tilde y) \ge 1$ for all $\tilde y\in \{0, 1\}^n$. We claim that in this case the profile $(x^*, y^*)$ where $x^* := ( \tilde x / m, 1 - \norm{\tilde x / m}_1)$ and $y^* :=  e_{n+2}$ (i.e., always play pure action $n+2$) is an exact equilibrium. Indeed, we have $U^{(1)}(x^*, y^*) = 0$,
        \begin{align*}
             \max_{x \in X} \,  U^{(1)}(x, y^*) = \max_{x \in X} \, -L \psi_m(x) = 0
        \end{align*}
        and
        \begin{align*}
           &\min_{y \in Y} \, U^{(1)}(x^*, y) \\
           &= \min_{y \in Y} \, \qty((1 - y_{n+2}) \qty(\phi'(\tilde x, \tilde y) - \frac{1}{2}) + L\psi_{n}(y)) \\
           &\ge \min_{y \in Y} \, (1 - y_{n+2}) \cdot \frac12 + 0 \ge 0.
        \end{align*}

    Hence, neither player can unilaterally improve on the outcome of $(x^*, y^*)$. Before we get to Claim 2, we will need two lemma-like observations.

    \paragraph{(I) Best Response and Integrality}
    We say that $x$ is {\em $\delta$-integral} for $0 < \delta < 1/2$ if $\tilde x_i \in [0, \delta] \cup [1-\delta, 1+\delta]$ for every $1 \le i \le m$. We define it analogously for $y$ where we now have to check for $1 \leq j \leq n$. We claim that if $x$ is an $\eps$-best response to $y$ (resp. $y$ is an $\eps$-best response to $x$), then $x$ (resp. $y$) is $\sqrt{\eps}$-integral (note that $\sqrt{\eps} \leq 1/28 < 1/2$). We show its contraposition. First, observe that $\phi'$ has $k$ terms of degree at most $3$, and that $\tilde x_i , \tilde y_j \in [0, \max \{ m, n \}]$ for all $i, j$. Hence, parameter $R$ was chosen large enough such that we have for all $(x,y) \in X \times Y$ that $0 \le \phi'(\tilde x, \tilde y) \le R$. For the contraposition, suppose $x$ is not $\sqrt{\eps}$-integral, that is, there is $i \in [m]$ such that $\tilde x_i \notin [0, \sqrt{\eps}] \cup [1 - \sqrt{\eps}, 1 + \sqrt{\eps}]$. Then
    $$
        \psi_m(x) \ge \tilde x_i^2 (1 - \tilde x_i)^2 \ge \qty(\frac12)^2 (\sqrt{\eps})^2 = \eps / 4
    $$
    where the second inequality is based on the fact that $\tilde x_i$ must have at least $1/2$ distance from either $0$ or $1$ (or both). Hence, by how we set $L$, we have
    \begin{align*}
        U^{(1)}(x, y) &\le 1 \cdot ( R - 1/2) - L \psi_m(x) + L \psi_n(y) 
        \\
        &\le R - \frac{ 8R}{ \eps } \cdot \frac{\eps}{4} + L \psi_n(y) 
        \\
        &= R - 2R + L \psi_n(y) 
        \\
        &\le L \psi_n(y) - 1 \, .
    \end{align*}

    But then P1 at least $1/2$ utility units of incentives to deviate to her pure action $e_{m+1}$ because
    \begin{align*}
        U^{(1)}(e_{m+1}, y) &\ge 1 \cdot ( 0 - 1/2) - L \cdot 0 + L \psi_n(y) 
        \\
        &= L \psi_n(y) - 1/2 \, .
    \end{align*}
    In particular, $x$ couldn't have been an $\eps$-best response to $y$ since we set $\eps < 1/2$. Almost analogous reasoning yields that if $y$ is an $\eps$-best response to $x$, then $y$ is $\sqrt{\eps}$-integral. The only difference is that we derive
    $$
        U^{(1)}(x, y) \ge 1 \cdot ( 0 - 1/2) - L \psi_m(x) + 2R \ge - L \psi_m(x) + 3/4
    $$
    and
    $$
        U^{(1)}(x, e_{n+2}) \le 0 \cdot ( \dots ) - L \psi_m(x) + 0 = - L \psi_m(x) \, .
    $$
    
    \paragraph{(II) Integrality and Value Approximation} 

    Suppose $(x',y')$ are $\sqrt{\eps}$-integral, and let $[\cdot]$ denote element-wise rounding. Recall that we denote $\tilde x' = m \cdot x_{1:m}'$ and analogous for $\tilde y'$. Then $\tilde x'$ and $[\tilde x']$ are at most $\sqrt{\eps}$ distant from each other in each entry $i$; and same for $\tilde y'$ and $[\tilde y']$. Hence, using how we set $\eps$, we can derive that 
    \begin{align*} 
        |\phi'(\tilde x', \tilde y') - \phi'([\tilde x'], [\tilde y'])| &\le k \qty((1+\sqrt{\eps})^3 - 1^3) 
        \\
        &\overset{\sqrt{\eps} < 1}{\le} k ( 1 + 3 \sqrt{\eps} + 3 \sqrt{\eps} + \sqrt{\eps} - 1) 
        \\
        &= 7k\sqrt{\eps} = 1/4 \, .
    \end{align*}
    The first inequality comes from the fact that that $\phi'$ has $k$ terms with unit coefficients and degree at most $3$, so that the maximal difference in a term is if $\tilde x_i'$ (resp. $\tilde y_j'$) is $1+\sqrt{\eps}$ and $[\tilde x_i']$ (resp. $[\tilde y_j']$) is $1$. The second inequality uses the Binomial Theorem.

    \paragraph{Starting on Claim 2} 
    Suppose $(x^*,y^*)$ is an $\eps$-equilibrium. By paragraph (I), both $x^*$ and $y^*$ are $\sqrt{\eps}$-integral. Thus, we can set $\tilde x_i := [\tilde x_i^*]$ and $\tilde y_j := [\tilde y_j^*]$ for $i \in [m]$ and $j \in [n]$, and obtain $\tilde x \in \{0,1\}^m$ and $\tilde y \in \{0,1\}^n$. That is, $(\tilde x, \tilde y)$ are Boolean assignments for $\phi'$. We will show that for all possible assignments $y \in \{0,1\}^n$, we have $\phi'(\tilde x, y) \ge 1$. To that end, we will first show an intermediary fact.

    \paragraph{Observation $\phi'(\tilde x, \tilde y) \ge 1$:} For the sake of contraposition, suppose $\phi'(\tilde x, \tilde y) < 1$, that is, since $(\tilde x, \tilde y)$ are exactly integral, $\phi'(\tilde x, \tilde y) = 0$. Then consider P2's alternative strategy $\hat y = (\tilde y / n, 1 - ||\tilde y / n||_1, 0)$. Since $y^*$ is an $\eps$-best response to $x^*$, we derive
    \begin{align*}
        &-\frac{3}{4}(1-y_{n+2}^*) - L \psi_m(x^*) 
        \\
        &= (1-y_{n+2}^*) (\phi'(\tilde x, \tilde y) -\frac{3}{4}) - L \psi_m(x^*)
        \\ 
        &\overset{\textnormal{(II)}}{\le} (1-y_{n+2}^*) (\phi'(\tilde x^*, \tilde y^*) +\frac{1}{4}-\frac{3}{4}) - L \psi_m(x^*) + L \psi_n(y^*)
        \\ 
        &= U^{(1)}(x^*,y^*) \le U^{(1)}(x^*, \hat y) + \eps
        \\
        &= 1 \cdot (\phi'(\tilde x^*, \tilde y) -\frac{1}{2}) - L \psi_m(x^*) + 0 + \eps
        \\
        &\overset{\textnormal{(II)}}{\le} \phi'(\tilde x, \tilde y) +\frac{1}{4} - \frac{1}{2} - L \psi_m(x^*) + \frac{1}{8}
        \\
        &= 0 - \frac{1}{8} - L \psi_m(x^*) \, .
    \end{align*}
    This simplifies to $-\frac{3}{4}(1-y_{n+2}^*) \leq - \frac{1}{8}$, that is,
    \begin{align}
        \label{1 minus y lower bound}
        1-y_{n+2}^* \geq \frac{1}{6} \, .
    \end{align}
    This allows us to show that $x^*$ could not have been an $\eps$-best response to $y^*$. Consider P1's alternative assignment $\tilde z = (\tilde x_{1 : m-1}, 1 - \tilde x_m) \in \{0,1\}^m$, which is $\tilde x$ except for the last bit flipped. Since we started with $\phi'(\tilde x, \tilde y) = 0$, we in particular have $\tilde x_m (1 - \tilde y_n) + (1 - \tilde x_m) \tilde y_n = 0$. But then $\tilde z_m (1 - \tilde y_n) + (1 - \tilde z_m) \tilde y_n = 1$, and thus $\phi'(\tilde z, \tilde y) = 1$. Moreover, we also have versus $\tilde y^*$ that
    \begin{align*}
        \phi'(\tilde z, \tilde y^*) &\overset{\textnormal{(II)}}{\ge} \phi'(\tilde z, \tilde y) - 1/4 = 3/4 + 0 = 3/4 + \phi'(\tilde x, \tilde y)
        \\
        &\overset{\textnormal{(II)}}{\ge} 3/4 + \phi'(\tilde x^*, \tilde y^*) - 1/4 = \phi'(\tilde x^*, \tilde y^*) + 1/2
    \end{align*}
    Finally, consider P1's alternative strategy $z := (\tilde z / m, 1 - ||\tilde z / m||_1)$. Then
    \begin{align*}
        &U^{(1)}(z,y^*) = (1-y_{n+2}^*) (\phi'(\tilde z, \tilde y^*) -\frac{1}{2}) - 0 + L \psi_n(y^*)
        \\
        &\geq (1-y_{n+2}^*) (\phi'(\tilde x^*, \tilde y^*) +\frac{1}{2} -\frac{1}{2}) - L \psi_m(x^*) + L \psi_n(y^*)
        \\
        &= (1-y_{n+2}^*) \cdot \frac{1}{2} + U^{(1)}(x^*,y^*) \overset{(\ref{1 minus y lower bound})}{\ge} \frac{1}{6} \cdot \frac{1}{2} + U^{(1)}(x^*,y^*)
        \\
        &\overset{ \frac{1}{12} > \eps }{>} U^{(1)}(x^*,y^*) + \eps \, .
    \end{align*}
    Therefore, $x^*$ could not have been an $\eps$-best response to $y^*$.
    
    \paragraph{Completing Claim 2} By above observation, we have $\phi'(\tilde x, \tilde y) \ge 1$. 
    Let $y \in \{0,1\}^n$ be any possible assignment in the $3$-DNF formula $\phi'(\tilde x, \cdot)$. Define $\hat y := (y/n, 1 - ||y/n||_1, 0)$. Then, since $y^*$ is an $\eps$-best response to $x^*$, we can derive
    \begin{align*}
        &\phi'(\tilde x, y) - \frac{1}{4} - L \psi_m(x^*) \overset{\textnormal{(II)}}{\ge} \phi'(\tilde x^*, y) - \frac{1}{4} - \frac{1}{4} - L \psi_m(x^*)
        \\
        &= U^{(1)}(x^*, \hat y) \ge U^{(1)}(x^*, y^*) - \eps
        \\
        &= (1-y_{n+2}^*) (\phi'(\tilde x^*, \tilde y^*) -\frac{1}{2}) - L \psi_m(x^*) + L \psi_n(y^*) - \eps
        \\
        &\ge (1-y_{n+2}^*) (\phi'(\tilde x, \tilde y) -\frac{1}{4} -\frac{1}{2}) - L \psi_m(x^*) + L \psi_n(y^*) - \eps
        \\
        &\ge (1-y_{n+2}^*) (1 -\frac{3}{4}) - L \psi_m(x^*) - \eps
        \\
        &\ge - L \psi_m(x^*) - \eps \, .
    \end{align*}
    But this implies
    \begin{align*}
        \phi'(\tilde x, y) \ge \frac{1}{4} - \eps > 0 \, .
    \end{align*}
    Since $\phi'(\tilde x, y)$ is an integer, we have $\phi'(\tilde x, y) \geq 1$, that is, $\phi'(\tilde x, y)$ is satisfied. Note that $y \in \{0,1\}^n$ was chosen arbitrarily, hence, we conclude that $\phi'$ is $\exists\forall$-satisfiable (with $\tilde x$).
\end{proof}

\subsection{On the Search of EDT Equilibria in Single-Player}
\label{app:EDT eq search in 1P}

\begin{thm*}[Restatement of \Cref{thm:EDT PLS-C}]
    {\sc 1P-EDT-S} is \PLS{}-complete when the branching factor is constant. Hardness holds even when the branching factor and the degree of absentmindedness are $2$.
\end{thm*}

We will prove this result over multiple steps. First, we consider the corresponding polynomial optimization problem to {\sc 1P-EDT-S}.

\begin{defn}
\label{defn:NE-PolyOverSimplices}
    An instance of the search problem {\sc NE-Poly-S} consists of
    \begin{enumerate}
        \item integers $\ninfs$ and $(m_j)_{j \in [\ninfs]}$ in binary, determining simplices $S_j := \Delta^{m_j - 1}$
        \item a polynomial function $p : \bigtimes_{j \in [\ninfs]} \R^{m_j} \to \R$ in the Turing (bit) model, and
        %\item a Lipschitz constant $L_{\infty} > 0$ w.r.t. $||\cdot||_{\infty}$, and
        \item a precision parameter $\epsilon > 0$ in binary.
    \end{enumerate}
    A solution consists of a point $x \in S := \bigtimes_{j \in [\ninfs]} S_j$ such that for all $j \in [\ninfs]$ and $y_j \in S_j$, we have $p(x) \geq p(y_j, x_{-j}) - \epsilon$.
\end{defn}

\begin{lemma}
\label{lem:1P EDT eqvl to NE poly}
    {\sc 1P-EDT-S} is computationally equivalent to {\sc NE-Poly-S}.
\end{lemma}
\begin{proof}
    This follows straightforwardly from the connection of imperfect-recall games and polynomial optimization described in \Cref{sec:ir games,app:poly fcts to IR game}. It only requires the realization that condition $p(x) \geq p(y_j, x_{-j}) - \epsilon$ corresponds to $\U^{(1)}(\mu^{(1)}) \geq \U^{(1)}( \mu_{I \mapsto \alpha}^{(1)}) - \eps$ in this connection.
\end{proof}

For \PLS{}-hardness, it is enough to work on the the hypercube as a domain. 

\begin{defn}
\label{defn:NE-PolyOverCube}
    An instance of the search problem {\sc Cube-NE-Poly-S} consists of
    \begin{enumerate}
        \item integer $\ninfs$ in binary, determining hypercube $[0,1]^{\ninfs}$
        \item a polynomial function $p : \R^{\ninfs} \to \R$ in the Turing (bit) model, and
        %\item a Lipschitz constant $L_{\infty} > 0$ w.r.t. $||\cdot||_{\infty}$, and
        \item a precision parameter $\epsilon > 0$ in binary.
    \end{enumerate}
    A solution consists of a point $x \in [0,1]^{\ninfs}$ such that for all $j \in [\ninfs]$ and $y_j \in [0,1]$, we have $p(x) \geq p(y_j, x_{-j}) - \epsilon$.
\end{defn}

\begin{lemma}
\label{lem:Cube NE Poly reduces to NE poly}
    {\sc Cube-NE-Poly-S} reduces to {\sc NE-Poly-S}.
\end{lemma}
\begin{proof}
    Take an instance $J = (\ninfs, p : \R^{\ninfs} \to \R, \, \epsilon)$ of {\sc Cube-NE-Poly-S}. Define the corresponding {\sc NE-Poly-S} instance as $\hat{J} = ( \ninfs, (m_j)_j, \hat{p}, \epsilon)$, where $\forall j \in [\ninfs] : m_j := 2$ and
    \begin{align*}
        &\hat{p} : \bigtimes_{j = 1}^\ninfs \R^2 &&\to \R
        \\
        &\quad \Big( (x_{j1}, x_{j2}) \Big)_{j = 1}^\ninfs &&\mapsto p(x_{11}, x_{21}, \ldots, x_{\ninfs 1}) \, .
    \end{align*}
    Then, if $\hat{x}^*$ is an $\epsilon$-Nash equilibrium of $\hat{J}$, then so will be $(\hat{x}_{j1}^*)_{j \in [\infs]}$ for $J$.
\end{proof}

Next, we show that {\sc Cube-NE-Poly-S} is \PLS{}-hard. For that, we introduce the \PLS{}-complete problem {\sc MaxCut/Flip}.

Let $G = (V,E,w)$ be an undirected graph, $w : E \to \N$ be positive edge weights, and $V = A \sqcup B$ be a vertex partition. Then, the cut of $A \sqcup B$ is defined as all the edges in between $A$ and $B$:
\begin{align*}
    &E \cap (A,B) :=
    \\
    &\{ \{u,v\} = e \in E : u \in A \wedge v \in B \, \textnormal{ or } \, u \in B \wedge v \in A \} \, .
\end{align*}
Its weight is defined as $w(A, B) := \sum_{e \in E \cap (A,B)} w( e )$. The FLIP neighbourhood of partition $A \sqcup B$ is the set of partitions that can be obtained from $(A,B)$ by just moving one vertex from one part to the other:
\begin{align*}
    &\textnormal{FLIP}(A,B) := 
    \\
    &\Big\{ (A \cup \{b\} ) \sqcup ( B \setminus \{b\} ) \Big\}_{b \in B} \cup \Big\{ ( A \setminus \{a\} ) \sqcup (B \cup \{a\} ) \Big\}_{a \in A} \, .
\end{align*}

\begin{defn}
    An instance of the search problem {\sc MaxCut/Flip} consists of an undirected graph $G = (V,E,w)$ with weights $w : E \to \N$. A solution consists of a partition $V = A \sqcup B$ that has maximal cut weight among its \textnormal{FLIP} neighbourhood.
\end{defn}
For problems involving weighted graphs $G = (V,E,w)$, we are interested in their computational complexities in terms of $|V|$, $|E|$, and a binary encoding of all weight values.
\begin{lemma}[\citet{Yannakakis2003},\citet{SchafferY91}]
\label{lem:maxcut PLS-c}
    {\sc MaxCut/Flip} is \PLS-complete.
\end{lemma}

This allows us to proof \PLS{}-hardness of our problems of interest.
\begin{lemma}
\label{lem:maxcut to cubeNEpoly}
    {\sc MaxCut/Flip} reduces to {\sc Cube-NE-Poly-S}.
\end{lemma}
\begin{cor}
\label{cor:PLS hardness of 1P EDT}
    {\sc Cube-NE-Poly-S}, {\sc NE-Poly-S}, and {\sc 1P-EDT-S} are \PLS-hard. Hardness holds even when the branching factor ($\max_j m_j$) and the degree of the polynomial / absentmindedness are $2$.
\end{cor}

\begin{proof}[Proof of \Cref{lem:maxcut to cubeNEpoly}]
    Let $G = (V,E,w)$ be an instance of {\sc MaxCut/Flip}. First, we create the associated {\sc Cube-NE-Poly-S} instance. Let $\ninfs = |V|$ such that each vertex $v \in V$ is associated to an entry $x_v$ in $x \in S = [0,1]^{\ninfs}$. We can define for point $x \in S$ and vertices $t,v \in V$ the function 
    \[
        d_{t,v}(x) := x_t (1-x_v) + (1-x_t) x_v
    \]
    which is maximized if one of the values $x_t$ and $x_v$ is $0$ and the other is $1$; corresponding to $t$ and $v$ belonging to different partitions. Set $W = \sum_{e \in E} w(e)$, $W' := 2 (W+1)$ and
    \begin{align*}
        p(x) = &W' \sum_{v \in V} (\frac{1}{2} - x_v)^2 + \sum_{\{t,v\} \in E} w(t,v) \cdot d_{t,v}(x) \, .
    \end{align*}
    The first summand has a large weight $W'$ and forces any solution $x^*$ to have values $x_v^*$ far away from $\frac{1}{2}$. We can get the Lipschitz constant $L_{\infty} = 15W$ for $p$ over $S$ by the method described in \Cref{app:lps constant}. Set $\epsilon = 1/(2L_{\infty} + 2) < \frac{1}{2}$.

    %Then, we claim: $p$ is $L$-Lipschitz over $S$ and therefore, a solution to this {\sc NE-PolyOverCube} instance will be an $\epsilon$-NE point $x^*$ of $p$. Moreover, $\forall v : x_v^* \neq \frac{1}{2}$. 
    Let $x^*$ be a solution to this {\sc Cube-NE-Poly-S} instance. Then we claim: (1) $L_{\infty}$ is actually a Lipschitz constant of $p$ over $S$, (2) We have $\forall v : x_v^* \leq \epsilon \vee x_v^* \geq 1-\epsilon$. Define $z^* \in \{0,1\}^{\ninfs}$ as $z_v^* = 0$ if $x_v^* \leq \epsilon$ and as $z_v^* = 1$ if $x_v^* \geq 1-\epsilon$. Then, (3) partition $V = \{v \in V: z_v^* = 0\} \sqcup \{v \in V: z_v^* = 1\}$ is a solution to the original {\sc MaxCut/Flip} instance.

    \paragraph{Claim (1):} We have for $u \in V$
    \begin{align*}
        \nabla_u \, p (x) &= - 2W' (\frac{1}{2} - x_u) + \sum_{\{u,v\} \in E} w(u,v) \cdot ( 1 - 2x_v )
        \\
        &= -W' + 2W' x_u + W - 2 \sum_{\{u,v\} \in E} w(u,v) x_v
    \end{align*}
    Using $W \geq 1$ and $W' \leq W$, these polynomial coefficients yield Lipschitz constant
    \begin{align*}
        W' + 2W' + W + 2W \leq 15W =: L_{\infty}
    \end{align*}
    for $p$ over the hypercube.
    
    \paragraph{Claim (2):} We start with $x^*$ being an $\epsilon$-Nash equilibrium. Suppose vertex $u \in V$ has $x_u^* \leq \frac{1}{2}$. Then
    \begin{align*}
        \epsilon &\geq p(0, x_{-u}^*) - p(x^*) 
        \\
        &= W' (\frac{1}{2} - 0)^2 - W' (\frac{1}{2} - x_u^*)^2 
        \\
        &\, \quad + \sum_{\{u,v\} \in E} w(u,v) \cdot d_{u,v}(0, x_{-u}^*) 
        \\
        &\, \quad - \sum_{\{u,v\} \in E} w(u,v) \cdot d_{u,v}(x^*)
        \\
        &= W' \big( \frac{1}{4}  - (\frac{1}{2} - x_u^*)^2 \big) + \sum_{\{u,v\} \in E} w(u,v) \cdot ( x_{v}^* - d_{u,v}(x^*))
        \\
        &= (\dagger)
    \end{align*}
    Note that with $x_u^* \leq \frac{1}{2}$, we have
    \[
        \frac{1}{4}  - (\frac{1}{2} - x_u^*)^2 = (1- x_u^*) x_u^* \geq \frac{1}{2} x_u^*
    \]
    and
    \begin{align*}
        x_{v}^* - d_{u,v}(x^*) = x_{v}^* - x_{u}^* + x_{u}^* x_{v}^* - x_{v}^* + x_{u}^* x_{v}^* \geq - x_{u}^* \, .
    \end{align*}
    Therefore, recalling that $u$ was a fixed vertex, we can continue with
    \begin{align*}
        (\dagger) &\geq W' \cdot \frac{1}{2} x_u^* + \sum_{\{u,v\} \in E} w(u,v) \cdot (-x_{u}^*)
        \\
        &= x_u^* \Big( \frac{1}{2}W' - \sum_{\{u,v\} \in E} w(u,v) \Big) \geq x_u^* \Big( W + 1 - W \Big) 
        \\
        &= x_u^*
    \end{align*}

    On the other hand, suppose that vertex $u \in V$ has $x_u^* > \frac{1}{2}$. Then
    \begin{align*}
        \epsilon &\geq p(1, x_{-u}^*) - p(x^*) 
        \\
        &= W' \big( \frac{1}{4}  - (\frac{1}{2} - x_u^*)^2 \big) 
        \\
        &\, \quad + \sum_{\{u,v\} \in E} w(u,v) \cdot ( 1 - x_{v}^* - d_{u,v}(x^*))
        \\
        &\overset{(*)}{\geq} W' \cdot \frac{1}{2} (1- x_u^*) + \sum_{\{u,v\} \in E} w(u,v) \cdot (-(1-x_{u}^*))
        \\
        &\geq (1 - x_u^*) \Big( \frac{1}{2} W' - \sum_{\{u,v\} \in E} w(u,v) \Big) 
        \\
        &\geq 1 - x_u^*
    \end{align*}
    which implies $x_u^* \geq 1 - \epsilon$. In $(*)$, we used
    \begin{align*}
        1 - x_{v}^* - d_{u,v}(x^*) &= 1 - x_{u}^* - 2 x_{v}^* + 2 x_{u}^* x_{v}^* 
        \\
        &= (1 - x_{u}^*)(1 - 2 x_{v}^*) \geq -(1 - x_{u}^*) \, .
    \end{align*}

    \paragraph{Claim (3):} Any point $z = \in \{0,1\}^{\ninfs}$ induces a partition 
    \[
        V = A(z) \sqcup B(z) := \{v \in V: z_v = 0\} \sqcup \{v \in V: z_v = 1\} \, .
    \]
    Moreover, for any such point $z$ and vertices $t,v \in V$, we have
    \[
        d_{t,v}(z) = 
        \begin{cases} 
            0 &\textnormal{if } z_t, z_v \in A(z) \textnormal{ or } z_t, z_v \in B(z) \\ 
            1 &\textnormal{else} 
            \end{cases} \, .
    \]
    Therefore, the cut weight associated to point $z = \in \{0,1\}^{\ninfs}$ has a relationship to polynomial $p$ in the form of
    \begin{align}
    \label{obj and cut weight}
        p(z) = W' \cdot \ninfs \cdot \frac{1}{4} + w\Big( A(z), B(z) \Big) \, .
    \end{align}
    Now define $z^*$ as
    \[
        z_v^* := \begin{cases} 0 &\textnormal{if } x_v \leq \epsilon \\ 1 &\textnormal{if } x_v \geq 1 - \epsilon \end{cases} \, .
    \]
    Let us now show that its induced partition is a solution to the original {\sc MaxCut/Flip} instance. Consider a vertex $u \in V$ that wants to change the part of the partition it is in. The new partition is induced by the point $(1 - z_u^*, z_{-u}^*)$. Using that $p$ is $L_{\infty}$-Lipschitz and that $x^*$ is an $\epsilon$-Nash equilibirum that is also $\epsilon$-close to $z^*$, we get
    \begin{align*}
        &w\Big( A(1 - z_u^*, z_{-u}^*), B(1 - z_u^*, z_{-u}^*) \Big) - w\Big( A(z^*), B(z^*) \Big)
        \\
        &\overset{(\ref{obj and cut weight})}{=} p(1 - z_u^*, z_{-u}^*) - p(z^*)
        \\
        &= p(1 - z_u^*, z_{-u}^*) - p(1 - z_u^*, x_{-u}^*) + p(1 - z_u^*, x_{-u}^*) 
        \\
        &\, \quad - p(x^*) + p(x^*) - p(z^*)
        \\
        &\leq L_{\infty} ||(1 - z_u^*, z_{-u}^*) - (1 - z_u^*, x_{-u}^*)||_{\infty}
        \\
        &\, \quad + \epsilon + L_{\infty} ||x^* - z^*||_{\infty}
        \\
        &\leq L_{\infty} \epsilon + \epsilon + L_{\infty} \epsilon = \epsilon (2L_{\infty} + 1)
        \\
        &< 1
    \end{align*}
    by the choice of $\epsilon$. Recall that edge weights are integers, and hence, also the weight of a cut. Therefore, the inequality chain above started with an integer that was shown to be strictly less than $1$ at the end. We get
    \[
        w\Big( A(1 - z_u^*, z_{-u}^*), B(1 - z_u^*, z_{-u}^*) \Big) \leq w\Big( A(z^*), B(z^*) \Big) \, ,
    \]
    proving that if vertex $u$ changes the part of the partition it is in, then the cut weight does not increase. Since $u \in V$ was arbitrary, we have shown that partition $V = A(z^*) \sqcup B(z^*)$ has maximal weight among its FLIP neighbourhood. 
\end{proof}

\begin{proof}[Proof of \Cref{cor:PLS hardness of 1P EDT}]
    Follows from \Cref{lem:maxcut PLS-c,lem:1P EDT eqvl to NE poly,lem:Cube NE Poly reduces to NE poly}. For the hardness restrictions, note that we started with a degree two polynomial and a hypercube. This is associated to a game tree of depth 3, with a depth of absentmindedness of at most 2, and with a number of actions per infoset of 2.
\end{proof}

Next, we show \PLS{}-membership.

\begin{lemma}
    {\sc 1P-EDT-S} and {\sc NE-Poly-S} when the branching factor is constant is in \PLS.
\end{lemma}

\begin{proof}
    By \Cref{lem:1P EDT eqvl to NE poly}, it suffices to show this for {\sc 1P-EDT-S}. We show it by giving a best response dynamics that can be run between the infosets in order to find an $\epsilon$-EDT equilibrium. So let $(\Gamma, \epsilon)$ be a {\sc 1P-EDT-S} instances. 

    \paragraph{Computing an $\eps$-best response: }
    We will now describe a method that, given a profile $\mu$ for $\Gamma$ and an infoset $I_j$, computes an $\eps/2$-best response $\alpha \in \Delta^{m_j^{(1)}-1} =: S_j$ of that infoset to strategy $\mu_{-j}^{(1)}$ at other infosets. The method is similar to the one described in the proof of \Cref{prop:Comp NE expo time}. This time, however, instead of working on the whole profile set $S$, we only work on the randomized action simplex $S_j$. We also initialize the hypercube as $B_j := [0,1]^{m_j^{(1)}}$, and describe it with bounds $(y_k, z_k)_{k}$. Then, the sentences we will have to solve are whether there exists ($\exists$) $\alpha \in \R^{m_j^{(1)}}$ such that
    \begin{align}
    \label{target achieved in box}
        S_j(\alpha) \wedge B_j(\alpha) \wedge \, \U^{(1)}(\alpha, \mu_{-j}^{(1)}) \geq t \, ,
    \end{align}
    where $t \in \Q$ is a target value. 
    
    As a preprocessing step, we shall first approximate the maximal utility value $u^* \in \R$ achievable with an exact best response. To that regard, initialize 
    \[
        \underline{u} := \min_{z \in \term} u^{(1)}(z) - 1 \textnormal{ and } \bar{u} := \min_{z \in \term} u^{(1)}(z) + 1 \, .
    \]    
    Then $\underline{u} < u^* < \bar{u}$. Therefore, sentence (\ref{target achieved in box}) is true and false for values $t = \underline{u}$ and $t = \bar{u}$ respectively. Hence, we can do binary search on $\R$ to pinpoint $u^*$ by updating the lower and upper bounds $\underline{u}$ and $\bar{u}$ accordingly such that $\underline{u} < u^* < \bar{u}$ says satisfied and until $| \bar{u} - \underline{u} | < \epsilon / 4$. Then, in particular, $\hat{u} := \underline{u}$ satisfies $| u^* - \hat{u} | < \epsilon / 4$.
    
    \begin{algorithm}[H]
    \caption{Subdivison Search for a Best Response}
    \label{alg:subdivision BR}
    \begin{algorithmic}[1]
        \WHILE{$\textnormal{diam} \geq \frac{\epsilon}{4 L_{\infty}}$ }
            \FOR{$k \in [m_j^{(i)}]$}
                \IF{$\exists \alpha : \, (\ref{target achieved in box}) \wedge \, \alpha_k \leq \frac{y_k + z_k}{2}$}
                    \STATE $z_k \gets \frac{y_k + z_k}{2}$
                \ELSE
                    \STATE $y_k \gets \frac{y_k + z_k}{2}$
                \ENDIF
                \STATE Update $B_j$ accordingly
            \ENDFOR
            \STATE $\textnormal{diam} \gets \textnormal{diam} / 2$
        \ENDWHILE
    \end{algorithmic}
    \end{algorithm}

    Next, we run \Cref{alg:subdivision BR} where (\ref{target achieved in box}) is always invoked for value $t=\hat{u}$. Upon termination, select any point $\alpha$ that satisfies the linear (in-)equality system $S_j(\alpha) \wedge B_j(\alpha)$. If $\alpha^*$ is the point that satisfies (\ref{target achieved in box}), then due to termination condition, we have $||\alpha - \alpha^*||_{\infty} < \frac{\epsilon}{4 L_{\infty}}$. This yields
    \begin{align*}
        &\U^{(1)}(\alpha, \mu_{-j}^{(1)}) 
        \\
        &= \U^{(1)}(\alpha, \mu_{-j}^{(1)}) - \U^{(1)}(\alpha^*, \mu_{-j}^{(1)}) + \U^{(1)}(\alpha^*, \mu_{-j}^{(1)})
        \\
        &= \U^{(1)}(\alpha^*, \mu_{-j}^{(1)}) - |\U^{(1)}(\alpha, \mu_{-j}^{(1)}) - \U^{(1)}(\alpha^*, \mu_{-j}^{(1)})|
        \\
        &\geq \hat{u} - L_{\infty} \cdot ||(\alpha, \mu_{-j}^{(1)}) - (\alpha^*, \mu_{-j}^{(1)})||_{\infty}
        \\
        &\geq u^* - \epsilon / 4 - L_{\infty} \cdot \frac{\epsilon}{4 L_{\infty}} = u^* - \epsilon/2 \, .
    \end{align*}
    
    Finally, the running time analysis works analogous to the one in the proof of \Cref{prop:Comp NE expo time}, except that the number of variables ``$m$'' now is $m_j^{(1)}$. Since it is constant by assumption, we get polytime computability of such an $\eps$-best response $\alpha$.

    \paragraph{$\eps$-Best Response Dynamics:}
    The best response dynamics can start at any profile $\mu \in S$, e.g., at the one that plays the first action of each infoset deterministically. The neighbourhood of an iterate $\mu \in S$ shall be all profiles of the form $(\alpha, \mu_{-j}^{(1)})$ where $j \in [\ninfs^{(1)}]$ and $\alpha$ is an $\eps/2$-best response to $\mu_{-j}^{(1)}$. This neighbourhood can be determined within polytime. Finally, we can evaluate the utility $\U^{(1)}(\pi)$ of any iterate $\pi = \mu$ or any neighbour $\pi = (\alpha, \mu_{-j}^{(1)})$ within polytime. If there is a neighbour $(\alpha, \mu_{-j}^{(1)})$ with utility $\U^{(1)}(\mu) \leq \U^{(1)}(\alpha, \mu_{-j}^{(1)}) - \eps/2$, then take that neighbour as the next iterate, otherwise, terminate and return $\mu$.

    Let $\pi$ be returned by this algorithm. Then $\pi$ is an $\epsilon$-EDT equilibrium of $\Gamma$: Take any infoset $I_j$. Let $(\alpha, \pi_{-j}^{(1)})$ be the neighbour associated to that infoset and $u^*$ be the maximal utility value achievable from that infoset with an exact best response. Then, for any alternative randomized action $\alpha' \in \Delta^{m_j^{(1)}-1}$, we have
    \begin{align*}
        \U^{(1)}(\pi) &\geq \U^{(1)}(\alpha, \pi_{-j}^{(1)}) - \eps/2
        \\
        &\geq u^* - \eps/2 - \eps/2 = u^* - \eps
        \\
        &\geq \U^{(1)}(\alpha', \pi_{-j}^{(1)}) - \eps/2 \, .
    \end{align*}
    This concludes the proof.
\end{proof}

\begin{cor*}[Restatement of \Cref{cor:EDT SPIR inv poly in P}]
    {\sc 1P-EDT-S} for \invpoly{} precision is in \P{} when the branching factor is constant.
\end{cor*}
\begin{proof}
    Let the reward range of the game be in $[0, 1]$ and the desired approximation error be $\eps$.
    Take the best response method described in the previous proof. When the branching factor is constant, it iteratively computes and transitions to a 
    %there exists a best-response function that for any infoset computes a 
    $\eps/2$-best response in time $\poly(|\Gamma|, \log(1/\eps))$. 
    %We repeatedly iterate over all information sets, invoking the best-response oracle with $\delta = \eps/2$. If there is some infoset $I$ at which a $\delta$-profitable deviation is found, we apply that deviation and restart; otherwise, we terminate. 
    %But if the reward range of the game is $[0, 1]$, 
    But this process can update the strategy at most $\mathcal{O}(1/\eps)$ times. Thus, if $\eps$ is of \invpoly{} size, then the method runs in time $\poly(|\Gamma|, 1/\eps)$ overall. 
    %Moreover, the strategy it returns must be an $\eps$-EDT equilibrium; otherwise, some infoset $I$ would have found an $\eps/2$-improvement. 
\end{proof}

\subsection{On the Search of CDT Equilibria}
\label{app:CDT eq search}

\begin{thm*}[Restatement of \Cref{thm:CDT is PPAD}]
    {\sc CDT-S} is \PPAD{}-complete. Hardness holds even for two-player perfect-recall games with one infoset per player and for \invpoly{} precision.
\end{thm*} 

We prove this in parts.
\begin{lemma}
    {\sc CDT-S} is \PPAD{}-hard, even for two-player perfect-recall games with one infoset per player and for \invpoly{} precision.
\end{lemma}
\begin{proof}
    For \PPAD{}-hardness, we can use a well-known reduction from normal-form games to perfect-recall extensive-form games. In particular, we reduce from the \PPAD{}-complete problem of computing an approximate Nash equilibrium, for inverse-polynomial precision, of a two-player normal-form game \cite{ChenDT09}. The representation of such a game is the number of pure actions of each player, and all utility payoffs encoded in binary.

    Starting from such an instance $(G,\epsilon)$, we can a corresponding instance $(\Gamma, \epsilon)$ of {\sc CDT-EQ} as follows: Assign the root node of $\Gamma$ to P1, with an infoset $I_1^{(1)}$, and the action set that P1 has in $G$. Each child of the root node shall be assigned to P2, grouped together to one infoset $I_1^{(2)}$, and with the action set that P2 has in $G$. Finally, each node of depth 3 is a terminal node, with utility payoffs equal to what the players would have received in $G$ if they played the same action there that lead to this terminal node. This is a polytime construction.

    Then, $\Gamma$ has perfect recall, hence no absentmindedness, and it has one infoset per player. By \Cref{rem:w/o absentmindedness edt equals cdt,rem: 1 infoset yields EDT eq NE}, $\epsilon$-CDT equilibria of $\Gamma$ equal its $\epsilon$-Nash equilibria which, in return, equal the $\epsilon$-Nash equilibria in $G$.
\end{proof}

Next, we show \PPAD{}-membership of {\sc CDT-S} by leveraging the general tool of \citet{EtessamiY10}[Section 2.3] to show that a fixed point problem is in \PPAD{}. First, we define the fixed point function that is similar in structure to the one given by \citet{Nash51}. Given a game $\Gamma$ with imperfect recall and a profile $\mu$ for it, define the advantage of a pure action $k \in [m_j^{(i)}]$ at infoset $j \in [\ninfs^{(i)}]$ of player $i \in [N]$ as
\[
    g_{jk}^{(i)}(\mu) := \U_{\CDT}^{(i)} (a_k \mid \mu, I_j) - \U^{(i)}(\mu) \, .
\]

If these advantages are at most $\eps$ for all $i,j,k$ at $\mu$, then $\mu$ is an $\eps$-CDT equilibrium by \Cref{defn:CDT eq} and \Cref{rem:CDT util linearity}. The fixed point mapping $F$ we will consider is one that sends profile $\mu$ of $\Gamma$ to profile $F(\mu)$ of $\Gamma$ with entries
\begin{align}
\label{fp mapping}
    F(\mu)_{jk}^{(i)} := \frac{ \mu_{jk}^{(i)} + \max\{0,g_{jk}^{(i)}(\mu)\} }{ 1 + \sum_{k' = 1}^{m_{j}^{(i)}} \max\{0,g_{jk'}^{(i)}(\mu)\} } \, .
\end{align}

Intuitively, from $\mu$ to $F(\mu)$ it increases the probabilities of those actions that have a positive advantage over the currently played randomized action $\mu_{j \cdot }^{(i)}$ at $\Delta(A_{I_j})$.

\begin{lemma}
\label{lem:CDT eq iff FP}
    Let $\Gamma$ be a game with imperfect recall. Then mapping $F$ as defined above maps the profile set $S$ onto itself. Moreover, a profile $\mu$ is an exact CDT equilibrium of $\Gamma$ if and only if it is a fixed point of $F$, that is, $F(\mu) = \mu$.
\end{lemma}

\begin{proof}
    If $\mu \geq 0$, then also $F(\mu) \geq 0$. Moreover, for all player $i$ and infosets $j$, we have that if $\sum_{k = 1}^{m_{j}^{(i)}} \mu_{jk}^{(i)}$ then also
    \[
        \sum_{k = 1}^{m_{j}^{(i)}} F(\mu)_{jk}^{(i)} = \frac{ \sum_{k = 1}^{m_{j}^{(i)}} \mu_{jk}^{(i)} + \sum_{k = 1}^{m_{j}^{(i)}} \max\{0,g_{jk}^{(i)}(\mu)\} }{ 1 + \sum_{k' = 1}^{m_{j}^{(i)}} \max\{0,g_{jk'}^{(i)}(\mu)\} } = 1 \, .
    \]
    Hence, $F(\mu)$ is a profile of $\Gamma$ if $\mu$ is.

    Next, suppose $\mu$ is an exact CDT equilibrium of $\Gamma$. Then, by definition, we have for all $i,j,k$ that the advantage $g_{jk}^{(i)}(\mu)$ is non-positive, hence $F(\mu)_{jk}^{(i)} = \frac{ \mu_{jk}^{(i)} + 0 }{ 1 + 0 } = \mu_{jk}^{(i)}$, yielding $F(\mu) = \mu$.

    Last but not least, suppose $F(\mu) = \mu$. Let us show that any player $i$ plays optimally at any infoset $I_j$ with randomized action $\mu_{j \cdot}^{(i)}$, by showing that no pure action $a_k$ does better at $I_j$, using \Cref{rem:CDT util linearity}. We show this by proving that advantage $g_{jk}^{(i)}(\mu)$ is non-positive for each action $a_k$. For the sake of contradiction, suppose that the subset $\mathcal{K} := \{ k \in [m_j^{(i)}] : g_{jk}^{(i)}(\mu) > 0 \}$ of actions with positive advantage is non-empty. Then observe that $\sum_{k' = 1}^{m_{j}^{(i)}} \max\{0,g_{jk'}^{(i)}(\mu)\} > 0$. Thus, all actions $\tilde{k} \notin \mathcal{K}$ with non-positive advantage satisfy
    
    \begin{align*}
    %\label{nonpos adv implies no prob}
        \mu_{j \tilde{k}}^{(i)} = F(\mu)_{j \tilde{k}}^{(i)} = \frac{ \mu_{j \tilde{k}}^{(i)} + 0 }{ 1 + \sum_{k' = 1}^{m_{j}^{(i)}} \max\{0,g_{jk'}^{(i)}(\mu)\} } \, ,
    \end{align*}

    which implies $\mu_{j \tilde{k}}^{(i)} = 0$. Hence, $\supp(\mu_{j \cdot }^{(i)}) \subseteq \mathcal{K}$. Knowing this, we can derive the contradiction

    \begin{align*}
        &\U_{\CDT}^{(i)} ( \mu_{j \cdot }^{(i)} \mid \mu, I_j) = \sum_{k \in [m_j^{(i)}] } \mu_{jk}^{(i)}  \cdot \U_{\CDT}^{(i)} (a_k \mid \mu, I_j)
        \\
        &= \sum_{k \in \supp(\mu_{j \cdot}^{(i)}) } \mu_{jk}^{(i)}  \cdot \U_{\CDT}^{(i)} (a_k \mid \mu, I_j)
        \\
        &= \sum_{k \in \supp(\mu_{j \cdot}^{(i)}) } \mu_{jk}^{(i)}  \cdot \Big( g_{jk}^{(i)}(\mu) + \U_{\CDT}^{(i)} ( \mu_{j \cdot }^{(i)} \mid \mu, I_j) \Big)
        \\
        &> \sum_{k \in \supp(\mu_{j \cdot}^{(i)}) } \mu_{jk}^{(i)}  \cdot \U_{\CDT}^{(i)} ( \mu_{j \cdot }^{(i)} \mid \mu, I_j) 
        \\
        &= \U_{\CDT}^{(i)} ( \mu_{j \cdot }^{(i)} \mid \mu, I_j)  \, .
    \end{align*}

    Thus, $\mathcal{K}$ must have been empty. Since this holds for each player and infoset, $\mu$ must have been an exact CDT equilibrium for $\Gamma$.
\end{proof}

Next, we show that $F$ is Lipschitz continuous for a moderately sized Lipschitz constant in terms of the game instance $\Gamma$. 

\begin{lemma}
\label{lem:FP fct is lps cont}
    Given a game $\Gamma$ with imperfect recall, its mapping $F$ from (\ref{fp mapping}) is Lipschitz continuous on the profile set $S$ with Lipschitz constant $L_F := 11 |\nds|^2 L_{\infty}$, where $L_{\infty}$ is the Lipschitz constant of $\Gamma$ as described in \Cref{app:lps constant}.
\end{lemma}
\begin{proof}
    We follow the proof outline of \cite{DaskalakisP11}[Lemma 3.4], and show that if profiles $\mu$ and $\pi$ have distance $||\mu - \pi||_{\infty} \leq \eps$, then $||F(\mu) - F(\pi)||_{\infty} \leq 11 |\nds|^2 L_{\infty} \eps$.

    First, consider $h_{j k}^{(i)} = \U_{\CDT}^{(i)} (a_k \mid \cdot, I_j)$ as a function in profile $\mu' \in S$, for a given $i,j,k$. 

    \paragraph{$h_{j k}^{(i)}$ is Lipschitz continuous:} We show this with the profiles $\mu$ and $\pi$ above. We have
    
    \begin{align*}
        &| h_{j k}^{(i)}(\mu) - h_{j k}^{(i)}(\pi) |
        \\
        &= | \U^{(i)}(\mu) + \nabla_{j k} \, \U^{(i)}(\mu) - \sum_{ k' \in [m_j^{(i)}]} \mu_{jk'}^{(i)} \cdot \nabla_{jk'} \, \U^{(i)}(\mu)
        \\
        &\, \quad - \U^{(i)}(\pi) - \nabla_{j k} \, \U^{(i)}(\pi) + \sum_{ k' \in [m_j^{(i)}]} \pi_{jk'}^{(i)} \cdot \nabla_{jk'} \, \U^{(i)}(\pi)  |
        \\
        &\leq | \U^{(i)}(\mu) - \U^{(i)}(\pi) | + | \nabla_{j k} \, \U^{(i)}(\mu) - \nabla_{j k} \, \U^{(i)}(\pi) |
        \\
        &\, \quad + | \sum_{ k' \in [m_j^{(i)}]} \mu_{jk'}^{(i)} \cdot \nabla_{jk'} \, \U^{(i)}(\mu) - \pi_{jk'}^{(i)} \cdot \nabla_{jk'} \, \U^{(i)}(\pi)  |
        \\
        &\leq L_{\infty} \eps + L_{\infty} \eps
        \\
        &\, \quad + \sum_{ k' \in [m_j^{(i)}]} | \mu_{jk'}^{(i)} \cdot \nabla_{jk'} \, \U^{(i)}(\mu) - \pi_{jk'}^{(i)} \cdot  \nabla_{jk'} \, \U^{(i)}(\mu)
        \\
        &\, \quad + \pi_{jk'}^{(i)} \cdot  \nabla_{jk'} \, \U^{(i)}(\mu) - \pi_{jk'}^{(i)} \cdot \nabla_{jk'} \, \U^{(i)}(\pi)  |
        \\
        &\leq 2 L_{\infty} \eps + \sum_{ k' \in [m_j^{(i)}]} | \nabla_{jk'} \, \U^{(i)}(\mu) | \eps + | \pi_{jk'}^{(i)} | L_{\infty} \eps 
        \\
        &\leq 2 L_{\infty} \eps + \sum_{ k' \in [m_j^{(i)}]} ( L_{\infty} \eps + 1 \cdot L_{\infty} \eps )
        \\
        &\leq 2 L_{\infty} \eps + 2 |\nds| L_{\infty} \eps = 4 |\nds| L_{\infty} \eps \, .
    \end{align*}

    \paragraph{$g_{jk}^{(i)}$ is Lipschitz continuous:} \,
    \begin{align*}
        | g_{jk}^{(i)}(\mu) - g_{jk}^{(i)}(\pi) | &= | h_{j k}^{(i)}(\mu) - \U^{(i)}(\mu) - h_{j k}^{(i)}(\pi) + \U^{(i)}(\pi) |
        \\
        &\leq | h_{j k}^{(i)}(\mu) - h_{j k}^{(i)}(\pi) | + |\U^{(i)}(\mu) - \U^{(i)}(\pi)| 
        \\
        &\leq 4 |\nds| L_{\infty} \eps + L_{\infty} \eps \leq 5 |\nds| L_{\infty} \eps \, .
    \end{align*}

    Therefore, by case distinction, we also get 
    \begin{align*}
        &| \max \{ 0, g_{jk}^{(i)}(\mu) \} - \max \{ 0, g_{jk}^{(i)}(\pi) \} | 
        \\
        &\leq | g_{jk}^{(i)}(\mu) - g_{jk}^{(i)}(\pi) | \leq 5 |\nds| L_{\infty} \eps \, .
    \end{align*}

    \paragraph{$F$ is Lipschitz continuous:} We show for each entry index $i,j,k$
    \begin{align*}
        &| F(\mu)_{jk}^{(i)} - F(\pi)_{jk}^{(i)} | 
        \\
        &\overset{(*)}{\leq} |\mu_{jk}^{(i)} - \pi_{jk}^{(i)}| + | \max \{ 0, g_{jk}^{(i)}(\mu) \} - \max \{ 0, g_{jk}^{(i)}(\pi) \} |
        \\
        &\, \quad | \sum_{k' = 1}^{m_{j}^{(i)}} \max \{ 0, g_{jk'}^{(i)}(\mu) \} - \sum_{k' = 1}^{m_{j}^{(i)}} \max \{ 0, g_{jk'}^{(i)}(\pi) \} |
        \\
        &\leq \eps + 5 |\nds| L_{\infty} \eps + |\nds| \cdot 5 |\nds| L_{\infty} \eps
        \\
        &\leq 11 |\nds|^2 L_{\infty} \eps \, ,
    \end{align*}
    where in $(*)$ we used \cite{DaskalakisP11}[Lemma~3.6]. Thus, 
    \[
        || F(\mu) - F(\pi) ||_{\infty} \leq 11 |\nds|^2 L_{\infty} \eps \, . \qedhere
    \]
\end{proof}

We call $\mu \in S$ an $\eps$-fixed point of $F$ if $||F(\mu) - \mu||_{\infty} < \eps$.

\begin{lemma}
\label{lem:eps FP in PPAD}
    It is in \PPAD{} to find an $\eps$-fixed point of associated mapping $F$ to a game $\Gamma$ with imperfect recall.
\end{lemma}
\begin{proof}
    We invoke \cite{EtessamiY10}[Proposition 2.2 (2)] for this, for which we need that mapping $F$ is \emph{polynomially continuous} and \emph{polynomially computable}. The former follows from \Cref{lem:FP fct is lps cont}. The latter follows because the profile set $S$ is easy to describe and $F$ is polytime computable.
\end{proof}

For the next (and last) result, note that the value
\[
    \theta := \max \Big\{ 1 \, , \, 3 |\nds| \cdot \max_{z \in \term, i \in \pls} | u^{(i)} (z) | \Big\}
\]
serves as an upper bound on values $g_{jk}^{(i)}(\mu)$. (We need the factor $|\nds|$ because of (\ref{CDT util bound})).

\begin{lemma}
\label{lem:eps FP to delta CDT eq}
    For any game $\Gamma$ with imperfect recall, if $\mu$ is an $\eps$-fixed point ($\eps < \frac{1}{4}$) of its associated mapping $F$, then $\mu$ is an $\eps'$-CDT equilibrium of $\Gamma$, where $\eps' := 2 \theta |\nds|^{3/2} \sqrt{\eps}$ and $\theta$ is defined as above.
\end{lemma}

\begin{proof}
    We follow the proof outline of \cite{EtessamiY10}[Proposition 2.3]. Let $\mu$ be an $\eps$-fixed point of $F$. Then we show that for all indices $i,j,k$, we have $\max \{ 0, g_{jk}^{(i)}(\mu) \} \leq \eps'$. This then implies that $\mu$ is an $\eps'$-CDT equilibrium by \Cref{defn:CDT eq} and \Cref{rem:CDT util linearity}.

    Take any player $i$ and infoset $j$. 
    %First, note that for any action $k \in [m_j^{(i)}]$, we always have $|g_{jk'}^{(i)}(\mu)| \leq \theta$. Thus, 
    Then $||F(\mu) - \mu||_{\infty} < \eps$ implies for any action $k \in [m_j^{(i)}]$:
    \begin{align}
    \label{eps FP condition rearranged}
    \begin{aligned}
        &| \max\{0,g_{jk}^{(i)}(\mu) \} -  \mu_{jk}^{(i)} \cdot \sum_{k' = 1}^{m_{j}^{(i)}} \max\{0,g_{jk'}^{(i)}(\mu)\} | 
        \\
        &\leq \eps \cdot \Big( 1 + \sum_{k' = 1}^{m_{j}^{(i)}} \max\{0,g_{jk'}^{(i)}(\mu)\} \Big) \leq \eps ( 1 + |\nds| \cdot \theta)
        \\
        &\leq 2 \theta |\nds| \eps \, .
    \end{aligned}
    \end{align}

    Next, define $\mathcal{K}^+ := \{ k \in [m_j^{(i)}] : g_{jk}^{(i)}(\mu) > 0 \}$ and $\mathcal{K}^- := \{ k \in [m_j^{(i)}] : g_{jk}^{(i)}(\mu) < 0 \}$. 

    Case 1: There is an index $\tilde{k} \in \mathcal{K}^-$ with $\mu_{j\tilde{k}}^{(i)} \geq \sqrt{\frac{\eps}{|\nds|}}$, then for any action $k \in [m_j^{(i)}]$:
    
    \begin{flalign*}
        &\max \{ 0, g_{jk}^{(i)}(\mu) \} \leq \sqrt{\frac{|\nds|}{\eps}} \cdot \sqrt{\frac{\eps}{|\nds|}} \sum_{k' = 1}^{m_{j}^{(i)}} \max\{0,g_{jk'}^{(i)}(\mu)\}
        \\
        &\leq \sqrt{\frac{|\nds|}{\eps}} \cdot \mu_{j\tilde{k}}^{(i)}  \sum_{k' = 1}^{m_{j}^{(i)}} \max\{0,g_{jk'}^{(i)}(\mu)\}
        \\
        &\leq \sqrt{\frac{|\nds|}{\eps}} \cdot \Big| 0 - \mu_{j\tilde{k}}^{(i)}  \sum_{k' = 1}^{m_{j}^{(i)}} \max\{0,g_{jk'}^{(i)}(\mu)\} \Big|
    \end{flalign*}
    \begin{flalign*}
        &\overset{\tilde{k} \in \mathcal{K}^-}{=} \sqrt{\frac{|\nds|}{\eps}} \cdot \Big|  \max\{0,g_{j\tilde{k}}^{(i)}(\mu)\}- \mu_{j\tilde{k}}^{(i)}  \sum_{k' = 1}^{m_{j}^{(i)}} \max\{0,g_{jk'}^{(i)}(\mu)\} \Big|
        \\
        &\overset{(\ref{eps FP condition rearranged})}{\leq} \sqrt{\frac{|\nds|}{\eps}} \cdot 2 \theta |\nds| \eps = 2 \theta |\nds|^{3/2} \sqrt{\eps} = \eps' \, .
    \end{flalign*}

    Case 2: For all indices $\tilde{k} \in \mathcal{K}^-$, we have $\mu_{j\tilde{k}}^{(i)} < \sqrt{\frac{\eps}{|\nds|}}$. We first have to observe that
    \begin{align*}
        &\sum_{k' = 1}^{m_{j}^{(i)}} \mu_{jk'}^{(i)} \cdot \max\{0,g_{jk'}^{(i)}(\mu)\} = \sum_{k' \in \mathcal{K}^+} \mu_{jk'}^{(i)} \cdot g_{jk'}^{(i)}(\mu)
        \\
        &= \sum_{k' = 1}^{m_{j}^{(i)}} \mu_{jk'}^{(i)} \cdot g_{jk'}^{(i)}(\mu) - 0 - \sum_{\tilde{k} \in \mathcal{K}^-} \mu_{j\tilde{k}}^{(i)} \cdot g_{j\tilde{k}}^{(i)}(\mu)
        \\
        &= \sum_{k' = 1}^{m_{j}^{(i)}} \mu_{jk'}^{(i)} \cdot \U_{\CDT}^{(i)} (a_k \mid \mu, I_j) - \sum_{k'' = 1}^{m_{j}^{(i)}} \mu_{jk''}^{(i)} \cdot \U^{(i)}(\mu) 
        \\
        &\, \quad - \sum_{\tilde{k} \in \mathcal{K}^-} \mu_{j\tilde{k}}^{(i)} \cdot g_{j\tilde{k}}^{(i)}(\mu)
        \\
        &= \U_{\CDT}^{(i)} ( \mu_{j \cdot}^{(i)} \mid \mu, I_j) - \U^{(i)}(\mu) - \sum_{\tilde{k} \in \mathcal{K}^-} \mu_{j\tilde{k}}^{(i)} \cdot g_{j\tilde{k}}^{(i)}(\mu)
        \\
        &= \sum_{\tilde{k} \in \mathcal{K}^-} \mu_{j\tilde{k}}^{(i)} \cdot (-g_{j\tilde{k}}^{(i)}(\mu)) \leq |\nds| \sqrt{\frac{\eps}{|\nds|}} \cdot \theta
        \\
        &\leq \theta \sqrt{|\nds|} \sqrt{\eps} \, .
    \end{align*}
    
    Next, set $k^* = \argmax_{k \in \mathcal{K}^+} g_{jk}^{(i)}(\mu)$.  
    
    Case 2.1: Probability $\mu_{jk^*}^{(i)} \geq \frac{1}{2 |\nds|}$. Then for any action $k \in [m_j^{(i)}]$:
    \begin{align*}
        &\max\{0,g_{jk}^{(i)}(\mu)\} \leq \max\{0,g_{jk^*}^{(i)}(\mu)\} 
        \\
        &= \frac{2 |\nds|}{2 |\nds|} \max\{0,g_{jk^*}^{(i)}(\mu)\} \leq 2 |\nds| \mu_{jk^*}^{(i)} \max\{0,g_{jk^*}^{(i)}(\mu)\} 
        \\
        &\leq 2 |\nds| \sum_{k' = 1}^{m_{j}^{(i)}} \mu_{jk'}^{(i)} \max\{0,g_{jk'}^{(i)}(\mu)\}
        \\
        &\leq 2 |\nds| \cdot \theta \sqrt{|\nds|} \sqrt{\eps} = 2 \theta |\nds|^{3/2} \sqrt{\eps} =  \eps' \, .
    \end{align*}    

    Case 2.2: Probability $\mu_{jk^*}^{(i)} < \frac{1}{2 |\nds|}$. Then for any action $k \in [m_j^{(i)}]$:
    \begin{flalign*}
        &\max\{0,g_{jk}^{(i)}(\mu)\} \leq \max\{0,g_{jk^*}^{(i)}(\mu)\} 
        \\
        &= 2 \cdot \Big( \max\{0,g_{jk^*}^{(i)}(\mu)\} - \frac{1}{2} \max\{0,g_{jk^*}^{(i)}(\mu)\} \Big)
        \\
        &= 2 \cdot \Big( \max\{0,g_{jk^*}^{(i)}(\mu)\} - \frac{1}{2 |\nds|} |\nds| \cdot \max\{0,g_{jk^*}^{(i)}(\mu)\} \Big)
        \\
        &\leq 2 \cdot \Big( \max\{0,g_{jk^*}^{(i)}(\mu)\} - \frac{1}{2 |\nds|} \sum_{k' = 1}^{m_{j}^{(i)}} \max\{0,g_{jk^*}^{(i)}(\mu)\}
    \end{flalign*}
    \begin{flalign*}
        &\leq 2 \cdot \Big( \max\{0,g_{jk^*}^{(i)}(\mu)\} - \mu_{jk^*}^{(i)}  \sum_{k' = 1}^{m_{j}^{(i)}} \max\{0,g_{jk'}^{(i)}(\mu)\} \Big)
        \\
        &\leq 2 \cdot \Big| \max\{0,g_{jk^*}^{(i)}(\mu)\} - \mu_{jk^*}^{(i)}  \sum_{k' = 1}^{m_{j}^{(i)}} \max\{0,g_{jk'}^{(i)}(\mu)\} \Big|
        \\
        &\overset{(\ref{eps FP condition rearranged})}{\leq} 2 \cdot 2 \theta |\nds| \eps \leq 2 \theta |\nds|^{3/2} \sqrt{\eps} \cdot 2 \sqrt{\eps}
        \\
        &\leq 2 \theta |\nds|^{3/2} \sqrt{\eps}  = \eps' \, ,
    \end{flalign*} 
    
    where we used at the last line that $\eps \leq \frac{1}{4}$.

    This covers all cases, and hence $\mu$ is an $\eps'$-CDT equilibrium of $\Gamma$.
\end{proof}

\begin{prop}
%\label{thm:CDT is PPAD}
    {\sc CDT-S} is in \PPAD{}.
\end{prop} 

\begin{proof}
    Given an instance $(\Gamma,\eps)$ of {\sc CDT-S}, construct mapping $F$ as in (\ref{fp mapping}) and $\delta := \Big( \frac{\eps}{2 \theta |\nds|^{3/2}} \Big)^2$. Use \Cref{lem:eps FP in PPAD}. Then, a $\delta$-fixed point of $F$ makes an $2 \theta |\nds|^{3/2} \sqrt{\delta} = \eps$-CDT equilibrium of $\Gamma$ by \Cref{lem:eps FP to delta CDT eq}.
\end{proof}

\section{On \Cref{sec:perf info IR}}
\label{app:perf info IR}

In this section, we prove the results in \Cref{sec:perf info IR}. To that end, we restate results taken from the main body, and give new numbers to results presented first in this appendix.

\subsection{On General Chance Node Removal}

Here, we will show:

\begin{thm*}[Restatement of \Cref{thm:restr to no chance possible}]
    All computational hardness results in this paper for the three equilibrium concepts \{Nash, EDT, CDT\} still hold even when the game has no chance nodes. They hold together with previously possible restrictions (\eg, on the branching factor), except that the restrictions on the number of infosets and the degree of absentmindedness increase by one and to $\mathcal{O}(\log|\nds|)$ respectively.
\end{thm*}

This result immediately follows from the following construction. For this subsection, let \emph{equilibrium} be any of the three equilibrium concepts \{Nash, EDT, CDT\}.

\begin{thm}
\label{thm:remove chance}
    For an $N$-player game $\Gamma$ with imperfect recall, we can create an $N$-player game $\Gamma'$ with imperfect recall and without chance nodes such that
    \begin{enumerate}[nolistsep]
        \item $\Gamma'$ has the same info and action sets as $\Gamma$ except an arbitrary player (PL1) has one additional infoset $I_c$ with absentmindedness which will induce randomness,
        \item there is a randomized action $\alpha$ for $I_c$ such that for exact equilibrium sets $E$ and $E'$ of $\Gamma$ and $\Gamma'$, we have identity $E' = E \times \{\alpha\}$,
        \item there is an polynomial relationship between precision errors of approximate equilibria in $\Gamma'$ and the approximate equilibria they induce in $\Gamma$.
    \end{enumerate}
\end{thm}

We prove \Cref{thm:remove chance} in parts. 

First, we show how to transition to a game with a single, polysized chance node $h_c$ at the root. 

\begin{lemma}
\label{lem:get to one chance node}
    A game $\Gamma$ with imperfect recall can be polytime reduced to a game $\Gamma'$ with imperfect recall such that $\Gamma'$ has the same strategy sets and utility functions as $\Gamma$, and $\Gamma'$ has only one chance node which is placed at the root. That chance node randomizes uniformly over $2^t$ actions for some integer $t$ in $\mathcal{O}(\log|\nds|)$ where $|\nds|$ is the number of nodes in $\Gamma$.

    In particular, $\Gamma$ and $\Gamma'$ have the same $\epsilon$-equilibria.
\end{lemma}

\begin{proof}
    Let $\Gamma$ be a game with imperfect recall. Get its utility functions $\U^{(1)}, \dots, \U^{(N)}$. Use \Cref{app:poly fcts to IR game} to create a game $\tilde{\Gamma}$ with imperfect recall out of it, that has utility functions $\U^{(1)}, \dots, \U^{(N)}$. Both of these steps take polytime. Notably, $\tilde{\Gamma}$ has only one chance node $h_0$ at the root, and that one is randomizing uniformly over a number of outgoing actions $r$ that is equal to the number of monomials with nonzero coefficients in the functions $\U^{(1)}, \dots, \U^{(N)}$. This is bounded by the number of terminal nodes $|\term|$ in $\Gamma$, which is bounded by the number of nodes $|\nds|$ in $\Gamma$. Next, we pad the number of outgoing edges at $h_0$ in $\tilde{\Gamma}$ to $2^{ \lceil \log(r) \rceil }$ to obtain the final game $\Gamma'$: Add $2^{ \lceil \log(r) \rceil } - r$ many actions to $h_0$, each leading to a terminal node with utility $0$ for all player. Next, make the probability distribution at $h_0$ uniform over these $2^{ \lceil \log(r) \rceil }$ actions, and rescale the payoffs at terminal nodes that were in $\tilde{\Gamma}$ before this padding action by $2^{ \lceil \log(r) \rceil } / r$. This padding procedure is polytime (we added at most $r$ additional actions at the root) and it ensures that the new game $\Gamma'$ has the same utility functions as $\tilde{\Gamma}$ which has the same utility function as original game $\Gamma$. Hence, $\Gamma'$ and $\Gamma$ have the same $\epsilon$-equilibria since those are defined in terms of the strategy sets and ex-ante utility functions. Last but not least, $\Gamma'$ has only one chance node at the root which randomizes uniformly over $2^{ \lceil \log(r) \rceil }$ actions where $\lceil \log(r) \rceil = \mathcal{O}(\log|\nds|)$.
\end{proof}

\begin{prop}
\label{prop:remove chance at root}
    Let $\Gamma$ be a game with imperfect recall that has only one chance node at the root which randomizes uniformly over $2^t$ actions for some $t \in \N$. Then $\Gamma$ can be polytime reduced to a game $\Gamma'$ with imperfect recall such that 
    \begin{enumerate}
        \item $\Gamma'$ has no chance nodes
        \item exact equilibria of $\Gamma$ correspond 1-1 to exact equilibria of $\Gamma'$
        \item $\delta$-equilibria of $\Gamma'$ give rise to $\eps$-equilibria of $\Gamma$, where we (might) set
        \[
            \delta = \min \{\frac{1}{4}, \frac{\eps}{2^t + t \cdot L_{\infty} } \}
        \]
        using a Lipschitz constant $L_{\infty}$ as described in \Cref{app:lps constant}.
    \end{enumerate}
\end{prop}

\begin{figure*}[t]
    \centering
    \tikzset{
        every path/.style={-},
        every node/.style={draw},
    }
    \forestset{
        subgame/.style={regular polygon,
        regular polygon sides=3,anchor=north, inner sep=0pt},
    }
    \begin{forest}
    [,p1,name=p0
        [,p1,name=p0a
            [\util1{-1},terminal,name=t1]
            [,p1,name=p0b
                [,p1,name=p0c
                    [\util1{0},terminal]
                    [,p1,name=p0d
                        [,p1,name=p0e
                            [\util1{0},terminal]
                            [$G_1$,subgame, yshift=10pt]
                        ]
                        [,p1,name=p0f
                            [$G_2$,subgame, yshift=10pt]
                            [\util1{0},terminal]
                        ]
                    ]
                ]   
                [,p1,,name=p0g
                    [,p1,,name=p0h
                        [,p1,,name=p0i
                            [\util1{0},terminal]
                            [$G_3$,subgame, yshift=10pt]
                        ]
                        [,p1,,name=p0j
                            [$G_4$,subgame, yshift=10pt]
                            [\util1{0},terminal]
                        ]
                    ]
                    [\util1{0},terminal]
                ]
            ]
        ]
        [,p1,name=p0k
            [,p1,name=p0l
                [,p1,name=p0m
                    [\util1{0},terminal]
                    [,p1,name=p0n
                        [,p1,name=p0o
                            [\util1{0},terminal]
                            [$G_5$,subgame, yshift=10pt]
                        ]
                        [,p1,name=p0p
                            [$G_6$,subgame, yshift=10pt]
                            [\util1{0},terminal]
                        ]
                    ]
                ]   
                [,p1,name=p0q
                    [,p1,name=p0r
                        [,p1,name=p0s
                            [\util1{0},terminal]
                            [$G_7$,subgame, yshift=10pt]
                        ]
                        [,p1,name=p0t
                            [$G_8$,subgame, yshift=10pt]
                            [\util1{0},terminal]
                        ]
                    ]
                    [\util1{0},terminal]
                ]
            ]
            [\util1{-1},terminal]
        ]
    ]     
    \node[below=10pt of p0,draw=none,p1color]{$I_c$};
    \draw[infoset1,bend right=30] (p0a) to (p0) to (p0k);
    \draw[infoset1,bend right=30] (p0k) to (p0l);
    \draw[infoset1,bend left=30] (p0l) to (p0m);
    \draw[infoset1,bend left=30] (p0m) to (p0n);
    \draw[infoset1,bend left=30] (p0n) to (p0o);
    \draw[infoset1,bend right=30] (p0n) to (p0p);
    \draw[infoset1,bend right=30] (p0l) to (p0q);
    \draw[infoset1,bend right=30] (p0q) to (p0r);
    \draw[infoset1,bend left=30] (p0r) to (p0s);
    \draw[infoset1,bend right=30] (p0r) to (p0t);
    \draw[infoset1,bend left=30] (p0a) to (p0b);
    \draw[infoset1,bend left=30] (p0b) to (p0c);
    \draw[infoset1,bend right=30] (p0b) to (p0g);
    \draw[infoset1,bend left=30] (p0c) to (p0d);
    \draw[infoset1,bend left=30] (p0d) to (p0e);
    \draw[infoset1,bend right=30] (p0d) to (p0f);
    \draw[infoset1,bend right=30] (p0g) to (p0h);
    \draw[infoset1,bend left=30] (p0h) to (p0i);
    \draw[infoset1,bend right=30] (p0h) to (p0j);
    \end{forest}
    \caption{ Another example for the construction of \Cref{prop:remove chance at root}, this time the chance node at the root randomizes uniformly over $8$ actions into subgames $G_1, \dots, G_8$. }
    \label{fig:chance removal multi actions}
\end{figure*}
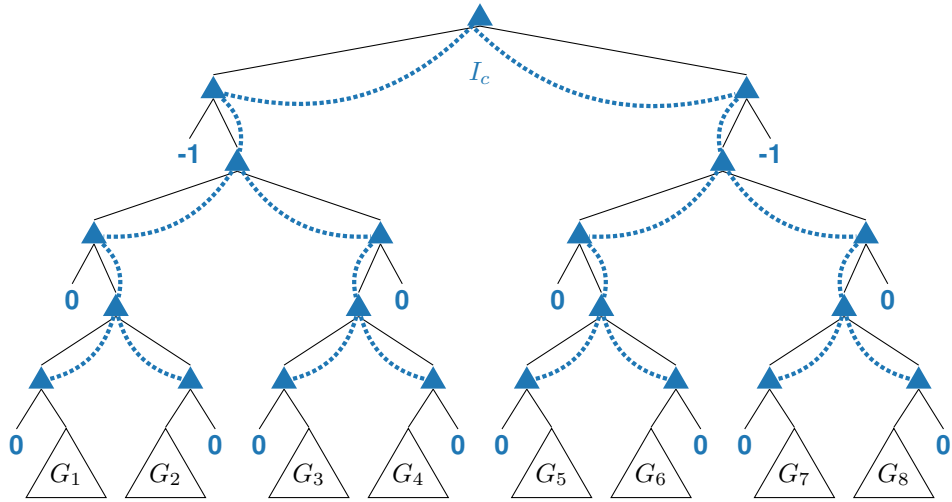

\begin{proof}
    Let $\Gamma$ be a game with imperfect recall with one chance node $h_0$ at the root with the above description. We assume w.l.o.g.\ that the payoffs in $\Gamma$ are $\geq 1$ (otherwise first shift the payoffs in $\Gamma$ by $1 - \min_{z \in \term, i \in \pls} u^{(i)}(z)$). Replace $h_0$ with one big infoset $I_c$ with $2$ actions $l$ (left) and $r$ (right) and a degree of absentmindedness of $2t$. It is irrelevant which player is assigned to $I_c$, so let it be P1. \Cref{fig:chance removal} gives an example for $t = 1$ and \Cref{fig:chance removal multi actions} gives an example for $t = 3$.

    \paragraph{The Construction:} \,
    \\
    Formally, we replace $h_0$ by a tree $\mathcal{T}$ of terminal nodes and nodes in $I_c$ as follows. $\mathcal{T}$ will have a depth of $2t+1$. Start at the root $h_0'$, assign it to $I_c$, and call its depth level $0$. Create two outgoing edges $l$ and $r$ to nodes $h_1'$ and $h_2'$. Assign those nodes to $I_c$ as well. Make nodes $h_1'l$ and $h_2'r$ terminal nodes with payoff $-1$ to all players. Assign nodes $h_1'r$ and $h_2'l$ to $I_c$. Next, we create depth levels $2$ and $3$ to $2t-3$ and $2t-2$ by induction. Let node $h'$ be in $I_c$ and at an even depth level. Assign its two children $h'l$ and $h'r$, as well as their respective child $h'lr$ and $h'rl$, to $I_c$. The nodes $h'll$ and $h'rr$ shall be terminal nodes with payoff $0$ to all players. Finally, depth levels $2t-1$ and $2t$ shall be created in the same way, except that the nodes at level $2t$ that would have been assigned to $I_c$ are now instead being replaced by the children of the original chance node $h_0$ (order of replacement is irrelevant). This works out number-wise because there are exactly $2^{2t / 2} = 2^t$ many nonterminal nodes at depth level $2t$.

    \paragraph{How $\Gamma'$ looks like:} \,
    \\
    If $S = \bigtimes_{i \in \pls} S^{(i)}$ is the strategy set of $\Gamma$, then 
    \[
        S' = S^{(1)} \times \Delta(\{l,r\}) \times \bigtimes_{i \in \pls \setminus \{1\}} S^{(i)} = S \times [0,1]
    \]
    is the strategy set of $\Gamma$, where the interval $[0,1]$ stands for the probability that P1 assigns to playing left $l$ at $I_c$. Take the utility function of any player $i \in \pls$ and write it as 
    \[
        \U^{(i)}(\mu) = \sum_{a \in [2^t]} \frac{1}{2^t} \U^{(i)}(\mu \mid h_0a)
    \]
    for strategy $\mu \in S$.

    Observe that each children $h_0a$ of $h_0$ in the corresponding game $\Gamma'$ has an action history with exactly $t$ appearances of $l$ and $t$ appearances of $r$. Therefore, they are all reached with equal probability $l^t (1-l)^t$, where by abuse of notation we use $l \in [0,1]$ for the probability put by P1 on action $l$ at $I_c$. Moreover, after the constructed tree $\mathcal{T}$ at the beginning, infoset $I_c$ does not occur again. Next, recall that all terminal nodes in $\mathcal{T}$, except for the two on depth level $2$, have a payoff of $0$. Hence, if $V^{(i)}$ denotes the utility functions in $\Gamma'$, we can rewrite them as
    \begin{align*}
        V^{(i)}(\mu,l) &= -l^2 -(1-l)^2 + \sum_{a \in [2^t]} l^t (1-l)^t \U^{(i)}(\mu \mid h_0a)
        \\
        &= -l^2 -(1-l)^2 + 2^t \big( l (1-l) \big)^t \U^{(i)}(\mu)
        \\
        &= -\frac{1}{2} - 2 (l -\frac{1}{2})^2 + 2^t \Big( \frac{1}{4} - (l -\frac{1}{2})^2 \Big)^t \U^{(i)}(\mu)
    \end{align*}
    for strategy $(\mu,l) \in S'$. Note that $0 \leq l \leq 1$ implies $0 \leq (l -\frac{1}{2})^2 \leq \frac{1}{4}$ and thus the factor in front of $\U^{(i)}(\mu)$ is always non-negative in a valid profile $(\mu,l)$. Moreover, using that utility $\U^{(i)}$ is positive, we observe that any profile $(\mu,l)$ for $l \neq \frac{1}{2}$ is strictly dominated by profile $(\mu,\frac{1}{2})$. In fact, the best response set of P1 to $\mu^{(-1)}$ in $\Gamma'$ is the best response set to $\mu^{(-1)}$ in $\Gamma$ and playing $\frac{1}{2}$ at $I_c$. This is because function $V^{(1)}(\cdot, \mu^{(-1)}, \frac{1}{2})$ is just a positive factor scaling and subsequent constant shift of $\U^{(1)}(\cdot, \mu^{(-1)})$. Analogous reasoning yields that the best response set of player $i \neq 1$ in $\Gamma'$ to $(\mu^{(-i)}, l)$ for $0 \neq l \neq 1$ is equal her best response set in $\Gamma$ to $\mu^{(-i)}$. 
    
    \paragraph{Exact Nash and EDT:} \,
    \\
    Suppose $\mu$ is an exact Nash equilibrium or EDT equilibrium of $\Gamma$. Then by above reasoning, profile $(\mu, \frac{1}{2})$ makes an exact Nash equilibrium or, respectively, EDT equilibrium in $\Gamma'$. 

    \paragraph{Approximate Nash:} \,
    \\
    We will now show that given $(\Gamma,\eps)$, we can set $\delta>0$ sufficiently small but of size $\poly(\eps, |\Gamma|)$, such that if $(\mu,l)$ is a $\delta$-Nash equilibrium in $\Gamma'$, then $\mu$ is an $\eps$-Nash equilibrium in $\Gamma$. 
    
    First, we bound how far away $l$ can be from $\frac{1}{2}$. A $\delta$-Nash equilibrium in particularly satisfies the $\delta$-EDT equilibrium condition that $(\mu,l)$ does not perform more than $\delta$ worse than $(\mu,\frac{1}{2})$ for P1. Thus, using that utility $\U^{(i)}$ is positive, we get
    \begin{align*}
        \delta &\geq V^{(1)}(\mu,\frac{1}{2}) - V^{(1)}(\mu,l)
        \\
        &= 2 (l -\frac{1}{2})^2 + 2^t \U^{(1)}(\mu) \Big[ \frac{1}{4^t} - \Big( \frac{1}{4} - (l -\frac{1}{2})^2 \Big)^t \Big]
        \\
        &\geq 2 (l -\frac{1}{2})^2 + 2^t \U^{(1)}(\mu) \cdot 0 \, ,
    \end{align*}

    which implies that $l$ must satisfy
    \begin{align}
    \label{nochance NE prob at Ic close to 1/2}
        (l -\frac{1}{2})^2 \leq \delta / 2 \, .
    \end{align}
    In particular, we will choose $\delta \leq \frac{1}{4}$, and have $0 \neq l \neq 1$. 
    
    Next, we show that $\mu$ is an $\eps$-Nash equilibrium of $\Gamma$. Consider any deviation strategy $\pi^{(i)}$ of player $i \in \pls$ in $\Gamma$. Then, we get
    \begin{flalign*}
        &\U^{(i)}(\pi^{(i)}, \mu^{(-i)}) - \U^{(i)}(\mu)&&
        \\
        &= \frac{2^t \Big( \frac{1}{4} - (l -\frac{1}{2})^2 \Big)^t}{2^t \Big( \frac{1}{4} - (l -\frac{1}{2})^2 \Big)^t} ( \U^{(i)}(\pi^{(i)}, \mu^{(-i)}) - \U^{(i)}(\mu) )&&
        \\
        &= \frac{ V^{(i)}(\pi^{(i)}, \mu^{(-i)},l) - V^{(i)}(\mu,l) }{ 2^t \Big( \frac{1}{4} - (l -\frac{1}{2})^2 \Big)^t }&&
        \\
        &\overset{(*)}{\leq} \frac{ \delta }{ 2^t \Big( \frac{1}{4} - (l -\frac{1}{2})^2 \Big)^t }&&
        % \\
    \end{flalign*}
    \begin{flalign*}
        &\overset{(\ref{nochance NE prob at Ic close to 1/2})}{\leq} \frac{ \delta }{ 2^t \Big( \frac{1}{4} - \frac{\delta}{2} \Big)^t }&&
        \\
        &\overset{(\dagger)}{\leq} \frac{ \delta }{ 2^{t} \Big( \frac{1}{4} - \frac{1}{8} \Big)^t }&&
        \\
        &= \frac{ \delta }{ 2^{t} \frac{1}{4^t} }&&
        \\
        &\overset{(\star)}{\leq} \eps \, ,&&
    \end{flalign*}
    
    where we use in $(*)$ that $(\mu,l)$ is a $\delta$-Nash equilibrium in $\Gamma'$, in $(\dagger)$ that we will choose $\delta \leq \frac{1}{4}$, and in $(\star)$ that we will choose $\delta \leq \frac{1}{2^t}\eps$. Hence, $\mu^{(i)}$ is an $\eps$-best response of player $i$ to $\mu^{(-i)}$ in $\Gamma$. All in all, if we set $\delta := \min \{\frac{1}{4}, \frac{1}{2^t}\eps \}$, then any $\delta$-Nash equilibrium $(\mu,l)$ in $\Gamma'$ induces $\mu$ to be an $\eps$-Nash equilibrium in $\Gamma$. 

    \paragraph{Approximate EDT:} \,
    \\
    Analogous reasoning as in approximate Nash. One merely has to consider each infoset $I$ in $\Gamma$ and only deviations $\mu_{I \mapsto \alpha}^{(i)}$.

    \paragraph{Exact CDT:} \,
    \\
    The KKT characterization \Cref{lem:CDT KKT and CLS} for game $\Gamma$ states that profile $\mu$ is an exact CDT equilibrium of $\Gamma$ if and only if there exist KKT multipliers $\{ \tau_{jk}^{(i)} \in \R \}_{i,j,k=1}^{\pls,\ninfs^{(i)},m_j^{(i)}}$ and $\{ \kappa_j^{(i)} \in \R \}_{i,j = 1}^{\pls,\ninfs^{(i)}}$ such that
    \begin{align}
    \label{no chance game KKT conditions}
        &\mu_{jk}^{(i)} \geq 0 \quad \forall i \in [N], \forall j \in [\ninfs^{(i)}], \forall k \in [m_j^{(i)}] \nonumber
        \\    
        &\sum_{k = 1}^{m_j^{(i)}} \mu_{jk}^{(i)} = 1 \quad \forall i \in [N], \forall j \in [\ninfs^{(i)}] \nonumber
        \\
        &\tau_{jk}^{(i)} \geq 0 \quad \forall i \in [N], \forall j \in [\ninfs^{(i)}], \forall k \in [m_j^{(i)}] \nonumber
        \\
        &\tau_{jk}^{(i)} = 0 \quad \textnormal{or} \quad \mu_{jk}^{(i)} = 0 \quad \forall i \in [N], \forall j \in [\ninfs^{(i)}], \forall k \in [m_j^{(i)}] \nonumber
        \\
        &\nabla_{jk} \, \U^{(i)}(\mu) + \tau_{jk}^{(i)} - \kappa_j^{(i)} = 0 \, \forall i \in [N], \forall j \in [\ninfs^{(i)}], \forall k \in [m_j^{(i)}] %\, .
    \end{align}

    The KKT characterization for a profile $(\mu,l)$ in game $\Gamma'$ is the same, except that we replace $\nabla_{jk} \, \U^{(i)}(\mu)$ in (\ref{no chance game KKT conditions}) with $\nabla_{jk} \, V^{(i)}(\mu,l)$, and that we need additional multipliers $\tau_l^-$ and $\tau_l^+$ such that 
    \begin{align}
    \label{no chance additional I_c KKT conditions}
        &l \geq 0  \quad \textnormal{and} \quad l \leq 1    \nonumber
        \\
        &\tau_l^- \geq 0  \quad \textnormal{and} \quad \tau_l^+ \geq 0  \nonumber
        \\
        &\tau_l^- = 0  \quad \textnormal{or} \quad l = 0    \nonumber
        \\
        &\tau_l^+ = 0  \quad \textnormal{or} \quad l = 1    \nonumber
        \\
        &\nabla_{l} \, V^{(i)}(\mu,l) + \tau_l^- - \tau_l^+ = 0 \, .
    \end{align} 

    We first observe that 
    \begin{align*}
        &\nabla_{l} \, V^{(i)}(\mu,l) 
        \\
        &= 2 (1 - 2l) + 2^t (1 - 2l) t \Big( \frac{1}{4} - (l -\frac{1}{2})^2 \Big)^{t-1} \U^{(i)}(\mu) 
        \\
        &= (1-2l)\bigg( 2 + 2^t t \Big( \frac{1}{4} - (l -\frac{1}{2})^2 \Big)^{t-1} \U^{(i)}(\mu) \bigg) \, .
    \end{align*}
    Here, using that utility $\U^{(i)}$ is positive, the second factor (the big bracket) will always be positive. Hence, after another look at the KKT conditions, boundary points $l=0$ and $l=1$ cannot satisfy the KKT conditions in $\Gamma'$ no matter the choice of $\tau_l^- \geq 0$ or, respectively, $\tau_l^+ \geq 0$. In the interior $(0,1)$, the KKT conditions on $l$ reduce to stationary condition $\nabla_{l} \, V^{(i)}(\mu,l) = 0$ which is only satisfied at $l = \frac{1}{2}$.

    Next, we observe that for indices $i,j,k$, we have
    \[
        \nabla_{jk} \, V^{(i)}(\mu,l) = 2^t \Big( \frac{1}{4} - (l -\frac{1}{2})^2 \Big)^t \nabla_{jk} \U^{(i)}(\mu) \, ,
    \]
    which, for $0 < l < 1$ implies that $\nabla_{jk} \, V^{(i)}(\mu,l)$ is simply a positive rescaling of $\U^{(i)}(\mu)$.
    
    Therefore, all in all, we get the equivalence that (1) a point $(\mu,l)$ satisfies the KKT conditions of $\Gamma'$ for multipliers $\{ \tau_{jk}^{(i)} \in \R \}_{i,j,k=1}^{\pls,\ninfs^{(i)},m_j^{(i)}}$, $\{ \kappa_j^{(i)} \in \R \}_{i,j = 1}^{\pls,\ninfs^{(i)}}$, $\tau_l^-$, and $\tau_l^+$ if and only if (2) $l = 1/2$, $\tau_l^- = 0 = \tau_l^+$, and $\mu$ satisfies the KKT conditions of $\Gamma$ for multipliers $\{ 2^t \tau_{jk}^{(i)} \in \R \}_{i,j,k=1}^{\pls,\ninfs^{(i)},m_j^{(i)}}$ and $\{ 2^t \kappa_j^{(i)} \in \R \}_{i,j = 1}^{\pls,\ninfs^{(i)}}$.

    \paragraph{Approximate CDT:} \,
    \\
    We will now show that given $(\Gamma,\eps)$, we can set $\tilde{\delta}>0$ sufficiently small but of size $\poly(\eps, |\Gamma|)$, such that if $(\tilde{\mu},\tilde{l})$ is a $\tilde{\delta}$-CDT equilibrium in $\Gamma'$, then we can compute a profile $\mu'$ from it in polytime such that $\mu$ is an $\eps$-CDT equilibrium in $\Gamma$. 

    First, we use \Cref{lem:approximation in CDT well supported and KKT} to transition from the $\tilde{\delta}$-CDT equilibrium in $\Gamma'$ to a $\delta$-well-supported CDT equilibrium $(\mu,l)$ in $\Gamma'$, where $\delta = 3 L_{\infty} |\nds| \sqrt{\tilde{\delta}}$ and $L_{\infty}$ is chosen as in \Cref{app:lps constant}. Next, we note that an approximate well-supported CDT equilibrium (for precision $\eps$ in $\Gamma$ or $\delta$ in $\Gamma'$) satisfies the exact KKT conditions above except that equality (\ref{no chance game KKT conditions}) is replaced by
    \[
        | \nabla_{jk} \, \U^{(i)}(\mu) + \tau_{jk}^{(i)} - \kappa_j^{(i)} | \leq \eps
    \]
    and that equality (\ref{no chance additional I_c KKT conditions}) is replaced by
    \[
        | \nabla_{l} \, V^{(i)}(\mu,l) + \tau_l^- - \tau_l^+ | \leq \delta \, .
    \]
    So let $\{ \tau_{jk}^{(i)} \in \R \}_{i,j,k=1}^{\pls,\ninfs^{(i)},m_j^{(i)}}$, $\{ \kappa_j^{(i)} \in \R \}_{i,j = 1}^{\pls,\ninfs^{(i)}}$, $\tau_l^-$, and $\tau_l^+$ be those KKT multipliers for $(\mu,l)$ in $\Gamma$. Then, we will show that multiplier $\{ 2^t \tau_{jk}^{(i)} \in \R \}_{i,j,k=1}^{\pls,\ninfs^{(i)},m_j^{(i)}}$ and $\{ 2^t \kappa_j^{(i)} \in \R \}_{i,j = 1}^{\pls,\ninfs^{(i)}}$ show that $\mu$ is an $\eps$-well-supported CDT equilibrium for $\Gamma$.
    
    First, we bound how far away $l$ can be from $\frac{1}{2}$. Analogous to the exact case, we will implicitly choose $\delta \leq 1$ and therefore, boundary points $l=0$ and $l=1$ cannot be the case in a $\delta$-well-supported CDT equilibrium. Hence, $\tau_l^- = 0 = \tau_l^+$ and the above inequality simplifies to
    \begin{align*}
        \delta &\geq | \nabla_{l} \, V^{(i)}(\mu,l) |
        \\
        &= |1-2l| \cdot \Big| 2 + 2^t t \Big( \frac{1}{4} - (l -\frac{1}{2})^2 \Big)^{t-1} \U^{(i)}(\mu) \Big|
        \\
        &\geq |1-2l| \cdot |2|
    \end{align*}
    that is
    
    \begin{align}
    \label{nochance CDT prob at Ic close to 1/2}
        (l -\frac{1}{2})^2 \leq \delta^2 / 4 \, .
    \end{align}
    Next, we bound 
    
    \begin{align*}
        0 &= \frac{1}{2^t} - 2^t (\frac{1}{4} - 0 )^t \leq \frac{1}{2^t} - 2^t \Big( \frac{1}{4} - (l -\frac{1}{2})^2 \Big)^{t}
        \\
        &\overset{(\ref{nochance CDT prob at Ic close to 1/2})}{\leq} \frac{1}{2^t} - 2^t \big( \frac{1}{4} - \frac{\delta^2}{4} \big)^{t} = \frac{1}{2^t} - 2^t  \frac{1}{4^t} ( 1 - \delta^2 )^{t}
        \\
        &\overset{(\star)}{\leq} \frac{1}{2^t} - \frac{1}{2^t} ( 1 - t \delta^2 ) = \frac{t}{2^t} \delta^2
    \end{align*}
    where in $(\star)$ we used Bernoulli's inequality. Hence
    
    \begin{align}
        \label{nochance reach close to 1/2}
        \Big| \frac{1}{2^t} - 2^t \Big( \frac{1}{4} - (l -\frac{1}{2})^2 \Big)^{t} \Big| \leq \frac{t}{2^t} \delta^2 \, .
    \end{align}
    
    Finally, we have all the tools to conclude that $\mu$ is an $\eps$-well-supported CDT equilibrium of $\Gamma$: Clearly, the domain and complementary conditions are still satisfied since those are the same in $\Gamma$ and $\Gamma'$, and because we simply rescaled the $\tau$ multiplier by a positive constant. Next, fix any coordinate $(i,j,k)$. Then, we get
    \begin{align*}
        &| \nabla_{jk} \, \U^{(i)}(\mu) + 2^t \tau_{jk}^{(i)} - 2^t \kappa_j^{(i)} |
        \\
        &\leq 2^t | \frac{1}{2^t} \nabla_{jk} \, \U^{(i)}(\mu) + \tau_{jk}^{(i)} - \kappa_j^{(i)} |
        \\
        &\leq 2^t \cdot \Big| \frac{1}{2^t} \nabla_{jk} \, \U^{(i)}(\mu) - 2^t \Big( \frac{1}{4} - (l -\frac{1}{2})^2 \Big)^t \nabla_{jk} \, \U^{(i)}(\mu) 
        \\
        &\, \quad \quad + 2^t \Big( \frac{1}{4} - (l -\frac{1}{2})^2 \Big)^t \nabla_{jk} \, \U^{(i)}(\mu)  + \tau_{jk}^{(i)} - \kappa_j^{(i)} \Big|
        \\
        &\leq 2^t \cdot \Big| \frac{1}{2^t} \nabla_{jk} \, \U^{(i)}(\mu) - 2^t \Big( \frac{1}{4} - (l -\frac{1}{2})^2 \Big)^t \nabla_{jk} \, \U^{(i)}(\mu)  \Big|
        \\
        &\, \quad \quad + 2^t \cdot \Big| 2^t \Big( \frac{1}{4} - (l -\frac{1}{2})^2 \Big)^t \nabla_{jk} \, \U^{(i)}(\mu)  + \tau_{jk}^{(i)} - \kappa_j^{(i)} \Big|
        \\
        &\leq 2^t \Big| \nabla_{jk} \, \U^{(i)}(\mu) \Big| \cdot \Big| \frac{1}{2^t}  - 2^t \Big( \frac{1}{4} - (l -\frac{1}{2})^2 \Big)^t  \Big|
        \\
        &\, \quad \quad + 2^t \cdot | \nabla_{jk} \, V^{(i)}(\mu,l)  + \tau_{jk}^{(i)} - \kappa_j^{(i)} |
        \\
        &\overset{(\ref{nochance reach close to 1/2}),(*)}{\leq} 2^t \big| \nabla_{jk} \, \U^{(i)}(\mu) \big| \cdot \frac{t}{2^t} \delta^2 + 2^t \delta
        \\
        &\leq \delta \Big( 2^t + t \delta \cdot \max_{\mu \in S} || \nabla \, \U^{(i)}(\mu) ||_{\infty} \Big)
        \\
        &\overset{(\dagger)}{\leq} \delta \Big( 2^t + t \cdot \max_{\mu \in S} L_{\infty} \Big)
        \\
        &\overset{(\star)}{\leq} \eps \, ,
    \end{align*}
    
    where we use in $(*)$ that $(\mu,l)$ is a $\delta$-well-supported CDT equilibrium in $\Gamma'$, in $(\dagger)$ that we will implicitly choose $\delta \leq 1$, and in $(\star)$ that we will implicitly choose $\delta \leq \frac{1}{2^t + t \cdot L_{\infty} }\eps$. All in all, if we set 
    \[
        \tilde{\delta} := \bigg( \frac{\min \{ 1 , \frac{\eps}{2^t + t \cdot L_{\infty} } \} }{ 3 L_{\infty}|\nds| } \bigg)^2 \  \, , 
    \]
    then any $\tilde{\delta}$-CDT equilibrium $(\tilde{\mu},\tilde{l})$ in $\Gamma'$ gives rise to a $\min \{ 1 , \frac{\eps}{2^t + t \cdot L_{\infty} } \}$-well-supported equilibrium $(\mu,l)$ in $\Gamma'$, which in turn induces $\mu$ to be an $\eps$-well-supported CDT equilibrium in $\Gamma$ and therefore, by \Cref{lem:approximation in CDT well supported and KKT}, an $\eps$-CDT equilibrium in $\Gamma$.
    
\end{proof}

We conclude with the main result of this section.

\begin{proof}[Proof of \Cref{{thm:restr to no chance possible},thm:remove chance}]
    Starting with a game $\Gamma$ with utility payoffs in the range of $[0,2]$, a precision parameter $\eps \geq 0$ and an equilibrium concept \emph{equilibrium} in \{Nash, EDT, CDT\}, apply \Cref{lem:get to one chance node} and then \Cref{prop:remove chance at root} on $\Gamma$ to get a game $\Gamma'$. Then $\Gamma'$ was constructed in polytime, and it has no chance nodes. It also has the same strategy set and game tree structure as $\Gamma$, except for one additional infoset $I_c$ at the beginning. $I_c$ has a degree of absentmindedness that is bounded by $2 \cdot \lceil \log|\term| \rceil = \mathcal{O}(\log|\nds|)$.

    For exact computational problems ($\eps = 0$) we have that $\mu$ is an equilibrium of $\Gamma$ if and only if $(\mu,\frac{1}{2})$ is an equilibrium in $\Gamma'$. For approximate computational problems ($\eps > 0$), we still have the correspondence with exact equilibria, but we can also choose $\delta >0$ as in \Cref{prop:remove chance at root} such that $\delta$-equilibria in $\Gamma'$ will be or give rise to $\eps$-equilibria of $\Gamma$. Note that $2^t = \poly(|\nds|)$ and that because of bounded utility payoffs in $\Gamma$, Lipschitz constant $L_{\infty}$ will be of size $\poly(|\nds|)$ as well. Thus, if $\eps$ was of \invpoly{} or \invexpo{} precision (in $|\Gamma$), then $\delta$ will continue to be so.
\end{proof}

\subsection{On Single-Player Games without Chance Nodes}
\label{app:SPIR w/o chance}

Recall that a Boolean formula $\phi$ is in conjunctive normal form (CNF) if it is a conjunction of a collection of $m$ clauses $C_1', \dots, C_m'$ each of which is a disjuntion of literals $\{ x_i, \neg x_i \}_{i}$. The problem {\sc MinSAT} takes a Boolean formula $\phi$ in CNF together with an integer threshold $0 \leq s^* \leq m$ as an instance, and asks whether there is a truth assignment for the variables in $\phi$ that satisfies at most $s^*$ clauses in $\phi$. The problem {\sc 2-MinSAT} restricts {\sc MinSAT} to those instances where each clause $C_j'$ of $\phi$ contains no more than two literals.
\begin{lemma}[\citet{KohliKM94}]
\label{lem:2minsat NPC}
     {\sc 2-MinSAT} is \NP{}-complete.
\end{lemma}
We will consider a variant of {\sc 2-MinSAT}. First, suppose a clause $C_j'$ uses the same variable $x$ in both of its literals. Then, it is either always satisfied ($x \lor \neg x)$ in which case it can be removed. Otherwise, it reduces to a singleton clause in which case it can be padded with a dummy variable $y$ that is only used in that clause. There are only at most $m$ such paddings needed, and for the minimization procedure it is sufficient to consider only those truth assignments that set $y$ to be false. Hence, with linear blowup we can assume w.l.o.g. that the {\sc 2-MinSAT} instances solely consist of clauses that use two distinct variables.

Next, observe that the negation $\neg \phi$ -- after distributing the negation into the clauses -- is a disjunction of the collection of clauses $C_1, \dots, C_m$ where $C_j$ is a conjunction of the negations of the literals in $C_j'$. Moreover, a truth assignment $\pi$ satisfies $M$ clauses in $\psi$ if and only if it satisfies (exactly the other) $m-M$ clauses in $\neg \phi$. Putting both of these together, we obtain from \Cref{lem:2minsat NPC}:
\begin{cor}
    The following problem {\sc 2-DNF-MaxSAT} is \NP{}-complete: Given a threshold DNF formula $\phi$ which uses exactly two distinct variables in each of its $m$ clauses, and given an integer threshold $0 \leq s^* \leq m$, does there exist a truth assignment for the variables in $\phi$ that satisfies at least $s^*$ clauses in $\phi$?
\end{cor}

We get to the main result of this section.

\begin{prop*}[Restatement of \Cref{prop:SPIR no chance still NP-hard}]
    {\sc Opt-D} is \NP{}-hard, even for games with no chance nodes, one infoset, a degree of absentmindedness of~$2$, and \invpoly{} precision.
\end{prop*}

\begin{proof}
    We reduce from {\sc 2-DNF-MaxSAT}. Let $(\phi,s^*)$ be one of its instances, that is, $\phi$ is a collection of clauses $C_1, \dots, C_m$ over variables $x_1,\dots,x_n$, where each clause is a conjunction of 2 literals of distinct variables. Construct a single-player game $\Gamma$ with imperfect recall from it as follows. It has one infoset $I$ with $2n$ actions $\{t_1,f_1,t_2,f_2,\dots,t_n,f_n\}$, where taking action $t_i$ or $f_i$ will correspond to setting $x_i$ to true or false respectively in a corresponding truth assignment. Root $h_0$ belongs to $I$, each of its $2n$ children belong to $I$, and each of their respective $2n$ children are terminal nodes. Hence, there are $4n^2$ terminal nodes in $\Gamma$. Each terminal node $z$ has a history that corresponds to setting some variable $x_i$ to truth value $v$, and setting some (possibly other) variable $x_{i'}$ to truth value $w$, where $v,w \in \{t,f\}$. The utility payoff at such $z$ shall be as follows:
    \begin{itemize}
    \item If $i=i'$, then $z$ yields a penalty payoff of $u(z) = -B$, where $B = (16mn^2)^3 \in \N$ is a sufficiently large value (but still polynomially large). We will later see that because of this penalty, the player will try to maximize the probability that the case $i \neq i'$ happens. That is, the player will be incentivized to allocate, for each $i$, approximately $1/n$ probability to the actions $t_i$ and $f_i$ together.
    \item If $i \neq i'$, then $x_i \neq x_{i'}$, hence the partial truth value assignment might already satisfy some clauses $C_j$ of $\phi$. Let $\calC(x_i \mapsto v, x_{i'} \mapsto w) \in \{0, 1, \dots, m\}$ be the number of such satisfied clauses. For example, a terminal node $z$ with history $(f_5, t_3)$ satisfies all the occurrences of the clauses $x_3 \land \neg x_5$ and $\neg x_5 \land x_3$ in formula $\phi$ (and no other clauses). Define the payoff $u(z)$ to be $\calC(x_i \mapsto v, x_{i'} \mapsto w)$. 
    \end{itemize}

    Finally, choose target value $t^* :=  -B\frac{1}{n} + 2 \frac{1}{n^2} s^*$ and precision $\eps := 2 \frac{1}{n^2} \cdot \frac{1}{4}$. This whole construction takes poly-time and $\eps$ indeed makes a \invpoly{} precision. 
    
    We claim that 
    \begin{enumerate}[leftmargin=1.5cm]
        \item[Claim 1:] if there is a truth assignment that satisfies at least $s^*$ clauses of $\phi$, then there  is also a strategy of $\Gamma$ with utility at least $t^*$,
        \item[Claim 2:] if $\Gamma$ has a strategy $\mu$ with utility $\geq t^* - \eps$, then from it, we can construct an assignment $\psi$ that satisfies at least $s^*$ clauses of $\phi$, and hence, $\Gamma$ admits a strategy with utility at least $t^*$.
    \end{enumerate}
    Those two claims imply that one can either achieve utility $t^*$ in $\Gamma$, or one cannot achieve $t^* - \eps$. In particular, $(\phi,s^*)$ will be a ``yes'' (and resp. ``no'') instance of {\sc 2-DNF-MaxSAT} if and only if the corresponding $(\Gamma,t^*, \epsilon)$ is a ``yes'' (and resp. ``no'') instance of {\sc Opt-D}. This concludes the reduction.

    \paragraph{Utility in $\Gamma$:} First, we characterize the utility function $\U$ of the single player in $\Gamma$. In general, a strategy $\mu$ contains action probabilities $\mu(x_i \mapsto f)$ and $\mu(x_i \mapsto t)$ on actions $f_i$ and $t_i$ respectively. Any such strategy can instead be described by values $p_i = \mu(x_i \mapsto f) + \mu(x_i \mapsto t) \in [0,1]$, which are the probabilities with which variables $x_i$ are chosen under $\mu$, and values $\alpha_i = \frac{\mu(x_i \mapsto f)}{p_i} \in [0,1]$, which are the fractions of times with which variable $x_i$ -- if chosen -- is set to false. If $p_i = 0$, then we can set $\alpha_i$ to an arbitrary value in $[0,1]$ instead. Since $\mu$ is a strategy, we have $\sum_i p_i = 1$. We get for any strategy $\mu$ the identity
    \begin{align*}
        \U(&\mu) = \sum_{i \in [n]} ( -B) \cdot \Big[ \mu(x_i \mapsto f) \cdot \mu(x_i \mapsto f) 
        \\
        &+ \mu(x_i \mapsto f) \cdot \mu(x_i \mapsto t) + \mu(x_i \mapsto t) \cdot \mu(x_i \mapsto f) 
        \\
        &+ \mu(x_i \mapsto t) \cdot \mu(x_i \mapsto t) + \Big]
        \\
        &+ \sum_{i \in [n]} \sum_{i' \neq i} \Big[ 
        \\
        &\mu(x_i \mapsto f) \cdot \mu(x_{i'} \mapsto f)  \cdot \calC(x_i \mapsto f, x_{i'} \mapsto f)
        \\
        &+ \mu(x_i \mapsto f) \cdot \mu(x_{i'} \mapsto t)  \cdot \calC(x_i \mapsto f, x_{i'} \mapsto t) 
        \\
        &+ \mu(x_i \mapsto t) \cdot \mu(x_{i'} \mapsto f)  \cdot \calC(x_i \mapsto t, x_{i'} \mapsto f)
        \\
        &+ \mu(x_i \mapsto t) \cdot \mu(x_{i'} \mapsto t)  \cdot \calC(x_i \mapsto t, x_{i'} \mapsto t) \Big] 
        \\
        &= -B\sum_{i \in [n]} p_i^2 + 2 \sum_{i \in [n]} \sum_{i' > i} p_i p_{i'} \Big[ 
        \\
        &\alpha_i \alpha_{i'}  \cdot \calC(x_i \mapsto f, x_{i'} \mapsto f)
        \\
        &+ \alpha_i (1-\alpha_{i'}) \cdot \calC(x_i \mapsto f, x_{i'} \mapsto t) 
        \\
        &+ (1-\alpha_i )\alpha_{i'}  \cdot \calC(x_i \mapsto t, x_{i'} \mapsto f)
        \\
        &+ (1-\alpha_i ) (1-\alpha_{i'})  \cdot \calC(x_i \mapsto t, x_{i'} \mapsto t) \Big]
        \\
        &= -B \sum_{i \in [n]} p_i^2 + 2 V(\mu) \, ,
    \end{align*}
    
    where $V$ stands for the second (big double) sum. We will later use the fact that
    \begin{align}
    \label{bounding V}
        0 \leq V(\pi) \leq \sum_{i \in [n]} \sum_{i' > i} 1 \cdot ( m + m + m + m ) \leq 4mn^2 \, .
    \end{align}

    \paragraph{Claim 1} Suppose $\psi$ is a truth value assignment for variables $x_1, \dots, x_n$. Let $s(\psi)$ be the number of clauses $\psi$ satisfies in $\phi$. Define $\psi$'s associated strategy in $\Gamma$ as 
    \[
        \mu_{\psi}(x_{i} \mapsto v) = 
        \begin{cases} 1/n &\text{if } \psi(x_i) = v \\ 0 &\text{if } \psi(x_i) = \neg v
        \end{cases} 
    \]
    for all $i \in [n]$ and $v \in \{f,t\}$. Then observe that $p_i(\mu_{\psi}) = 1/n$, and $\alpha_i(\mu_{\psi}) = \psi(x_i)$, and
    \begin{align}
    \label{V for truth assgnm}
    \begin{aligned}
        V(\mu_{\psi}) &= \sum_{i \in [n]} \sum_{i' > i'} \frac{1}{n^2} \calC(x_i \mapsto \psi(x_i), x_{i'} \mapsto \psi(x_{i'}))
        \\
        &= \frac{1}{n^2} s(\psi) \, .
    \end{aligned}
    \end{align}
    Hence,
    \begin{align*}
        \U(\mu_{\psi}) &= -B \frac{1}{n} + 2 \frac{1}{n^2} s(\psi) =  t^* + 2 \frac{1}{n^2} (s(\psi) - s^*)
    \end{align*}

    Therefore, overall, if there is truth value assignment $\psi$ for $\phi$ with $s(\psi) \geq s^*$, then $\mu_{\psi}$ will achieve a utility of at least $t^*$.

    \paragraph{Observation 1 for Claim 2} First, we show that in an (exactly) optimal strategy $\pi$ for $\Gamma$, the probabilities $p_i$ have distance at most 
    \begin{align}
    \label{delta choice in B}
        \delta := \sqrt{16mn^2 / B} = \frac{1}{16mn^2} < 1
    \end{align}
    from value $\frac{1}{n}$. That is because by optimality, it in particular performs better than a strategy $\mu$ defined as follows: Give it the same distribution $\alpha$, and almost the same distribution $q$ as $p$ in $\pi$. The only difference is that for $i^* \in \argmax_i p_i$ and $i_* \in \argmin_i p_i$, we define $q_{i^*} = (p_{i^*} + p_{i_*})/2 = q_{i_*}$  instead. Then
    \begin{align*}
        0 &\leq \U(\pi) - \U(\mu) = -B \sum_{i \in [n]} p_i^2 + 2 V(\pi) + B \sum_{i \in [n]} q_i^2 - 2 V(\mu) 
        \\
        &\overset{(\ref{bounding V})}{\leq} -B \sum_{i \in [n]} ( p_i^2 - q_i^2) + 2 \cdot 4mn^2 - 0
        \\
        &= -B \Big( p_{i^*}^2 + p_{i_*}^2 - 2 \cdot (p_{i^*} + p_{i_*})^2/4 \Big) + 8mn^2
        \\
        &= -B \Big( p_{i^*}^2 / 2 + - p_{i^*} p_{i_*} + p_{i_*}^2 / 2 \Big) + 8mn^2
        \\
        &= - \frac{1}{2} B ( p_{i^*} - p_{i_*})^2 + 8mn^2 \, ,
    \end{align*}
    which implies $( p_{i^*} - p_{i_*})^2 \leq 16mn^2 / B$, and hence,
    \[
        |p_{i^*} - p_{i_*}| \leq \sqrt{16mn^2 / B} = \delta \, .
    \]
    Note that $i^*$ and $i_*$ were chosen as extreme values, and thus,
    \[
        1/n = 1/n \sum_{i \in [n]} p_i \in [1/n \sum_{i \in [n]} p_{i_*} , 1/n \sum_{i \in [n]} p_{i^*}] = [p_{i_*}, p_{i^*}] \, ,
    \]
    where this interval has a length of at most $\delta$. Putting the last two derivations together, we obtain for any $i \in [n]$:
    \[  
        p_i \in [p_{i_*}, p_{i^*}] \subset [1/n - \delta, 1/n + \delta] \, .
    \]
    As another consequence, we want to observe for later that for $i, i' \in [n]$, we have
    \begin{align}
    \label{bounding pi and pi'}
    \begin{aligned}
        p_i p_{i'} &\leq (1/n + \delta)^2 = 1/n^2 + 2 \delta / n + \delta^2 
        \\
        &\leq 1/n^2 + 2 \delta + \delta = 1/n^2 + 3 \delta \, .
    \end{aligned}
    \end{align}

    \paragraph{Observation 2 for Claim 2} Next, we shall argue that $-B \sum_{i \in [n]} p_i^2$ is maximized at $p_i = 1/n \, \forall i$. Recall that for any strategy $\mu$ we have $p \in \Delta^{n-1}$. Hence, minimizing the term above is equivalent to $\max_{p  \in \Delta^{n-1}} ||p||_2^2$. This is a uniformly convex function over a convex, compact polytope, hence it attains its global minimum in the relative interior of the simplex (i.e. the inequality constraints are slack). A global minimum also satisfies the KKT conditions, and the KKT conditions for a relative interior point become that $p \in \Delta^{n-1}$ and that $p = \frac{1}{2} \kappa \1$ for some $\kappa \in \R$ and the vector $\1$ that consists of $1$'s in each entry. This condition is only satisfied at $p^* = \frac{1}{n} \1$. In particular, for all $p \in \Delta^{n-1}$, we therefore obtain 
    \begin{align}
    \label{bounding B term}
        -B \sum_{i \in [n]} p_i^2 \leq -B \sum_{i \in [n]} \Big( \frac{1}{n} \Big)^2 = -B \frac{1}{n} \, .
    \end{align}

    \paragraph{Claim 2} Now suppose $\Gamma$ has a strategy $\mu$ with utility $\geq t^* - \eps$. Then, an optimal strategy $\pi'$ also achieves $\U(\pi') \geq t^* - \eps$. Let us create another optimal strategy $\pi$ from $\pi'$ that satisfies $\alpha \in \{0,1\}^n$, that is, its distribution $\alpha$ to truth and false values make a proper truth value assignment of variables $x_1, \dots, x_n$. To that end, note that $\U(\pi')$ is linear in $\alpha_i'$ of $\pi'$ for any given values $p$ and $\alpha_{-i}'$, and hence it is maximized at the boundary $\alpha_i' = 0$ or $\alpha_i' = 1$. Therefore, starting from $\pi'$, we can iterative over $i = 1, \dots, n$ and set $\alpha_i$ to one of these extreme values without decreasing the utility value. Denote the resulting strategy with $\pi$. It is also optimal for $\Gamma$ and its $\alpha \in \{0,1\}^n$. For that strategy, we can derive
    \begin{align*}
        &-B\frac{1}{n} + 2 \frac{1}{n^2} (s^* - \frac{1}{4}) =  t^* - \eps \leq \U(\pi') 
        \\
        &= \U(\pi)  = -B \sum_{i \in [n]} p_i^2 + 2 V(\pi) \overset{(\ref{bounding B term})}{\leq}  -B \frac{1}{n} + 2 V(\pi) 
        \\
        &\overset{(\ref{V for truth assgnm})}{=} -B \frac{1}{n} + 2 \sum_{i \in [n]} \sum_{i' > i'} p_i p_{i'} \calC(x_i \mapsto \alpha_i, x_{i'} \mapsto \alpha_{i'})
        \\
        &\overset{(\ref{bounding pi and pi'})}{=} -B \frac{1}{n} + 2 \sum_{i \in [n]} \sum_{i' > i'} (1/n^2 + 3 \delta) \cdot \calC(x_i \mapsto \alpha_i, x_{i'} \mapsto \alpha_{i'})
        \\
        &\overset{(\ref{V for truth assgnm})}{=} -B \frac{1}{n} + 2 \cdot (1/n^2 + 3 \delta) \cdot s(\alpha) \, .
    \end{align*}

    Rearranging yields
    \begin{align*}
        s^* &\leq \frac{1}{4} + s(\alpha) + s(\alpha) \cdot 3 \delta n^2 \leq s(\alpha) + \frac{1}{4} + \delta \cdot 3 m n^2
        \\
        &\overset{(\ref{delta choice in B})}{=} s(\alpha) + \frac{1}{4} + \frac{1}{16mn^2} \cdot 3 m n^2 \leq s(\alpha) + \frac{1}{2} \, .
    \end{align*}
    Since both $s^*$ and $s(\alpha)$ are integers, this can only be the case if $s(\alpha) \geq s^*$, that is, there is a truth value assignment that satisfies at least $s^*$ clauses of $\phi$. Using Claim 1 for $\psi = \alpha$, we obtain that there must also be a strategy in $\Gamma$ that achieves a utility of at least $t^*$ in $\Gamma$.
\end{proof}

\begin{cor}
    It is \NP{}-hard to distinguish between whether all EDT equilibria $\mu$ in a single-player game have an utility $\U^{(1)}(\mu) \geq t$ from whether there is an EDT equilibrium $\mu$ that satisfies $\U^{(1)}(\mu) \leq t - \epsilon$. Hardness holds even for games with no chance nodes, one infoset, a degree of absentmindedness of~$2$, and \invpoly{} precision.
\end{cor}
\begin{proof}
    This follows from \Cref{prop:SPIR no chance still NP-hard}, and from the fact that in single-infoset games all EDT equilibria are optimal strategies due to \Cref{rem: 1 infoset yields EDT eq NE}.
\end{proof}

\end{document}